\documentclass[a4paper,11pt]{article}
\usepackage{jheppub} 
\usepackage{nicefrac}
\usepackage{bm}
\usepackage{booktabs}
\usepackage{multirow}
\usepackage[normalem]{ulem}
\usepackage{comment}
\allowdisplaybreaks
\usepackage[frozencache,cachedir=.]{minted}
\usepackage{listings}
\usepackage{comment}

\lstdefinestyle{mystyle}{
    backgroundcolor=\color{backcolour},   
    commentstyle=\color{codegreen},
    keywordstyle=\color{magenta},
    numberstyle=\tiny\color{codegray},
    stringstyle=\color{codepurple},
    basicstyle=\ttfamily\footnotesize,
    breakatwhitespace=false,         
    breaklines=true,                 
    captionpos=b,                    
    keepspaces=true,                 
    numbers=left,                    
    numbersep=5pt,                  
    showspaces=false,                
    showstringspaces=false,
    showtabs=false,                  
    tabsize=2
}

\newcommand{\Vcb}{\ensuremath{|V_{cb}|} }
\newcommand{\mbMS}{\ensuremath{\overline{m}_{b}(\overline{m}_{b})}}
\newcommand{\mcMS}{\ensuremath{\overline{m}_{c}(3\, \mathrm{GeV})} }
\newcommand{\alphasZ}{\ensuremath{\alpha_{s}(M_{Z^0})} }
\newcommand{\scalembkin}{\ensuremath{\mu_{m_{b}^\mathrm{kin}}} }
\newcommand{\scalemcMS}{\ensuremath{\mu_{\overline{m}_{c} }}} 
\newcommand{\scaleas}{\ensuremath{\mu_{\alpha_s}} }
\newcommand{\mupi}{\ensuremath{\mu_{\pi}^2(1\,\mathrm{GeV})} }
\newcommand{\muG}{\ensuremath{\mu_{G}^2}(2.3\,\mathrm{GeV}) }
\newcommand{\rhoD}{\ensuremath{\rho_{D}^3(1\,\mathrm{GeV})} }
\newcommand{\rhoLS}{\ensuremath{\rho_{LS}^3} }

\newcommand{\mbkin}{\ensuremath{m_{b}^\mathrm{kin}} }

\newcommand{\Xclnu}{\ensuremath{\bar{B} \to X_c l \bar{\nu}_l} }

\usepackage{cancel}
\usepackage{enumitem}
\usepackage{xcolor}

\definecolor{rawnumcolor}{RGB}{200,0,0}

\newcommand{\numVcb}{42.32}
\newcommand{\numVcbErr}{0.28}
\newcommand{\numrhoD}{0.110}
\newcommand{\numscalembkin}{1.216}
\newcommand{\numscalemcMS}{1.860}
\newcommand{\numscaleas}{9.13}

\newcommand{\numPvalueDefault}{\ensuremath{8.4\times10^{-5}}}
\newcommand{\numPvalueNoqsq}{0.0996}
\newcommand{\numPvalueNoBabarEl}{0.0372}
\newcommand{\numPvalueNoBelleMX}{0.0112}
\newcommand{\numPvalueNoBelleqsq}{0.2257}

\newcommand{\numPvalueFixedDefault}{\ensuremath{1.4\times10^{-8}}}
\newcommand{\numPvalueFixedNoBelleqsq}{\ensuremath{3.8\times10^{-4}}}

\newcommand{\numScaleasMbThreeRatio}{8.0}
\newcommand{\numRhoDMbThreeRatio}{3.7}
\newcommand{\numRelSpreadVcb}{0.2}
\newcommand{\numRelSpreadMupi}{4.8}
\newcommand{\numRelSpreadMuG}{43.3}
\newcommand{\numRelStepRhoD}{13}

\newcommand{\numBoxMbkinHigh}{1.25}
\newcommand{\numLimitMbkinHigh}{1.30}
\newcommand{\numBoxMcMSLow}{1.80}
\newcommand{\numLimitMcMSLow}{1.50}

\newcommand{\numGammaSLBzero}{0.15440}
\newcommand{\numGammaSLBzeroTotalUp}{0.00280}
\newcommand{\numGammaSLBzeroTotalDown}{0.00279}
\newcommand{\numGammaNLBzero}{0.4188}
\newcommand{\numGammaNLBzeroFitUp}{0.0049}
\newcommand{\numGammaNLBzeroFitDown}{0.0047}
\newcommand{\numGammaNLBzeroScaleUp}{0.0805}
\newcommand{\numGammaNLBzeroScaleDown}{0.0036}
\newcommand{\numGammaNLBzeroBagUp}{0.0020}
\newcommand{\numGammaNLBzeroBagDown}{0.0020}
\newcommand{\numGammaNLBzeroHO}{0.0061}
\newcommand{\numGammaNLBzeroTotalUp}{0.0809}
\newcommand{\numGammaNLBzeroTotalDown}{0.0087}
\newcommand{\numTauInvBzero}{0.5732}
\newcommand{\numTauInvBzeroTotalUp}{0.0812}
\newcommand{\numTauInvBzeroTotalDown}{0.0111}
\newcommand{\numTauBzero}{1.7446}
\newcommand{\numTauBzeroTotalUp}{0.0343}
\newcommand{\numTauBzeroTotalDown}{0.2262}
\newcommand{\numTauBzeroShort}{1.74}
\newcommand{\numTauBzeroShortUp}{0.03}
\newcommand{\numTauBzeroShortDown}{0.23}
\newcommand{\numGammaNLBplus}{0.3754}
\newcommand{\numGammaNLBplusFitUp}{0.0044}
\newcommand{\numGammaNLBplusFitDown}{0.0041}
\newcommand{\numGammaNLBplusScaleUp}{0.0637}
\newcommand{\numGammaNLBplusScaleDown}{0.0034}
\newcommand{\numGammaNLBplusBagUp}{0.0050}
\newcommand{\numGammaNLBplusBagDown}{0.0048}
\newcommand{\numGammaNLBplusHO}{0.0055}
\newcommand{\numGammaNLBplusTotalUp}{0.0643}
\newcommand{\numGammaNLBplusTotalDown}{0.0091}
\newcommand{\numTauInvBplus}{0.5298}
\newcommand{\numTauInvBplusTotalUp}{0.0646}
\newcommand{\numTauInvBplusTotalDown}{0.0113}
\newcommand{\numTauBplus}{1.8875}
\newcommand{\numTauBplusTotalUp}{0.0407}
\newcommand{\numTauBplusTotalDown}{0.2151}
\newcommand{\numTauBplusShort}{1.89}
\newcommand{\numTauBplusShortUp}{0.04}
\newcommand{\numTauBplusShortDown}{0.22}
\newcommand{\numGammaClnu}{0.06909}
\newcommand{\numGammaClnuFitUp}{0.00073}
\newcommand{\numGammaClnuFitDown}{0.00073}
\newcommand{\numGammaClnuHO}{0.00130}
\newcommand{\numGammaUlnu}{0.9560}
\newcommand{\numGammaUlnuFitUp}{0.0054}
\newcommand{\numGammaUlnuFitDown}{0.0050}
\newcommand{\numGammaUlnuHO}{0.0162}
\newcommand{\numGammaCtaunu}{0.01430}
\newcommand{\numGammaCtaunuFitUp}{0.00016}
\newcommand{\numGammaCtaunuFitDown}{0.00017}
\newcommand{\numGammaCtaunuHO}{0.00035}
\newcommand{\numTauRatio}{1.0819}
\newcommand{\numTauRatioFitUp}{0.0007}
\newcommand{\numTauRatioFitDown}{0.0007}
\newcommand{\numTauRatioScaleUp}{0.0277}
\newcommand{\numTauRatioScaleDown}{0.0026}
\newcommand{\numTauRatioBagUp}{0.0128}
\newcommand{\numTauRatioBagDown}{0.0132}
\newcommand{\numTauRatioHO}{0.0010}
\newcommand{\numTauRatioTotalUp}{0.0306}
\newcommand{\numTauRatioTotalDown}{0.0135}
\newcommand{\numTauRatioShort}{1.082}
\newcommand{\numTauRatioShortUp}{0.031}
\newcommand{\numTauRatioShortDown}{0.013}
\newcommand{\numBRTheoryShort}{12.55}
\newcommand{\numBRTheoryShortUp}{0.11}
\newcommand{\numBRTheoryShortDown}{1.50}
\newcommand{\numTauBzeroScaleRelDown}{12.8}
\newcommand{\numTauRatioScaleRelUp}{2.6}

\newcommand{\numBRMbThree}{10.906}
\newcommand{\numBRMbFour}{10.912}
\newcommand{\numBRMbFive}{10.701}
\newcommand{\numBRSigmaHO}{0.21}

\newcommand{\numWidthDeficitBzero}{13}
\newcommand{\numWidthPullBzero}{1.1}
\newcommand{\numEgnerPullBzero}{0.7}
\newcommand{\numWidthPullBplus}{1.2}
\newcommand{\numEgnerPullBplus}{0.8}

\newcommand{\numVcbSigmaHO}{0.10}
\newcommand{\numVcbTotal}{0.30}
\newcommand{\numBRCentral}{10.91}
\newcommand{\numBRErr}{0.11}

\newcommand{\asym}[3]{\ensuremath{#1^{+#2}_{-#3}}}

\renewcommand\vec{\boldsymbol}

\title{\boldmath Inclusive $B$ decays, lifetimes, and $|V_{cb}|$: a unified heavy-quark expansion analysis}

\author[a,b]{Matteo Fael,}
\author[c]{Ilija S. Milutin,}
\author[d]{Markus Prim,}
\author[e,f]{K. Keri Vos}

\affiliation[a]{
Dipartimento di Fisica e Astronomia ``G.~Galilei,'' Università di Padova, \\
via F.~Marzolo 8, 35131 Padova, Italy
}
\affiliation[b]{
Istituto Nazionale di Fisica Nucleare, Sezione di Padova, \\
via F.~Marzolo 8, 35131 Padova, Italy
}
\affiliation[c]{Theoretische Physik 1,
  Naturwiss.\ techn.\ Fakult\"at, Universit\"at Siegen, 
  57068 Siegen, Germany}
\affiliation[d]{Physikalisches Institut, Nußallee 12, University of Bonn, 53115 Bonn, Germany}
\affiliation[e]{Gravitational 
Waves and Fundamental Physics (GWFP),
Maastricht University, Duboisdomein 30,
NL-6229 GT Maastricht, the
Netherlands}
\affiliation[f]{Nikhef, Science Park 105,
NL-1098 XG Amsterdam, the Netherlands}

\emailAdd{matteo.fael@pd.infn.it, 
Ilija.Milutin@uni-siegen.de,
markus.prim@uni-bonn.de,
k.vos@maastrichtuniversity.nl}

\abstract{

We present a global HQE analysis of inclusive semileptonic $\bar{B} \to X_c l \bar \nu_\ell$ moments, partial branching fractions, and $B^+$, $B^0$ lifetime observables. Our fit setup follows a new approach where perturbative scales are treated as correlated nuisance parameters, while missing higher-order power corrections are assessed through order-by-order fits up to $O(1/m_b^5)$.
In the default scenario, we find
$
    |V_{cb}| = (\numVcb \pm \numVcbErr|_{\rm fit} \pm \numVcbSigmaHO|_{\rm h.o.})\times 10^{-3} = (\numVcb \pm \numVcbTotal) \times 10^{-3}
$
and 
$
\mathcal{B}(\bar B \to X_c \ell \bar \nu_\ell) = (\numBRCentral \pm \numBRErr|_{\rm fit} \pm \numBRSigmaHO|_{\rm h.o.})\%
$. 
For the lifetimes we obtain $\tau_{B^+} = \asym{\numTauBplusShort}{\numTauBplusShortUp}{\numTauBplusShortDown}$~ps and $\tau_{B^0} = \asym{\numTauBzeroShort}{\numTauBzeroShortUp}{\numTauBzeroShortDown}$~ps, where the strongly asymmetric uncertainty is dominated by the perturbative scale variation of the non-leptonic width. Our predictions are in agreement with previous predictions and the corresponding widths lie $\numWidthPullBzero\text{--}\numWidthPullBplus\,\sigma$ below the measurements. For the lifetime ratio, we find $\tau_{B^+}/\tau_{B^0} = \asym{\numTauRatioShort}{\numTauRatioShortUp}{\numTauRatioShortDown}$ which agrees with experiment.
For the full dataset of $q^2, E_\ell$ and $M_X$ moments, we obtain a poor fit quality, which we investigate using a leave-one-out study. In light of the persisting $|V_{cb}|$ puzzle, new correlated moment measurements as well as updated branching ratio measurements are highly encouraged. 
}

\begin{document}
\maketitle
\flushbottom

\section{Introduction}

Inclusive decays of heavy hadrons provide a unique laboratory for testing the
Standard Model (SM) and for determining fundamental parameters of the flavour 
sector. In particular, inclusive semileptonic decays of $B$ mesons play a 
central role in the extraction of the Cabibbo-Kobayashi-Maskawa (CKM)
matrix element $|V_{cb}|$, as well as in constraining non-perturbative heavy-quark dynamics. 
The theoretical description of these processes is based on the Heavy Quark Expansion (HQE), 
which allows decay rates and kinematic moments to be expressed as a double expansion 
in the strong coupling constant $\alpha_s$ 
and inverse powers of the bottom-quark mass $m_b$~\cite{Bigi:1992su,Bigi:1993fe,Blok:1993va,Manohar:1993qn}.

Over the past two decades, global fits of inclusive semileptonic moments~\cite{Bauer:2004ve,Gambino:2013rza,Alberti:2014yda,Bordone:2021oof,Bernlochner:2022ucr,Finauri:2023kte,Carvunis:2025vab}
have achieved remarkable precision, benefiting from high-quality experimental 
measurements from CDF~\cite{CDF:2005xlh}, DELPHI~\cite{DELPHI:2005mot}, CLEO~\cite{CLEO:2004bqt}, BaBar~\cite{BaBar:2009zpz}, Belle~\cite{Belle:2006jtu,Belle:2006kgy,Belle:2021idw} and Belle~II~\cite{Belle-II:2022evt} and continuous progress in perturbative QCD 
calculations~\cite{Czarnecki:1997sz,Aquila:2005hq,Dassinger:2006md,Pak:2008qt,Pak:2008cp,Dowling:2008mc,Biswas:2009rb,Mannel:2010wj,Alberti:2012dn,Alberti:2013kxa,Fael:2018vsp,Fael:2020iea,Fael:2020njb,Fael:2020tow,Fael:2022frj,Mannel:2023yqf,Egner:2023kxw,Mannel:2021zzr,Fael:2024gyw,Finauri:2025ost}. 
These analyses have established the HQE as a quantitatively successful framework 
and have led to precise determinations of $|V_{cb}|$ and of the leading 
non-perturbative HQE parameters. 

At the same time, the steady improvement of theoretical accuracy together with
new experimental data for the $q^2$ moments~\cite{Belle:2021idw,Belle-II:2022evt} has exposed 
tensions among different moment measurements and has raised questions about the
treatment of theoretical uncertainties, 
the convergence of the HQE, and the  robustness of global-fit strategies.

Previous fits typically modelled missing $\alpha_s$ and $1/m_b$ corrections by introducing ad-hoc 
uncertainties on a subset of HQE matrix elements. 
This strategy was acceptable in theoretical setups with limited perturbative accuracy. 
However, with the inclusion of increasingly higher-order QCD corrections, 
this approach has become difficult to justify. 
The residual perturbative uncertainty estimated from scale variations 
is now often subdominant compared to the imposed ad-hoc uncertainties, 
which masks the true sensitivity of the data to higher-order power corrections 
and complicates a meaningful assessment of the convergence and validity of the HQE.

In this work, we pursue a complementary strategy. Rather than assigning additional
theoretical uncertainties for missing higher-order power corrections, we perform global fits 
at \textit{fixed order} of the HQE and study the order-by-order convergence of 
the extracted parameters.
Starting at $\mathcal{O}(1/m_b^3)$ and extending up to $\mathcal{O}(1/m_b^5)$, we repeat the fit
while keeping the perturbative accuracy. 
If the HQE is well behaved, the fitted parameters are expected to stabilize 
as higher-order terms are included, providing a direct test of 
convergence without the need for ad-hoc uncertainty models.

A second new aspect of our analysis is the treatment of perturbative scale dependence. 
Instead of estimating perturbative uncertainties through independent scale 
variations for each observable, we treat the relevant unphysical scales as 
nuisance parameters and include them directly in the fit (see also Ref.~\cite{Tackmann:2024kci}). 
This procedure reflects the fact that scale variations induce coherent shifts 
across all inclusive observables, as they originate from the same underlying 
differential decay rates. 
The impact of these scales is then assessed by profiling over them, 
allowing the data themselves to constrain the \textit{preferred} scale regions.

Treating the scales as fitted nuisance parameters captures the logarithmic, 
re\-nor\-mal\-i\-sa\-tion-group-predictable part of the missing higher-order terms, 
however non-logarithmic constants or the new cut dependence that genuinely 
arise at the next order are not captured. 

In addition to semileptonic decay observables, we extend the analysis 
to include $B$ meson lifetimes, 
which are also described within the HQE framework but receive sizeable contributions 
from four-quark operators. 
Comparison with the lifetimes provides an important consistency check of the HQE parameters extracted from semileptonic moments and allows us to explore correlations between different decay modes of the $B$ meson. 
To this end, we implement lifetime predictions presented in Ref.~\cite{Egner:2024lay} in the publicly available Python package \texttt{kolya}~\cite{Fael:2024fkt}, 
which we employ throughout this work to compute rates and moments in the kinetic scheme.
The focus of this paper is the following: 
\begin{enumerate}
\item We perform a comprehensive global analysis of inclusive semileptonic 
moments and partial rates employing the HQE predictions up to $\mathcal{O}(1/m_b^5)$.  
\item We assess the internal consistency of the experimental dataset using leave-one-out validation techniques and study the impact of individual measurements on the fit quality and extracted parameters.
\item We investigate the convergence properties of the HQE and the interplay between semileptonic observables and lifetimes within a unified fitting framework.
\end{enumerate}

The paper is organised as follows. In Sec.~\ref{sec:semilep} we summarise the HQE for inclusive semileptonic decays, in Sec.~\ref{sec:lifetimes} we discuss the setup for the lifetimes. Section~\ref{sec:fitting} describes the fit setup and the treatment of the perturbative uncertainties and higher-order power corrections. The fit results are presented in Sec.~\ref{sec:fit_results}. Here we discuss in detail the leave-one-out assessment of the internal consistency of the dataset and the order-by-order study of HQE convergence. The lifetime predictions are given in Sec.~\ref{sec:lifetimepred}. We conclude in Sec.~\ref{sec:conl}. The definitions of the $1/m_b^{4,5}$ matrix elements, the implementation of the lifetimes in \texttt{kolya}, and the renormalisation-group evolution of $\mu_G^2$, are collected in the appendices.

\section{General theory setup}
\subsection{Semileptonic decay}\label{sec:semilep}
We consider the inclusive semileptonic decay
\begin{equation}
        \bar B(p_B) \to X_c(p_X) l(p_l) \bar \nu_l(p_\nu) \text{ with } l = e, \mu,
\end{equation}
where we define the momentum of the lepton pair as $q=p_l+p_\nu$. The total semileptonic rate $\Gamma_{\rm sl}$ and the spectral moments $\langle (O)^n\rangle$ can be obtained by a phase-space integration of the differential rate, schematically represented as
\begin{align}
    [ (O)^n] =\int\text{d}\Phi\, (O)^nW^{\mu\nu}L_{\mu\nu}\ , 
\end{align}
where $O$ is the kinematic variable of interest (raised to an integer power $n$), which in our case will be the dilepton invariant mass $q^2$, lepton energy $E_l$ and hadronic invariant mass $M_X^2=(M_Bv -q)^2$, where $v^\mu$ is defined as $v=p_B/M_B$.  To obtain the prediction for the total semileptonic rate $\Gamma_{\rm sl}$, one sets $O$ to 1. The leptonic tensor $L^{\mu\nu}$ and hadronic tensor $W^{\mu\nu}$ are integrated over the phase space $\text{d}\Phi$.

In the Standard Model, this decay is described by the effective Hamiltonian,
\begin{align}
    \mathcal{H}_{\rm eff}^{\rm SL}=\frac{4G_F}{\sqrt{2}}V_{\rm cb}\left(\bar{c}\gamma^\mu P_Lb\right)\left(\bar{l}\gamma_\mu P_L\nu_l\right)+\rm{h.c.}\ ,\label{eq:SLham}
\end{align}
where $P_L=(1-\gamma^5)/2$. The hadronic tensor $W^{\mu\nu}$ can be obtained through the optical theorem, which relates it to the imaginary part of the forward scattering amplitude of two hadronic currents:
\begin{align}
    W^{\mu\nu}&=\sum\limits_{X_c}(2\pi)^4\delta^4(p_B-q-p_X)\langle B|\bar{b}\gamma^\mu P_Lc|X_c\rangle\langle X_c|\bar{c}\gamma^\nu P_Lb|B\rangle\nonumber\\
    &=2\,\text{Im}\, \langle B|i\int\text{d}^4x\, e^{-im_b S\cdot x}T\left\{\bar{b}_v(x)\gamma^\mu P_Lc(x)\bar{c}(0)\gamma^\nu P_Lb_v(0)\right\}|B\rangle\ ,\label{eq:hadr_tens}
\end{align}
where $S=v-q/m_b$ and $b_v(x)=\text{exp}(im_bv\cdot x)b(x)$ is the re-phased $b$-quark field. The Heavy Quark Expansion (HQE) is obtained by applying an Operator Product Expansion (OPE) to the time-ordered product in \eqref{eq:hadr_tens}. We obtain the HQE by splitting the $b$-quark momentum as $p_b=m_bv+k$, and then
expanding the hadronic tensor in the small residual momentum $k\sim \Lambda_{\rm QCD}\sim iD$, yielding operators written as chains of covariant derivatives. For details, see for example Refs.~\cite{Manohar:1993qn,Blok:1993va}. The HQE allows for a prediction of the observables as a series in powers of $1/m_b$, which can be schematically denoted by
\begin{align}
    \sum\limits_{k=0}^{\infty}\frac{C_{\mu_1\dots\mu_k}^{(k)}}{m_b^{k}}\otimes\langle B|\bar{b}_v(iD^{\mu_1}\dots iD^{\mu_k})b_v|B\rangle\ ,
\end{align}
where $\otimes$ denotes the correct contraction of the spinor indices. The coefficients $C^{(k)}_{\mu_1\dots\mu_k}$ describe the short-distance effects, which can be computed in perturbative QCD as a series in the strong coupling constant $\alpha_s$. The operators $\langle B|\bar{b}_v(iD^{\mu_1}\dots iD^{\mu_k})b_v|B\rangle$ contain the non-perturbative long-distance physics, encoded into the HQE parameters. We will employ the notation $\langle \bar{b}_v\,...\,b_v\rangle\equiv\langle B(v)|\bar{b}_v\,...\,b_v|B(v)\rangle$ from here on. At leading order ($k=0$), we obtain the partonic result, while the $1/m_b$ contribution vanishes due to Heavy Quark Symmetry. The first HQE parameters arise at order $1/m_b^2$. Throughout this paper, we will employ the \textit{historical basis} for the HQE parameters (sometimes referred to as the ``perp'' basis). 

This basis is defined using the spatial covariant derivative $iD^\perp_\mu=g^\perp_{\mu\nu}iD^\nu$, where $ g^\perp_{\mu\nu}=g_{\mu\nu}-v_\mu v_\nu$ as in Refs.~\cite{Mannel:2010wj, Gambino:2016jkc}. At $1/m_b^2$, we have
\begin{align}
    2M_B(\mu_\pi^2)^\perp&=-\langle \bar{b}_v\, (iD^\rho)\, (iD^\sigma)\, b_v\rangle\, g_{\rho\sigma}^\perp\ ,\nonumber\\
    2M_B(\mu_G^2)^\perp&=\frac{1}{2}\langle\bar{b}_v \, \big[(iD^\rho),\, (iD^\sigma)\big]\, (-i\sigma^{\alpha\beta})\, b_v\rangle\, g_{\rho\alpha}^\perp g_{\sigma\beta}^\perp\ ,
\end{align}
where $\gamma^\mu\gamma^\nu=g^{\mu\nu}+(-i\sigma^{\mu\nu})$, which are known as the kinetic and chromomagnetic operator respectively. At $1/m_b^3$, we have the Darwin and spin-orbit operators:
\begin{align}
    2M_B(\rho_D^3)^\perp&= \frac{1}{2}\langle \bar{b}_v\, \Big[(iD^\rho),\, \big[(iD^\sigma),\, (iD^\lambda)\big]\Big]\,b_v\rangle\, g^\perp_{\rho\lambda} v_\sigma\ ,\nonumber\\
    2M_B(\rho_{LS}^3)^\perp&=\frac{1}{2}\langle \bar{b}_v\, \Big\{(iD^\rho),\, \big[(iD^\sigma),\, (iD^\lambda)\big]\Big\}\, (-i\sigma^{\alpha\beta})\,b_v\rangle\, g^\perp_{\rho\alpha}g^\perp_{\lambda\beta}v_\sigma\ .
\end{align}
The definitions of the HQE matrix elements at $1/m_b^4$ and $1/m_b^5$ can be found in Appendix \ref{app:HQE_def}. Since in this work, we only employ this basis, we will drop the superscript ``$\perp$'' when referring to the HQE parameters from here on. 

It is also possible to use another basis of HQE elements, based on Reparametrization Invariance (RPI). This basis exploits the fact that the choice of the four-velocity $v$, used to set up the HQE, is not unique. By applying the RPI transformation $v\to v'=v+\delta v$ with $v\cdot \delta v=0$, combinations of HQE parameters can be constructed which are RPI, which reduces the number of relevant parameters. These parameters have been identified in Ref.~\cite{Mannel:2018mqv} up to $\mathcal{O}(1/m_b^4)$ and in Ref.~\cite{Mannel:2023yqf} up to $\mathcal{O}(1/m_b^5)$. However, the reduction of the number HQE parameters only comes into effect for RPI quantities like the total rate $\Gamma$ and the $q^2$ moments. Since the $E_l$ and $M_X^2$ moments are not RPI, the full basis is required to describe them. Therefore, we use the historical basis for the HQE parameters.\\

Below, we extract the non-perturbative HQE parameters from experimental data. In the future, lattice QCD may also be able to predict these matrix elements or the $B\to X_c \ell\nu$ spectrum directly, see Refs.~\cite{Gambino:2017vkx,Gambino:2020crt,DeSantis:2025yfm,Kellermann:2025pzt}. An estimate of the size of the HQE parameters at $\mathcal{O}(1/m_b^4)$ and $\mathcal{O}(1/m_b^5)$ can be obtained using the ``lowest-lying state saturation ansatz'' (LLSA), as discussed in Ref.~\cite{Heinonen:2014dxa} and employed in the fit in Ref.~\cite{Gambino:2016jkc}. The LLSA assumes that the higher order matrix elements can be expressed in terms of $\mu_\pi^2, \mu_G^2, m_b$ and the excitation energies $\epsilon_{1/2}=M_{1/2}-M_B$ and $\epsilon_{3/2}=M_{3/2}-M_B$. For more details on the LLSA, we refer to Ref.~\cite{Heinonen:2014dxa} (see also Ref.~\cite{Mannel:2023yqf}). Perturbative corrections to these relations are not known. As only the lowest-lying state is included, theory uncertainties of the estimates may be 50\% or higher \cite{Heinonen:2014dxa}, and the correlations between these estimates are unknown. In Ref.~\cite{Heinonen:2014dxa}, only the leading order $1/m_b$ contributions are included for the estimates of $m_i$ and $r_i$. In Ref.\cite{Mannel:2023yqf}, the $1/m_b$ corrections were studied for the RPI-subset of HQE parameters. We apply this method to estimate the LLSA predictions for the HQE parameters in the historical basis. In Sec.~\ref{sec:convergence}, we use these estimates to compare to our fit results.

The observables of the semileptonic \Xclnu decay we will use in this paper are the partial branching fractions, lepton energy moments, hadronic invariant mass moments and dilepton invariant mass moments. The partial branching ratio is defined as
\begin{align}
    \Delta \mathcal{B}_\mathrm{sl} (E_\mathrm{cut}) &= 
    \tau_B\int_{E_l \ge E_\mathrm{cut}} 
    \frac{\mathrm{d} \Gamma}{ \mathrm{d} E_l }\, 
    \mathrm{d} E_l\ ,
\end{align}
where $\tau_B$ is the mean lifetime of the $B$ meson and $E_{\rm cut}$ denotes the restriction applied to the phase-space integration. 

The kinematic moments are defined as ratios, where a cut has been applied in a certain section of the phase space:
\begin{equation}
    \langle (O)^n \rangle_{\mathrm{cut}} =
    \int_{\mathrm{cut}}
    (O)^n 
    \frac{\mathrm{d} \Gamma}{\mathrm{d} O} \, \mathrm{d} O
    \Bigg/
    \int_{\mathrm{cut}}
    \frac{\mathrm{d} \Gamma}{\mathrm{d} O} \, \mathrm{d} O\ ,
    \label{eqn:defOmoments}
\end{equation}
where $O\in\{E_l,M_X^2,q^2\}$ and ``cut'' denotes either a cut on the lepton energy $E_l$ or on the dilepton invariant mass $q^2$. We use centralised moments, which are defined as linear combinations of the kinematic moments and are defined as
\begin{align}
    \ell_1 (E_\mathrm{cut}) &= \langle E_l \rangle_{E_l \ge E_\mathrm{cut}}\ , & 
    \ell_n (E_\mathrm{cut}) &= \Big\langle (E_l - \langle E_l\rangle )^n \Big\rangle_{E_l \ge E_\mathrm{cut}} 
    &\text{ for } n\ge2\ ,
    \label{eqn:defellcentralized}\\
    h_1(E_\mathrm{cut}) &= \langle M_X^2 \rangle_{E_l \ge E_\mathrm{cut}}\ , & 
    h_n (E_\mathrm{cut}) &= \Big\langle (M_X^2 - \langle M_X^2 \rangle )^n \Big\rangle_{E_l \ge E_\mathrm{cut}}
    &\text{ for } n\ge2\ ,
    \label{eqn:defhicentralized}\\
    q_1(q^2_\mathrm{cut}) &= \langle q^2 \rangle_{q^2 \ge q^2_\mathrm{cut}}\ , & 
    q_n (q^2_\mathrm{cut}) &= \Big\langle (q^2 - \langle q^2 \rangle )^n \Big\rangle_{q^2 \ge q^2_\mathrm{cut}}
    &\text{ for } n\ge2\ .
    \label{eqn:defqicentralized}
\end{align}
The double expansion in $1/m_b$ and $\alpha_s$ of these observables can be schematically be written as
\begin{align}
    X =& 
    (m_b)^{N_X} 
    \Bigg[ 
    \Bigg(X^{(0)} + 
    X^{(1)} \frac{\alpha_s(\scaleas)}{\pi} + 
    X^{(2)} \left( \frac{\alpha_s(\scaleas)}{\pi} \right)^2+\mathcal{O}(\alpha_s^3)\Bigg)\nonumber\\
    &+
    \frac{\mu_\pi^2}{m_b^2}
    \Bigg(
        X_{ \pi}^{(0)} + 
        X_{ \pi}^{(1)} \frac{\alpha_s(\scaleas)}{\pi} +\mathcal{O}(\alpha_s^2)
    \Bigg)+
    \frac{\mu_G^2}{m_b^2}
    \Bigg(
        X_{  G}^{(0)} + 
        X_{  G}^{(1)} \frac{\alpha_s(\scaleas)}{\pi} +\mathcal{O}(\alpha_s^2)
    \Bigg)\nonumber\\
    &+\frac{\rho_D^3}{m_b^3}
    \Bigg(
        X_{  D}^{(0)} + 
        X_{  D}^{(1)} \frac{\alpha_s(\scaleas)}{\pi} +\mathcal{O}(\alpha_s^2)
    \Bigg)
    +\frac{\rho_{LS}^3}{m_b^3}
    \Bigg(
       X_{  LS}^{(0)} + 
        X_{  LS}^{(1)} \frac{\alpha_s(\scaleas)}{\pi} +\mathcal{O}(\alpha_s^2)
    \Bigg)
    \nonumber\\
    &+\mathcal{O}\left(\frac{1}{m_b^4}\right)
    \Bigg]\ ,
    \label{eqn:defX}
\end{align}
where $X\in\{\Delta\mathcal{B}_{\rm sl},\ell_n,h_n,q_n\}$, $\alpha_s(\scaleas)\equiv\alpha_s^{(4)}(\scaleas)$ is taken at renormalisation scale $\scaleas$ and the dependence on the kinematic cuts and quark masses is implied in the $X_i^{(n)}$. Moreover, $N_X$ is the appropriate power corresponding to the observable $X$. In the kinematic moments, cross-terms between $\mu_\pi^2$, $\mu_G^2$, $\rho_D^3$ and \rhoLS appear when including $1/m_b^4$ and $1/m_b^5$ corrections in the kinematic moments, e.g. $\mu_\pi^2\times \mu_G^2$ at $\mathcal{O}(1/m_b^4)$ and $\mu_\pi^2\times\rho_D^3$ at $\mathcal{O}(1/m_b^5)$. 

We employ the publicly available code \verb|kolya| \cite{Fael:2024fkt,fael_2024_10818195} to obtain the predictions for the (partial) rates and kinematic moments. An overview of the known $1/m_b$ and $\alpha_s$ corrections is given in Tab.~\ref{tab:overview}, which are all implemented in \verb|kolya|. As discussed in Ref.~\cite{Fael:2024fkt}, the non-BLM contributions for the $\ell_n$ and $h_n$ moments are only known at several values for $\rho\equiv m_c/m_b$ and $E_{\rm cut}$~\cite{Biswas:2009rb}. A function $f(\rho, E_{\rm cut})$ with a certain ansatz for its shape is fitted to these known values to extrapolate the non-BLM contributions to other values of $(\rho, E_{\rm cut})$. We include these contributions following Ref.~\cite{Fael:2024fkt} for $\ell_{1,2,3}$ and $h_1$. For $h_{2,3}$ we do not include non-BLM contributions as not all required moments are calculated. For a more detailed discussion, we refer to Ref.~\cite{Fael:2024fkt}. Note that we do not include $h_3$ as an observable in the fit, as explained in Sec.~\ref{sec:inputs}. As such, not all moments are known with the same precision.

\begin{table}
\centering
\begin{tabular}{c|ccccl}
  $\Gamma_\mathrm{sl} $ & tree  & $\alpha_s$ & $\alpha_s^2$ & $\alpha_s^3$  \\
  \cline{1-5}
  \multirow{ 2}{*}{Partonic} &
  \multirow{ 2}{*}{} & 
  \multirow{ 2}{*}{\cite{Nir:1989rm}} & 
  \multirow{ 2}{*}{\cite{Pak:2008qt,Pak:2008cp,Dowling:2008mc,Egner:2023kxw}} & 
  \multirow{ 2}{*}{\cite{Fael:2020tow}} & \\
  \multirow{ 2}{*}{$\mu_\pi^2,\mu_G^2$} &
  \multirow{ 2}{*}{\cite{Manohar:1993qn,Blok:1993va}} & 
  \multirow{ 2}{*}{\cite{Becher:2006qw,Alberti:2012dn,Alberti:2013kxa,Mannel:2015jka}} & 
  \multirow{ 2}{*}{} & & \\
  \multirow{ 2}{*}{$\rho_D^3,\rho_{LS}^3$} &
  \multirow{ 2}{*}{\cite{Gremm:1996df}} & 
  \multirow{ 2}{*}{\cite{Mannel:2021zzr}} & 
  \multirow{ 2}{*}{} & & \\
  \multirow{ 2}{*}{$1/m_b^4,1/m_b^5$} &
  \multirow{ 2}{*}{\cite{Dassinger:2006md,Mannel:2010wj,Fael:2018vsp,Mannel:2023yqf,Finauri:2025ost}} & 
  \multirow{ 2}{*}{} & 
  \multirow{ 2}{*}{} & & \\[10pt] 
  \cline{1-5}
  $q_n(q^2_\mathrm{cut})$ & tree  & $\alpha_s$ & $\alpha_s^2$ & 
  \\
  \cline{1-5}
  \multirow{ 2}{*}{Partonic} &
  \multirow{ 2}{*}{} & 
  \multirow{ 2}{*}{\cite{Aquila:2005hq,Mannel:2021zzr}} & 
  \multirow{ 2}{*}{\cite{Fael:2024gyw}} & 
  \multirow{ 2}{*}{} & \\
  \multirow{ 2}{*}{$\mu_G^2,\mu_\pi^2$} &
  \multirow{ 2}{*}{\cite{Manohar:1993qn,Blok:1993va}} & 
  \multirow{ 2}{*}{\cite{Alberti:2012dn,Alberti:2013kxa}} & 
  \multirow{ 2}{*}{} & & \\
  \multirow{ 2}{*}{$\rho_D^3,\rho_{LS}^3$} &
  \multirow{ 2}{*}{\cite{Gremm:1996df}} & 
  \multirow{ 2}{*}{\cite{Mannel:2021zzr}} & 
  \multirow{ 2}{*}{} & & \\
  \multirow{ 2}{*}{$1/m_b^4,1/m_b^5$} &
  \multirow{ 2}{*}{\cite{Fael:2018vsp,Mannel:2023yqf,Finauri:2025ost}} & 
  \multirow{ 2}{*}{} & 
  \multirow{ 2}{*}{} & & \\[10pt]
  \cline{1-5}
  $\ell_n(E_\mathrm{cut}), h_n(E_\mathrm{cut})$ & tree  & $\alpha_s$ & $\alpha_s^2 \beta_0$ & $\alpha_s^2$  \\
  \cline{1-5}
  \multirow{ 2}{*}{Partonic} &
  \multirow{ 2}{*}{} & 
  \multirow{ 2}{*}{\cite{Trott:2004xc,Aquila:2005hq,NNLO_El}} & 
  \multirow{ 2}{*}{\cite{Aquila:2005hq}} & 
  \multirow{ 2}{*}{\cite{Biswas:2009rb}}$*$ & \\
  \multirow{ 2}{*}{$\mu_G^2,\mu_\pi^2$} &
  \multirow{ 2}{*}{\cite{Manohar:1993qn,Blok:1993va}} & 
  \multirow{ 2}{*}{\cite{Becher:2007tk,Alberti:2013kxa}} & 
  \multirow{ 2}{*}{} & & \\
  \multirow{ 2}{*}{$\rho_D^3, \rho_{LS}^3$} &
  \multirow{ 2}{*}{\cite{Gremm:1996df}} & 
  \multirow{ 2}{*}{} & 
  \multirow{ 2}{*}{} & & \\
  \multirow{ 2}{*}{$1/m_b^4, 1/m_b^5$} &
  \multirow{ 2}{*}{\cite{Dassinger:2006md,Mannel:2010wj,Mannel:2023yqf,Finauri:2025ost}} & 
  \multirow{ 2}{*}{} & 
  \multirow{ 2}{*}{} & & \\
\end{tabular} 
\caption{Overview of the different $1/m_b$ and perturbative $\alpha_s$ corrections included in the fit for the total rate $\Gamma_{\rm sl}$ and the $q^2$, $E_l$, and $M_X^2$ moments. (*) The non-BLM $\alpha_s^2$ corrections for the $E_l$ and $M_X^2$ moments are only known at several values of $\rho$ and $E_{\rm cut}$.}
\label{tab:overview}
\end{table}

The expressions in \eqref{eqn:defX} are computed in the pole scheme for the quark masses. However, these are plagued by renormalon ambiguities, which results in a poorly behaved perturbative series \cite{Bigi:1994em,Beneke:1994sw}. This can be resolved by switching to a short-distance mass scheme. We employ the $\overline{\rm MS}$ scheme for the charm quark mass $\overline{m}_c$ at the scale \scalemcMS, and the kinetic scheme for the bottom quark, following \cite{Bigi:1994ga,Bigi:1996si}:
\begin{equation}
    m_b^\mathrm{pole} = 
    m_b^\mathrm{kin}(\scalembkin) + [\overline \Lambda (\scalembkin)]_\mathrm{pert}
    +\frac{[\mu_\pi^2(\scalembkin)]_\mathrm{pert}}{2 m_b^\mathrm{kin}(\scalembkin)}
    +\mathcal{O}\left( \frac{1}{m_b^2} \right),
    \label{eqn:mpole2mkin}
\end{equation}
where \scalembkin is the Wilsonian cut-off with $\Lambda_{\rm QCD}\ll\scalembkin\ll m_b$. We implement this relation including $\mu_\pi^2$ (see Ref.~\cite{Mannel:2026vsz} for a recent discussion on higher order terms in a cut-off scheme). 
In addition, the HQE parameters are converted using
\begin{align}\label{eq:rhodpert}
    \mu_\pi^2(0) &= \mu_\pi^2(\scalembkin) - [\mu_\pi^2(\scalembkin)]_\mathrm{pert}\ , &
    \rho_D^3(0) &= \rho_D^3(\scalembkin) - [\rho_D^3(\scalembkin)]_\mathrm{pert}\ .
\end{align}
These conversions to the kinetic and $\overline{\rm MS}$ scheme are automatically implemented in \verb|kolya|. The perturbative corrections are known at NNLO \cite{Czarnecki:1997sz} and N$^{3}$LO \cite{Fael:2020iea,Fael:2020njb}.\\
Newly implemented in \verb|kolya| is the RGE evolution of $\mu_G^2$, for which we employ a reference scale of $2.3$ GeV. The expressions and explanation of the implementation in \verb|kolya| are presented in Appendix \ref{ap:muGrun}. There are no perturbative corrections to $\mu_G^2$ at leading order in $\scalembkin$ (see Ref.~\cite{Mannel:2026vsz} for a detailed discussion).

\subsection{Lifetimes}\label{sec:lifetimes}
The total decay width of a $B$ meson can also be predicted within the framework of the HQE. By applying the optical theorem, one can write the total decay width of a $B_q$ meson as
\begin{align}
    \Gamma(B_q)=\frac{1}{2M_{B_q}}\text{Im}\langle B_q|\mathcal{T}|B_q\rangle\ ,
\end{align}
where the transition operator is defined as
\begin{align}
    \mathcal{T}=i\int\text{d}^4x\, T\{\mathcal{H}_{\rm eff}(x),\mathcal{H}_{\rm eff}(0)\}\ .
\end{align}
The effective Hamiltonian is given by
\begin{align}
    \mathcal{H}_{\rm eff}=\mathcal{H}_{\rm eff}^{\rm NL}+\mathcal{H}_{\rm eff}^{\rm SL}\ ,\label{eq:effham}
\end{align}
where $\mathcal{H}_{\rm eff}^{\rm NL}$ describes non-leptonic (NL) decays and $\mathcal{H}_{\rm eff}^{\rm SL}$ describes semileptonic (SL) decays (see e.g.~Ref.~\cite{Buchalla:1995vs}). We neglect tiny contributions from rare decays.

The non-leptonic effective Hamiltonian is given by
\begin{align}
    \mathcal{H}_{\rm eff}^{\rm NL}=\frac{G_F}{\sqrt{2}}\sum\limits_{q_3=d,s}\Bigg[\sum\limits_{q_{1,2}=u,c}&\lambda_{q_1q_2q_3}\left(C_1(\mu_b)Q_1^{q_1q_2q_3}+C_2(\mu_b)Q_2^{q_1q_2q_3}\right)
    \nonumber\\    &
    -\lambda_{q_3}\sum\limits_{j=3,\dots,6,8}C_j(\mu_b)Q_j^{q_3}\Bigg]+\text{h.c.}\ ,
\end{align}
where $\lambda_{q_1q_2q_3}=V_{q_1b}^*V_{q_2q_3}$ and $\lambda_{q_3}=V_{tb}^*V_{tq_3}$ are the relevant CKM factors. For the definitions of the current-current operators $Q_{1,2}^{q_1q_2q_3}$, penguin operators $Q_{3,\dots,6}^{q_3}$ and the chromo magnetic operator $Q_8^{q_3}$, we refer to Ref.~\cite{Buchalla:1995vs,Egner:2024lay}.  The Wilson coefficients $C_i(\mu_b)$ of the $\Delta B=1$ operators are taken at renormalisation scale $\mu_b\sim m_b$.

The semileptonic part in \eqref{eq:effham} is given by
\begin{align}
    \mathcal{H}_{\rm eff}^{\rm SL}&=\frac{G_F}{\sqrt{2}}\sum\limits_{q=u,c}\sum\limits_{l=e,\mu,\tau}V_{qb}^*Q^{ql}+\text{h.c.}\ ,
\end{align}
where $Q^{ql}=(\bar b\gamma_\mu(1-\gamma_5)q)(\bar \nu_l\gamma^\mu(1-\gamma_5)l)$.
As for the semileptonic decays in Sec.~\ref{sec:semilep}, we split the $b$-quark momentum as $p_b=m_bv+k$ where the residual momentum $k$ is of order $\Lambda_{\rm QCD}\ll m_b$. Switching to the rephased quark field $b_v$ and expanding in $k$ results in the OPE for the total width. We obtain:
\begin{align}
    \Gamma(B_q)=\Gamma_3+\Gamma_5\frac{\langle \mathcal{O}_5\rangle}{m_b^2}+\Gamma_6\frac{\langle \mathcal{O}_6\rangle}{m_b^3}+\dots+16\pi^2\Bigg(\tilde{\Gamma}_6\frac{\langle \tilde{\mathcal{O}}_6\rangle}{m_b^3}+\tilde{\Gamma}_7\frac{\langle \tilde{\mathcal{O}}_7\rangle}{m_b^4}+\dots\Bigg)\ ,
\end{align}
where $\Gamma_n$ contain the short-distance physics, which can be calculated perturbatively in QCD:
\begin{align}
    \Gamma_n&=\Gamma_n^{(0)}+\frac{\alpha_s}{\pi}\Gamma_n^{(1)}+\left(\frac{\alpha_s}{\pi}\right)^2\Gamma_n^{(2)}+\dots\ .
\end{align}
Moreover, $\langle \mathcal{O}_d\rangle\equiv \langle B_q|\mathcal{O}_d|B_q\rangle/(2M_{B_q})$ are the matrix elements of the $\Delta B=0$ effective theory operators $\mathcal{O}_d$ of mass dimension $d$. Here, $\Gamma_3$ is the partonic result of a free $b$-quark decaying. At mass dimension 5 and 6, we find the contributions from ($\mu_\pi^2$, $\mu_G^2$) and ($\rho_D^3$, $\rhoLS$) respectively, and denoted by $\mathcal{O}_5$ and $\mathcal{O}_6$. At dimension 6, also four-quark operators $\tilde{\mathcal{O}}_6$ contribute, which are loop enhanced, explicitly shown by the factor of $16\pi^2$. 

At order $1/m_b^3$ there are four operators:
\begin{align}
		{\tilde O}_1^q  
		& =  
		(\bar{h}_v\,\gamma_\mu (1-\gamma_5) q)\,(\bar{q}\,\gamma^\mu (1-\gamma_5) h_v) ,
        &
		{\tilde O}_2^q  
		& =  
		(\bar{h}_v (1 - \gamma_5) q)\,(\bar{q} (1 + \gamma_5) h_v) ,
		  \notag \\[1mm]
		{\tilde O}_3^q  
		& =  
		(\bar{h}_v \, \gamma_\mu (1-\gamma_5) \, t^a q) 
		\, (\bar{q} \, \gamma^\mu (1-\gamma_5) \, t^a  h_v) ,
        &
		{\tilde O}_4^q 
		& =  
		(\bar{h}_v (1-\gamma_5) t^a q)\,(\bar{q}(1 + \gamma_5) t^a h_v),
		\label{eq:T2-HQET}
\end{align}
where $h_v$ is the HQET field.
They can be parameterized as~\cite{King:2021jsq}
\begin{align}
    \langle B_q|\tilde{O}_i^q|B_q\rangle&=F_q^2(\mu_b)M_{B_q}\tilde{B}_i^q(\mu_b)\ ,\nonumber\\
     \langle B_q|\tilde{O}_i^{q'}|B_q\rangle&=F_q^2(\mu_b)M_{B_q}\tilde{\delta}_i^{q'q}(\mu_b)\quad\quad (q\neq q')\ ,
\end{align}
where  $F_q (\mu_b)$ is the HQET decay constant, related to the 
$B$ meson constant $f_B$ evaluated in QCD by~\cite{Neubert:1992fk}
\begin{equation}
   		f_{B_q} = \frac{F_q (\mu_b)}{\sqrt{M_{B_q}}} \left[1 +  
		\frac{\alpha_s(\mu_b)}{2\pi} 
		\left(\ln \left(\frac{m_b^2}{\mu_b^2} \right)
		- \frac 4 3 \right) + 
		{\cal O}\left(\frac{1}{m_b}\right) \right] \,,
\end{equation}
Moreover, $\tilde{B}_i^q(\mu_b)$ are the Bag parameters and $\tilde{\delta}_i^{q'q}(\mu_b)$ are the ``eye-contractions'', all at renormalisation scale $\mu_b$ of the $\Delta B=0$ operators. We refer to Refs.~\cite{Lenz:2014jha,Lenz:2022rbq, Albrecht:2024oyn,Egner:2024lay} for more comprehensive discussions of the lifetime calculation. In this work, we do not consider the contributions to the lifetimes at order $1/m_b^4$ or higher.

We extend the Python package \texttt{kolya} to include predictions for the meson lifetimes, with the kinetic scheme for the $b$ quark mass and $\overline{\text{MS}}$ scheme for the charm quark.
In Appendix \ref{ap:kolya}, we present the implementation and overview of the QCD and power corrections included.

\section{Fit Setup}
\label{sec:fitting}

\subsection{Experimental Inputs}\label{sec:inputs}
The experimental landscape of inclusive $\bar B \to X_c l \bar\nu_l$ decays is defined by a high-precision dataset of spectral moments, which serve as the foundation for extracting $|V_{cb}|$ and non-perturbative parameters. These measurements were pioneered by the CLEO~\cite{CLEO:2004bqt} collaboration, whose early measurements of the lepton energy ($E_l$) and hadronic invariant mass ($M_X^2$) moments provided the first crucial validation of the Operator Product Expansion (OPE) in $b$-hadron decays. These foundations were subsequently built upon by the BaBar~\cite{BaBar:2009zpz} and Belle~\cite{Belle:2006jtu,Belle:2006kgy} $B$-factory experiments, which produced the bulk of the high-statistics data that dominate current global averages, supplemented by results from DELPHI~\cite{DELPHI:2005mot} at LEP and CDF~\cite{CDF:2005xlh} at the Tevatron. More recently, Belle~\cite{Belle:2021idw} and Belle~II~\cite{Belle-II:2022evt} have expanded this program with novel measurements of $q^2$ (dilepton invariant mass) moments.

While many spectral moments have been published by the aforementioned collaborations, the present analysis utilizes only a targeted subset of the available data. This selective approach is necessitated by the large experimental correlations existing between moments, particularly those measured within the same kinematic category (e.g., higher-order $M_X^2$ moments) or with overlapping lepton energy requirements. Inverting the full covariance matrix in such cases frequently leads to severe numerical instabilities, as the matrix becomes near-singular, resulting in unreliable $\chi^2$ minimizations and inflated uncertainties. Consequently, to ensure a robust and statistically sound fit, we follow the specific selection criteria in Ref.~\cite{Finauri:2023kte}. The selected subset of the experimental data is shown in Tab.~\ref{tab:data}.
We note that also the centralized 4th $q^2$ moments are available. Since in our default fit, we only include $1/m_b^3$ corrections, we do not include these moments. In addition, we also do not include the third centralized $M_X^2$ moments as this moment is very sensitive to higher-order $\alpha_s$ corrections which are currently unknown (see also Ref.~\cite{Fael:2024fkt} for a discussion).

\begin{table}[ht]
\centering
\small

\caption{Summary of cut values for different experiments, split into $\Delta \mathcal{B}_{\rm sl}$, $E_l$, $M_X^2$, and $q^2$ selections. Dashes (--) indicate no entries. The selection of data points follows the previous works in Ref.~\cite{Finauri:2023kte}. The $E_l$-cuts for $\Delta \mathcal{B}_{\rm sl}$, the $E_l$ and the $M_X^2$ moments are given in GeV and the $q^2$-cuts for the $q^2$ moments in GeV$^2$.}
\label{tab:data}%
\begin{tabular}{lllll}

\toprule
Experiment & $\Delta \mathcal{B}_{\rm sl}$ & $\langle E_l \rangle$ & $\langle (E_l - \langle E_l \rangle)^2 \rangle$ & $\langle (E_l - \langle E_l \rangle)^3 \rangle$ \\
\midrule
Babar   & 0.6, 1.2, 1.5 & 0.6, 0.8, 1.0, 1.2, 1.5 & 0.6, 1.0, 1.5 & 0.8, 1.2 \\
Belle   & 0.6, 1.0, 1.4 & 1.0, 1.4 & 0.6, 1.4 & 0.8, 1.2 \\
Belle II & -- & -- & -- & -- \\
CDF     & -- & -- & -- & -- \\
Cleo    & -- & -- & -- & -- \\
Delphi  & -- & 0.0 & 0.0 & 0.0 \\
\bottomrule

\\
\toprule
Experiment & $\langle M_X^{2} \rangle$ & $\langle (M_X^{2} - \langle M_X^{2} \rangle)^2 \rangle$ & \\ 
\midrule
Babar   & 0.9, 1.1, 1.3, 1.5 & 0.8, 1.0, 1.2, 1.4 &  \\
Belle   & 0.7, 1.1, 1.3, 1.5 & 0.7, 0.9, 1.3 &  \\
Belle II & -- & -- &  \\
CDF     & 0.7 & 0.7 &  \\
Cleo    & 1.0, 1.5 & 1.0, 1.5 &  \\
Delphi  & 0.0 & 0.0 &  \\
\bottomrule
\\

\toprule
Experiment & $\langle q^{2} \rangle$ & $\langle (q^{2} - \langle q^{2} \rangle)^2 \rangle$ & $\langle (q^{2} - \langle q^{2} \rangle)^3 \rangle$  \\
\midrule
Babar   & -- & -- & -- \\
Belle   & 3.0, 4.5, 6.0, 7.5 & 3.0, 4.5, 6.0, 7.5 & 3.0, 4.5, 6.0, 7.5 \\
Belle II & 1.5, 3.0, 4.5, 6.0, 7.5 & 1.5, 3.0, 4.5, 6.0, 7.5 & 1.5, 3.0, 4.5, 6.0, 7.5 \\
CDF     & -- & -- & -- \\
Cleo    & -- & -- & -- \\
Delphi  & -- & -- & -- \\
\bottomrule
\end{tabular}
\end{table}

\subsection{Treatment of uncertainties}\label{sec:scalevar}%
We define a new setup to deal with the uncertainties originating from missing higher-order contributions in the HQE. We distinguish between two main sources and first briefly revise how these uncertainties are typically treated. 

\vspace{0.4cm}
\textbf{Truncation of the perturbative expansion.} 
The predictions for the moments are computed only up to a finite order in the $\alpha_s$ expansion. This truncation introduces a dependence on unphysical renormalization scales,
such as the scale of the strong coupling $\scaleas$ 
and the quark masses $\scalembkin$ and $\scalemcMS$. 
In the all-orders limit, physical observables would be independent of such scales. 

The associated uncertainty is conventionally estimated by varying the scales within a reasonable range.\footnote{By ``reasonable range'' we mean values that do not introduce large logarithms of the form $\log(\mu/Q)$, where $Q$ is a characteristic scale of the process.}
This approach probes logarithmic/RG-predictable pieces at higher orders in QCD 
and correlated shifts but generically it does not catch unknown constants 
and new cut dependence that can  arise. 
The envelope of predictions obtained by varying unphysical scales over the allowed
ranges is assumed to contain the true all-orders prediction.

\vspace{0.4cm}
\textbf{Truncation of the power expansion.} 
This source of uncertainty reflects our limited knowledge of higher-order corrections 
in the $1/m_b$ expansion. Unlike the perturbative case, there are no associated unphysical parameters that can be varied, and therefore 
no canonical procedure for a quantitative estimate of this uncertainty.

A commonly adopted approach~\cite{Gambino:2013rza,Bordone:2021oof,Finauri:2023kte,Bernlochner:2022ucr, Carvunis:2025vab} 
is to vary the matrix elements $\rho^3_D$ and $\mu_G^2$ by $20\text{--}30\%$, where the latter are expected to also account for missing perturbative corrections to the HQE elements. For some moments this prescription leads to uncertainties that are significantly larger than those from perturbative truncation. In addition, this requires introducing
heuristic correlation models between the different moments at different energy cuts. 
It was shown in Ref.~\cite{Finauri:2023kte} that the specific choice of the theoretical covariance matrix has only a limited effect on the extracted results. 

\vspace{0.4cm}
\textbf{Fit rationale and assumptions of this work.} 
The truncation of the perturbative series
induces a residual dependence on the unphysical scales  $\scaleas, \scalembkin$ and $\scalemcMS$. 
Instead of a single SM prediction, one obtains a set 
of predictions parametrized by these scales. In our new setup, we exploit the parametric freedom associated 
with these scales to achieve a good description of semileptonic data (i.e., an acceptable $\chi^2/\mathrm{d.o.f.}$), without introducing ad hoc uncertainties in the power corrections. To this end, we introduce these three scales as nuisance parameters directly in the $\chi^2$ function and profile over them to assess their impact.\footnote{We note that there has been recent progress in the treatment of theoretical uncertainties beyond conventional scale variations~\cite{Tackmann:2024kci}. A more systematic implementation of such approaches is beyond the scope of this work and is left for future study.} 

Our procedure effectively treats the uncertainties arising from scale variations as being 100\% correlated across all predictions. The rationale for this choice is that,
for a given process, scale variations are expected to induce coherent shifts across all observables, rather than independent fluctuations, since all predictions originate from the same function, the triple differential rate $\text{d}^3\Gamma_{\rm sl}$, evaluated at a 
specific set of the scales. 
To further illustrate this point, consider a function $f(\mathbf{x}, \mathbf{\omega})$ which depends on a set of scales $\mathbf{x}=(x_1,x_2,\dots)$ and kinematic variables $\mathbf{\omega}=(\omega_1,\omega_2,\dots)$. 
The observables $\mathcal{O}$ that we consider are integrals of this 
function $f$ multiplied by some weight functions $\Omega(\mathbf{\omega})$:
\begin{align}
   \mathcal{O}_\Omega (\mathbf{x}) \equiv \int\text{d}\omega\, \Omega(\omega)f(\mathbf{x},\omega)\ .
\end{align}
For any choice of $\Omega(\omega)$, the dependency on $\mathbf{x}$ is fully determined by $f$, which plays the role of differential rate. 
The function $\Omega(\omega)$ can also contain a veto function to restrict the 
integration to a subregion of the domain.
Therefore, the uncertainties arising from $\mathbf{x}$ are fully correlated between different choices of $\Omega(\omega)$, i.e.\ across all theoretical predictions.

In our approach, by including $\scaleas, \scalembkin$ and $\scalemcMS$ as nuisance 
parameters, we search for the triplet which describes the experimental data best. This is our central assumption.
Therefore, at the best fit point, the triplet $\scaleas, \scalembkin$ and $\scalemcMS$
 which best describes the semileptonic moments will be used also to evaluate
 the rate.

In addition, different from previous analyses, we do not add a theoretical uncertainty to account for missing terms in the HQE to the covariance matrix. Instead, we determine the HQE parameters of interest and the nuisance scale parameters at a fixed order in the $1/m_b$ expansion. Rather than estimating the impact of higher orders, we perform fits sequentially, order by order, from $\mathcal{O}(1/m_b^3)$ to $\mathcal{O}(1/m_b^5)$, and examine the convergence of the fitted parameters. If the HQE is convergent, the corrections to the fitted parameters are expected to decrease with increasing order, and the central values should converge toward their true values in the limit $\mathcal{O}(1/m_b^\infty)$. When quoting our principal results, an additional uncertainty to account for missing higher orders is added, with the procedure described in Sec.~\ref{sec:convergence}.

\subsection{Implementation }\label{sec:implement}
We determine \Vcb , the HQE parameters $\vec{h} = \left(\mupi, \muG, \rhoD, \rhoLS \right)$, the auxiliary parameters $\vec{\theta} = \left(\mbMS, \mcMS, \alphasZ \right)$, and the scale parameters $\vec{\mu} = \left( \scalembkin, \scalemcMS, \scaleas \right)$ in a simultaneous fit. The $\chi^2$ function is defined as follows:
\begin{align}
    \chi^2 =&\ \chi^2\left(\Vcb^2, \vec{h}, \vec{\theta}, \vec{\mu} \right) \nonumber\\
    =&\ \left( \vec{m}(\vec{h}, \vec{\theta}, \vec{\mu}) - \vec{m}_\mathrm{meas} \right)T C^{-1}  \left( \vec{m}(\vec{h}, \vec{\theta}, \vec{\mu}) - \vec{m}_\mathrm{meas} \right) \nonumber\\
    &+ \sum_{i=0}^{2} \frac{\left(\theta_i - \bar{\theta}_i\right)^2}{\sigma_{\theta_i}^{2}}\nonumber\\
    &+ \sum_{i=0}^{2}     \begin{cases}
      0, & \forall \mu_i \in [\bar{\mu}_i \pm \zeta \sigma_i] \\
      \left( \frac{\mu_i - \bar{\mu}_i}{\kappa \sigma_i} \right)^2 - \left( \frac{\zeta}{\kappa} \right)^{2}, & \forall \mu_i \notin [\bar{\mu}_i \pm \zeta \sigma_i]
    \end{cases}\,.
\end{align}
The vector $\vec{m}$ contains the theoretical predictions for the partial rate and central moments at the experimentally measured thresholds, corresponding to the measured partial rate moments $\vec{m}_\mathrm{meas}$ listed in Tab.~\ref{tab:data}. The covariance matrix $C$ includes both statistical and systematic uncertainties associated with the measured moments. 

The auxiliary parameters $\vec{\theta}$ are constrained via Gaussian penalty terms, using external inputs for their central values $\bar{\theta}_i$ and associated uncertainties $\sigma_{\theta_i}$. We use $\overline{m}_b(\overline{m}_b)=4.203(11)\, \mathrm{GeV}$~\cite{FlavourLatticeAveragingGroupFLAG:2024oxs}, $\overline{m}_c(3\, \mathrm{GeV}) = 0.989(10)\, \mathrm{GeV}$~\cite{FlavourLatticeAveragingGroupFLAG:2024oxs}, and $\alphasZ = 0.1180(9)$~\cite{ParticleDataGroup:2024cfk}.
We take these three inputs as independent due to the lack of known correlations from the shared ensembles, scale setting and, in several cases, the same correlator fits.
\begin{figure}[t]
    \centering
    \includegraphics[width=\linewidth]{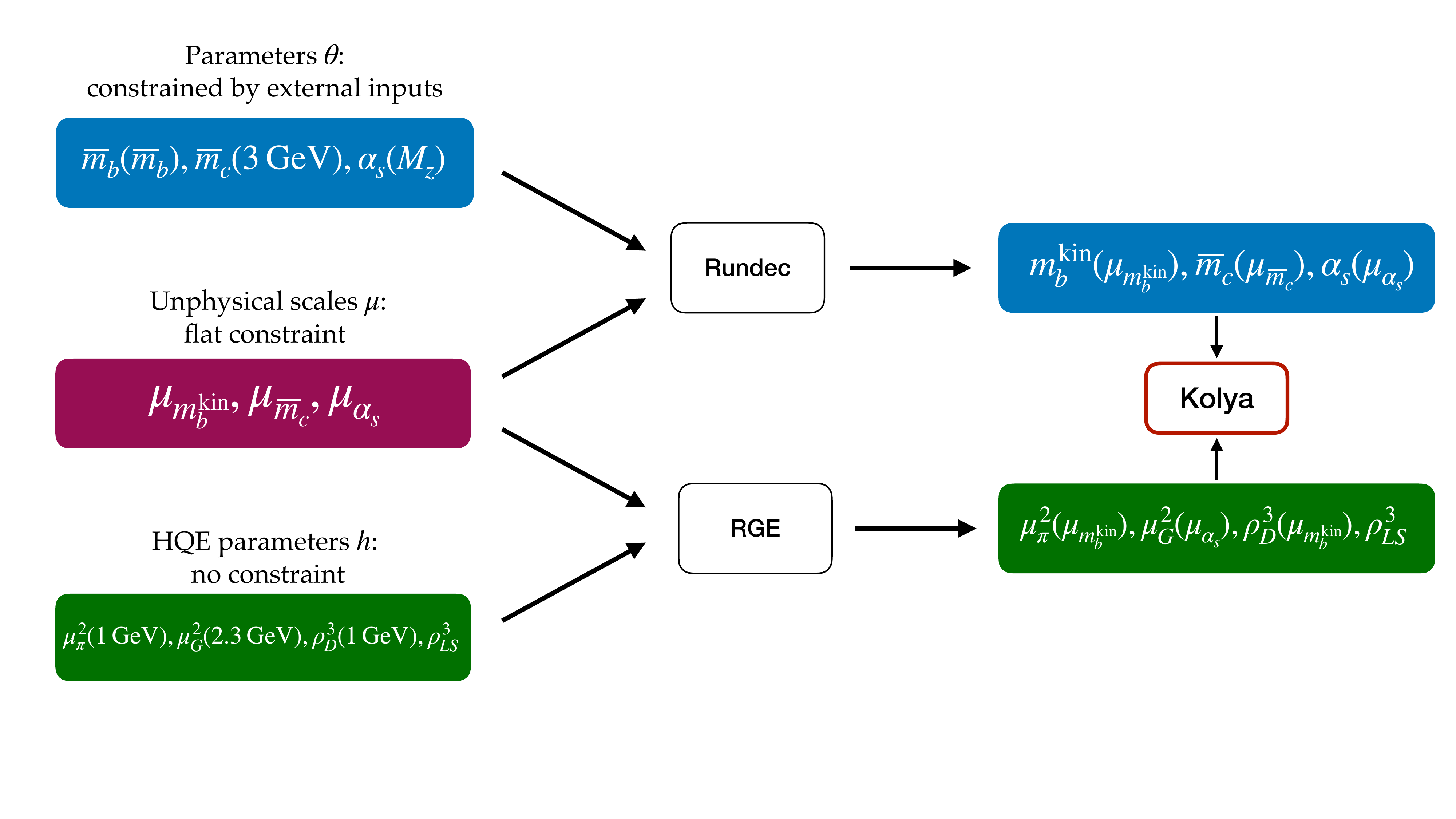}
    \caption{Schematic representation of the fit setup.
    The parameters of the fit (left) are used
    to evaluate the parameters on the right.
    The latter are used in \texttt{kolya} to evaluate
    the prediction for the moments.
    }
    \label{fig:scalesetting}
\end{figure}
The scale parameters $\vec{\mu}$ are constrained using the \textit{R}fit~\cite{Hocker:2001xe} approach, in particular the \textit{ER}fit implementation, using the parameter choices $\kappa = 0.8$ and $\zeta = 1$ to model the penalty potential. This allows the fit to explore any point within the predefined range without incurring a penalty, while a gradually increasing penalty is applied once the boundaries are exceeded to ensure a continuous $\chi^2$ function.
The chosen boundaries are $\scalembkin = [0.75, 1.25]\, \mathrm{GeV}$, $\scalemcMS = [1.8, 3.2]\, \mathrm{GeV}$, and $\scaleas = [m_b^\mathrm{kin}/2 ,2 m_b^\mathrm{kin}]$.
These ranges correspond to the conventional \emph{a priori} choices where the perturbative series is assumed to be well behaved. 
We extract all scale dependent parameters at (arbitrarily) chosen reference scales described below. To ensure this, we perform the following scheme conversions and the evolution of perturbative parameters to the scale values selected during the $\chi^2$ minimization using \texttt{RunDec}~\cite{Chetyrkin:2000yt,Schmidt:2012az,Herren:2017osy}:
\begin{enumerate}
    \item $\alpha_s^{(5)}(M_{Z^0}) \to \alpha_s^{(4)}(\scaleas)$: we
    evolve the value of $\alpha_s$ with five-loop accuracy down to the 
    decoupling scale equal to $2 \, \mbMS$, we decouple $\alpha_s^{(5)} \to\alpha_s^{(4)} $ and evolve again to the scale $\scaleas$ sampled,
    to obtain $\alpha_s^{(4)}(\scaleas)$;
    \item $\mbMS \to \mbkin(\scalembkin)$: we calculate $\mbkin$, the bottom quark mass in the kinetic scheme at the scale $\scalembkin$,
    using scheme B from Ref.~\cite{Fael:2020njb} using as inputs $\mbMS$, 
    $\mcMS$ and $\alpha_s^{(4)}(\scaleas)$;
    \item $\mcMS \to \overline{m}_{c}(\scalemcMS)$:
    we calculate the charm mass with $n_f=4$ active flavors  
    at the sampled scale $\scalemcMS$ and with four-loop accuracy;

    \item $\mu_\pi^2(1\, \mathrm{GeV}) \to \mu_\pi^2(\scalembkin)$ and 
    $\rho_D^3(1\, \mathrm{GeV}) \to \rho_D^3(\scalembkin)$:
    we evolve $\mu_\pi^2$ and $\rho_D^3$ from the reference scale of
    1~GeV to the sampled scale $\scalembkin$. To this end, we exploit the 
    fact that the l.h.s.\ of the equations in~\eqref{eq:rhodpert}
    are independent on $\scalembkin$ and calculate:
    \begin{align}
        \mu_\pi^2(\scalembkin) &= 
        \mu_\pi^2(1 \, \mathrm{GeV}) -
        [\mu_\pi^2(1 \, \mathrm{GeV})]_\mathrm{pert}
        +[\mu_\pi^2(\scalembkin)]_\mathrm{pert}, \notag \\
        \rho_D^3(\scalembkin) &= 
        \rho_D^3(1 \, \mathrm{GeV}) -
        [\rho_D^3(1 \, \mathrm{GeV})]_\mathrm{pert}
        +[\rho_D^3(\scalembkin)]_\mathrm{pert} .
    \end{align}
    \item $\mu_G^2(2.3\, \text{GeV})\to \mu_G^2(\scaleas)$: we employ the RGE evolution upto next-to-leading-logarithm (NLL) of $\mu_G^2$ from the reference scale of $2.3$ GeV to the scale $\scaleas$. We refer to Appendix \ref{ap:muGrun} for the details.
\end{enumerate}

This procedure, summarized in Fig.~\ref{fig:scalesetting}, ensures that the $\chi^2$ function is allowed to vary the scales $\vec{\mu}$ freely during the fit and the best-fit parameters do not have to be converted to compare different fit setups, e.g.\ the best-fit $\mu_\pi^2$ in any scenario will always be at $\scalembkin=1\,\mathrm{GeV}$ ensuring comparability independent of the best-fit point for $\scalembkin$. We do not apply any external constraints on the HQE parameters $\mupi$, $\muG$, $\rhoD$, and $\rhoLS$.

Throughout this work we count the degrees of freedom 
\begin{align}
    n_\mathrm{d.o.f.} = N_\mathrm{obs} - N_\mathrm{phys}\ ,
\end{align}
where $N_\mathrm{obs}$ is the number of measured moments and partial rates entering the first line of the $\chi^2$ and $N_\mathrm{phys}$ counts those fitted parameters that the data alone have to determine, namely \Vcb{}, $\vec h$ and, when they float, the scales $\vec\mu$. The auxiliary parameters $\vec\theta$ are constrained by auxiliary data and therefore not counted. The scales are counted because the penalty is flat inside the a-priori range: there it supplies no constraint, so that the scales are ordinary free parameters fixed by the data alone, and $\scalembkin$ and $\scalemcMS$ are indeed determined well inside their ranges, see Sec.~\ref{sec:profiles}. This is the conservative choice. The $\chi^2$ minimum of the fit with free scales lies between the one of a fit with the scales held fixed and the one of a fit with the scales unconstrained, so that the $p$ value quoted with this counting is a lower bound on the true one, the latter requiring a toy Monte Carlo distribution of $\chi^2_\mathrm{min}$~\cite[Sec.~3.3]{Hocker:2001xe}. With the scales fixed to their nominal values the count reverts to $N_\mathrm{obs} - 5$, the convention of Refs.~\cite{Bordone:2021oof,Bernlochner:2022ucr,Finauri:2023kte}. At $\mathcal{O}(1/m_b^3)$ this leaves $N_\mathrm{phys} = 8$ with free scales; at $\mathcal{O}(1/m_b^{4,5})$ the coefficients $m_i$ and $r_i$ again carry their own priors and cancel in the same way, so $N_\mathrm{phys}$ is unchanged.

We note that the covariance matrix at the minimum propagates the experimental uncertainties and the Gaussian constraints on $\vec\theta$, together with the local curvature in $\vec\mu$. The uncertainty does \emph{not} contain a perturbative truncation uncertainty in the usual sense: the scales are determined rather than varied, so the customary scale error has been absorbed into the central value and is profiled over. The fit uncertainty also does not contain uncertainties due to unknown higher-power corrections, since the fit is performed at fixed order in $1/m_b$. We discuss this in more detail below. These points should be considered when comparing our results with the inclusive semileptonic $B$ analysis in Refs.~\cite{Bordone:2021oof,Bernlochner:2022ucr,Finauri:2023kte,Carvunis:2025vab}.

\section{Fit Results} 
\label{sec:fit_results}

\newcommand{\pValueDefault}{\numPvalueDefault}
\newcommand{\pValueNoBelleqsq}{\numPvalueNoBelleqsq}
\newcommand{\pValueFixedDefault}{\numPvalueFixedDefault}
\newcommand{\pValueFixedNoBelleqsq}{\numPvalueFixedNoBelleqsq}
\newcommand{\pValueNoBabarEl}{\numPvalueNoBabarEl}
\newcommand{\pValueNoBelleMX}{\numPvalueNoBelleMX}
\newcommand{\pValueNoqsq}{\numPvalueNoqsq}

\subsection{Default fit scenario}
\begin{figure}
    \centering
    \includegraphics[width=0.32\linewidth]{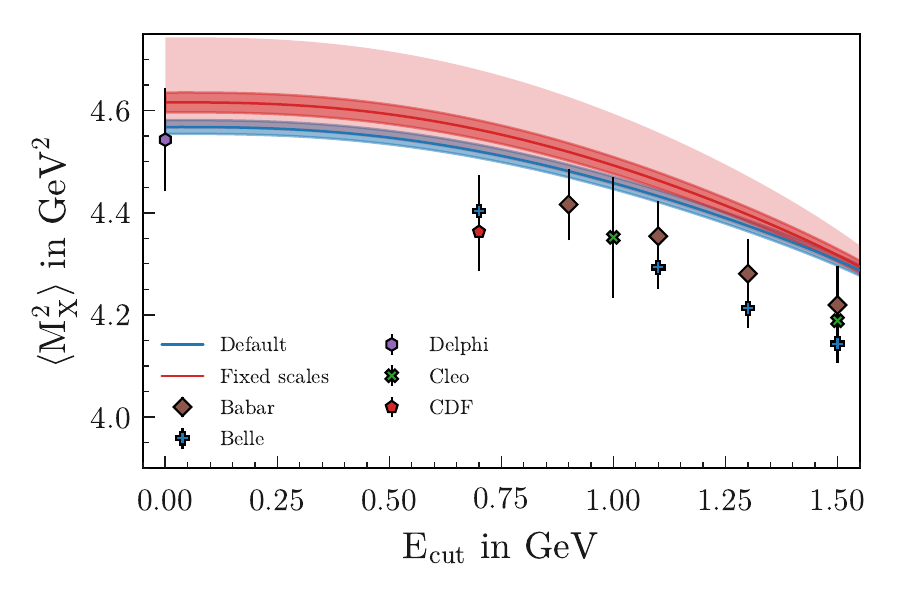}
    \includegraphics[width=0.32\linewidth]{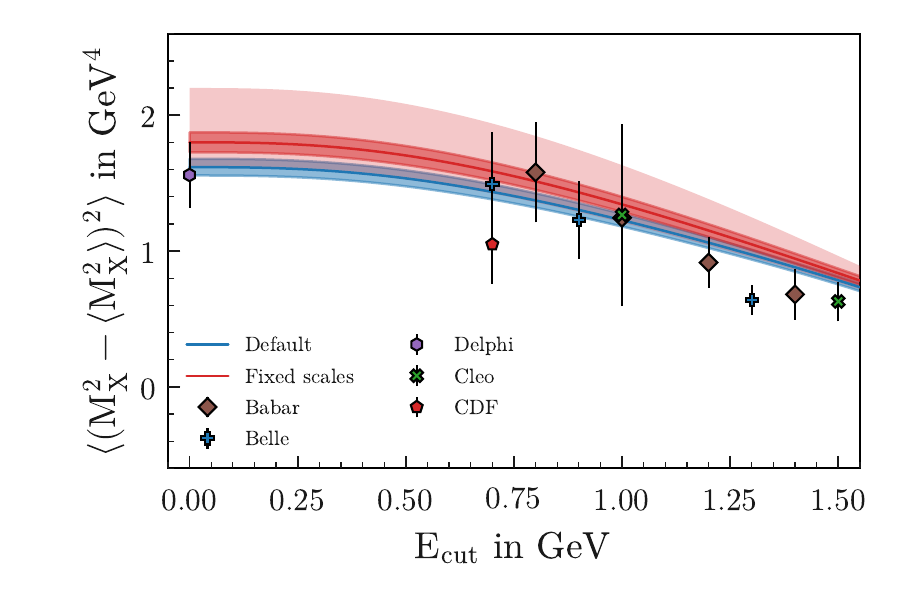}
    \caption{Fit projections for the mean (left) and second central (right) $M_X^2$ moments as a function of the lepton-energy threshold. Only data points included in the $\chi^2$ minimization are displayed.}
    \label{fig:fitted-mx}
\end{figure}

\begin{figure}
    \centering
    \includegraphics[width=0.32\linewidth]{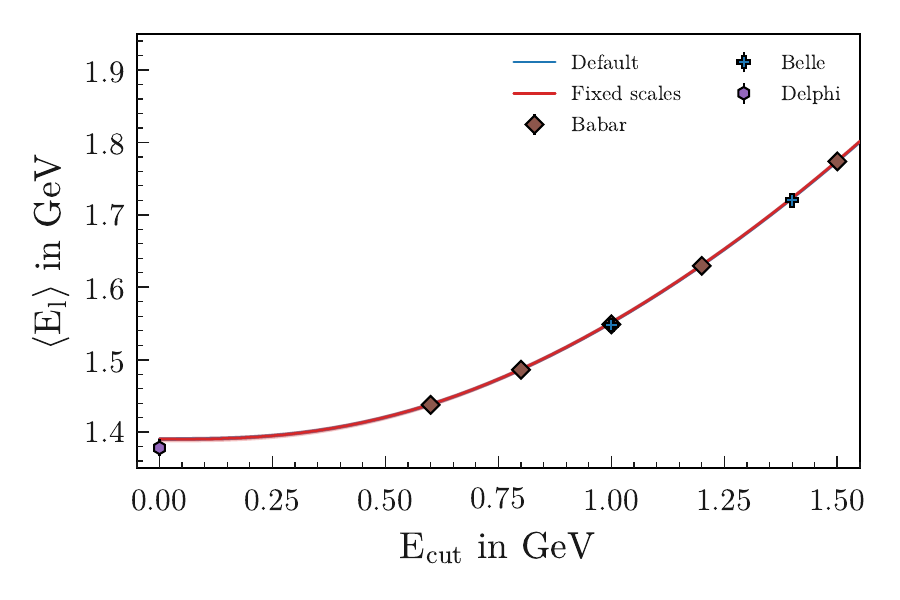}
    \includegraphics[width=0.32\linewidth]{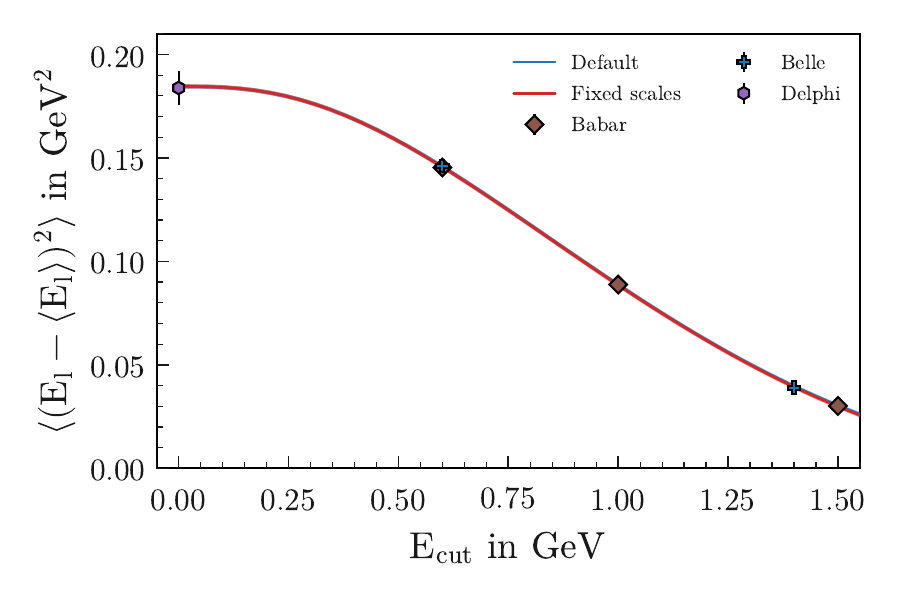}
    \includegraphics[width=0.32\linewidth]{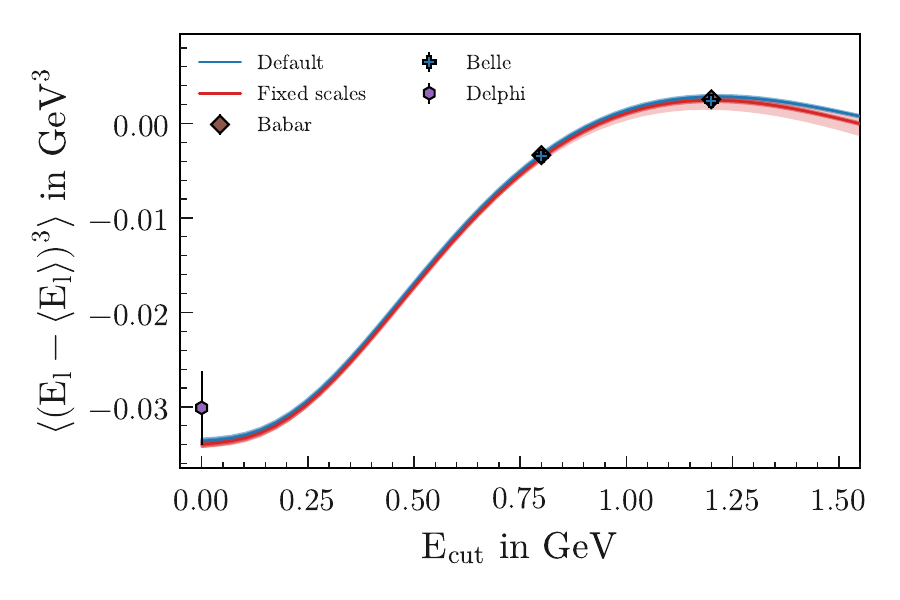}
    \caption{Fit projections for the 
    mean (left), second (center), and third (right) central $E_l$ moments as a function of the lepton-energy threshold. Only data points included in the $\chi^2$ minimization are displayed.}    
    \label{fig:fitted-el}
\end{figure}

\begin{figure}
    \centering
    \includegraphics[width=0.32\linewidth]{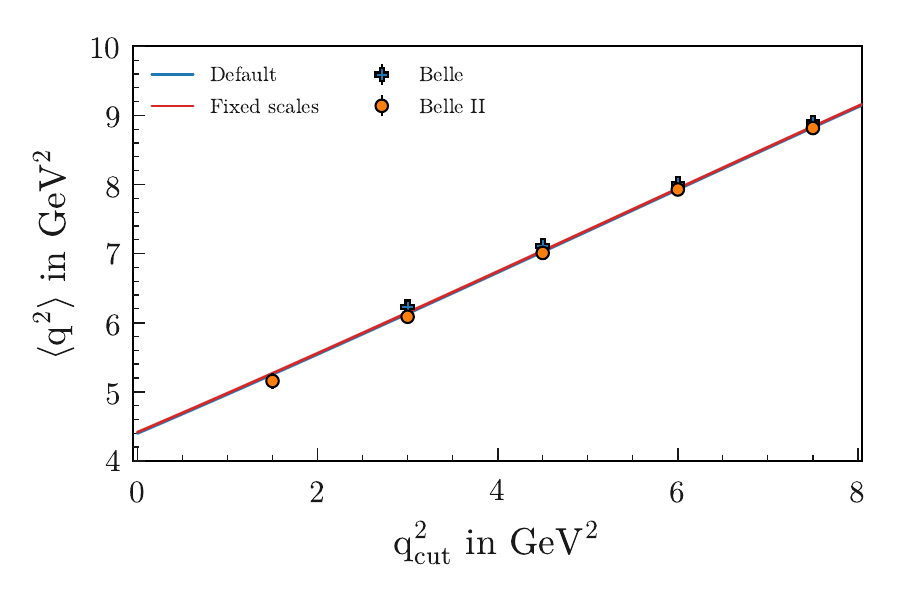}
    \includegraphics[width=0.32\linewidth]{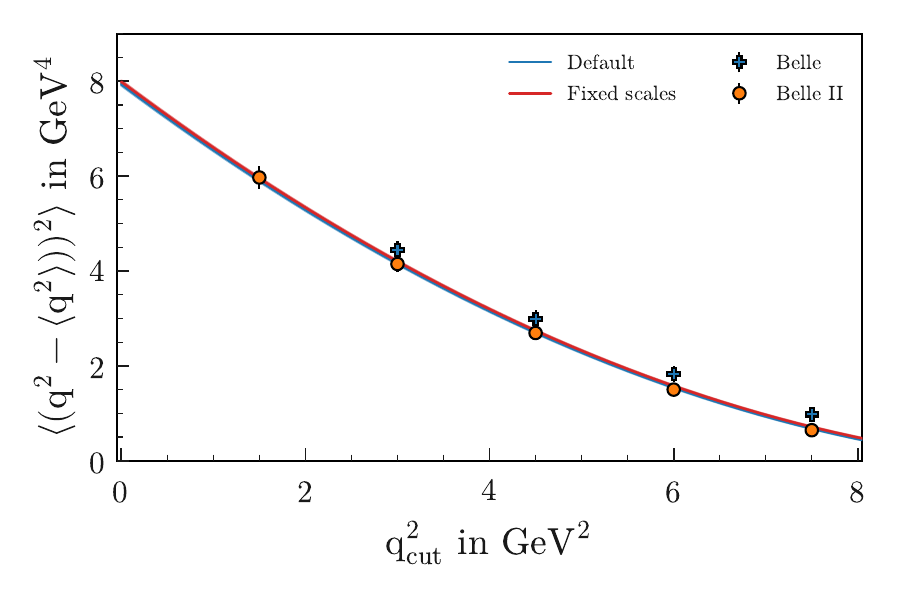}
    \includegraphics[width=0.32\linewidth]{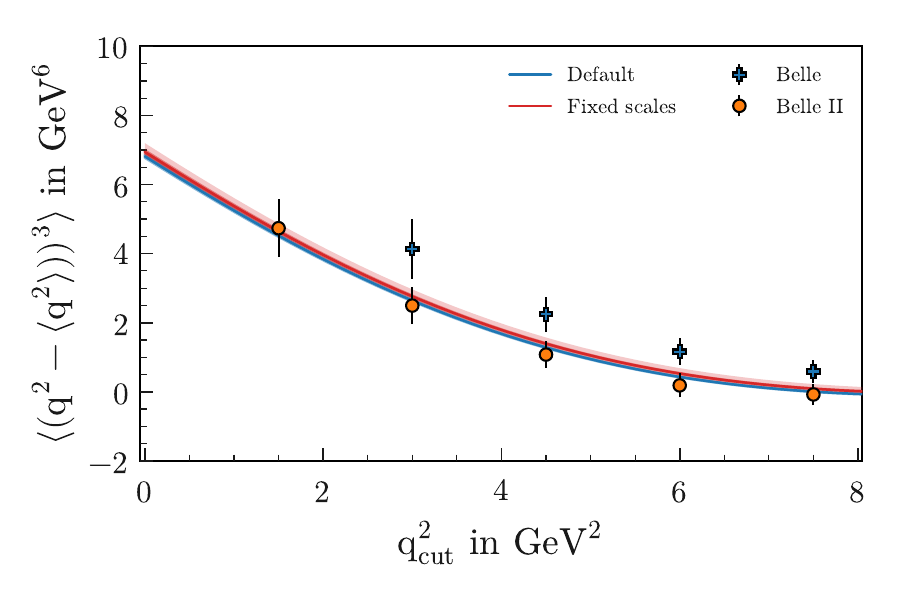}
    \caption{Fit projections for the 
    mean (left), second (center), and third (right) central $q^2$ moments as a function of the lepton-energy threshold. Only data points included in the $\chi^2$ minimization are displayed.}        
    \label{fig:fitted-q2}
\end{figure}

\begin{figure}
    \centering
    \includegraphics[width=0.32\linewidth]{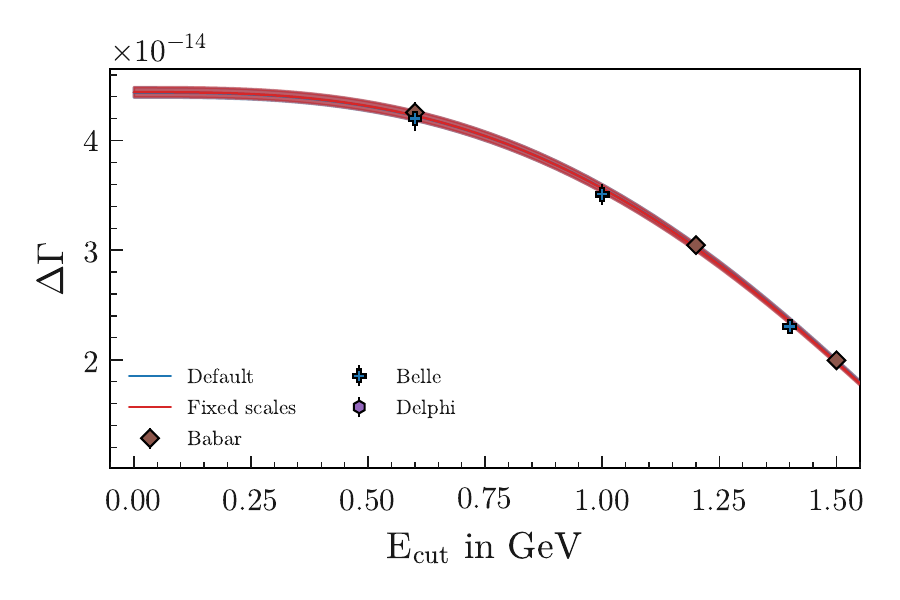}
    \caption{Fit projections for the 
    partial rate $\Delta \Gamma_{\rm sl}$ as a function of the lepton-energy threshold. Only data points included in the $\chi^2$ minimization are displayed.}      
    \label{fig:fitted-partial-rate}
\end{figure}

The results of our default fit scenario with free scales up to $1/m_b^3$ are summarized in Tab.~\ref{tab:default_fit_result} (top row). The $p$-value of the fit is \numPvalueDefault{}. 
The fitted data points together with the corresponding inclusive predictions are shown in Fig.~\ref{fig:fitted-mx} for the $M_X^2$ moments, Fig.~\ref{fig:fitted-el} for the $E_l$ moments, Fig.~\ref{fig:fitted-q2} for the $q^2$ moments, and Fig.~\ref{fig:fitted-partial-rate} for the partial decay rates. 

Table~\ref{tab:default_fit_result} lists the physical parameters only. The scales come out at
$\scalembkin = \numscalembkin \,\mathrm{GeV}$,
$\scalemcMS = \numscalemcMS \,\mathrm{GeV}$ and
$\scaleas = \numscaleas \,\mathrm{GeV}$,
and thus deviate from the commonly used values $\scalembkin = 1\,\mathrm{GeV}$, $\scalemcMS = 2\,\mathrm{GeV}$, and $\scaleas = 4.565\,\mathrm{GeV}$. 
We list both the 
Minos interval and the Hessian uncertainty, which differ for the parameters that
correlate with the scales. We discuss this point in more detail in Sec.~\ref{sec:profiles}. In the following, we quote as the fit uncertainty the symmetrized Minos interval.

The scale $\scalemcMS$ is found slightly below the standard choice, 
and the fit shows clear sensitivity to this scale parameter. 
We find a large value of $\scaleas$, corresponding to a smaller value of
$\alpha_s(\scaleas)$. The fit preference for a large value of $\scaleas$ should not be 
interpreted as a physical hard scale. 
Rather, it reflects the fact that, with our treatment, $\scaleas$ acts as a correlated 
nuisance parameter. Varying this scale changes the perturbative corrections coherently in all moment predictions. 

The fit selects the scale parameter that best aligns theoretical predictions with the measured moments across the full dataset. Lower values of $\scaleas$ corresponding to a larger coupling induce larger perturbative effects. Instead, the data favour a scale much larger than the conventional $m_b/2$ or $m_b$, corresponding to a regime with milder, smaller perturbative variations. We stress that the minimum sits on the boundary, therefore the data have not determined a scale, they have only indicated a direction. The preference for a large $\scaleas$ either reflects a preference for reduced perturbative corrections or acts as an effective parameter absorbing residual experimental tensions within the default fit setup.

The fitted $\scaleas$ should thus be viewed as an effective parameter 
describing missing correlated higher-order effects.

To study the impact of our treatment of the scale parameters in the fit, we also perform the minimization with the scale parameters fixed to commonly used values of $\scalembkin=1\,\mathrm{GeV}$, $\scalemcMS=2\,\mathrm{GeV}$, and $\scaleas=4.565\,\mathrm{GeV}$. This fit yields a $p$-value of \pValueFixedDefault{}. 
The fit coefficients obtained are summarized in Tab.~\ref{tab:default_fit_result} together with our nominal results. For this analysis, we quote the uncertainty from the fit and vary the scales $\scalembkin, \scalemcMS$ and $\scaleas$ within the ranges described above independently. The total uncertainty is obtained by adding these in quadrature. The results of this fit are also added to Figs.~\ref{fig:fitted-mx} - \ref{fig:fitted-partial-rate}, where the largest effect is observed in the $M_X$ moments and high $q^2$ and $E_\ell$ moments. 

The deterioration of the $p$-value is expected when moving away from the best-fit values. Fixing the scale parameters has a significant impact on the extracted values of $\Vcb$ and $\muG$, shifting their central values by a few standard deviations. However, due to the scale variation the uncertainties are significantly larger, such that the results are in agreement with our default analysis. The shift in central values can be understood from the correlation structure between the physical and unphysical parameters in the fit.

The results above should be interpreted as analysis at a fixed order in the HQE expansion, specifically up to $1/m_b^3$. As such, no theoretical uncertainty associated with missing higher-orders in the HQE expansion are included. As discussed above, our fit setup thus significantly differs from the approach of Refs.~\cite{Bernlochner:2022ucr, Carvunis:2025vab,Finauri:2023kte}. We investigate the effect of the missing higher orders in Sec.~\ref{sec:convergence}.

\begin{table}[]
    \centering
    \caption{Determined fit parameters and uncertainties 
    with free scale parameters (top rows) and with fixed scale parameters (bottom rows).}
    \resizebox{\textwidth}{!}{\begin{tabular}{lrrrrrrrr}
\toprule
 & $|V_{cb}|$ & $\overline{m}_{b}(\overline{m}_{b})$ & $\overline{m}_{c}(3\,\mathrm{GeV})$ & $\alpha_s(m_{Z^0})$ & $\mu_\pi^{2}$ & $\mu_G^{2}$ & $\rho_D^{3}$ & $\rho_{LS}^{3}$ \\
\midrule
Central & 42.316 & 4.199 & 1.000 & 0.118 & 0.383 & 0.134 & 0.110 & 0.007 \\
Fit Uncertainty & 0.284 & 0.010 & 0.010 & 0.001 & 0.014 & 0.032 & 0.004 & 0.070 \\
Hesse & 0.246 & 0.005 & 0.006 & 0.001 & 0.012 & 0.019 & 0.001 & 0.055 \\
Minos $-$ & -0.287 & -0.010 & -0.010 & -0.001 & -0.014 & -0.032 & -0.004 & -0.070 \\
Minos $+$ & 0.281 & 0.010 & 0.010 & 0.001 & 0.014 & 0.032 & 0.004 & 0.070 \\
\midrule
Central & 43.181 & 4.174 & 1.022 & 0.118 & 0.413 & 0.025 & 0.133 & 0.139 \\
Fit Uncertainty & 0.271 & 0.009 & 0.005 & 0.001 & 0.014 & 0.026 & 0.001 & 0.063 \\
Hesse & 0.271 & 0.009 & 0.005 & 0.001 & 0.014 & 0.026 & 0.001 & 0.063 \\
Minos $-$ & -0.270 & -0.009 & -0.005 & -0.001 & -0.014 & -0.026 & -0.001 & -0.063 \\
Minos $+$ & 0.272 & 0.009 & 0.005 & 0.001 & 0.014 & 0.026 & 0.001 & 0.063 \\
$\mu_{m_b}$ + & 0.630 & 0.019 & 0.017 & 0.001 & 0.042 & 0.081 & 0.015 & 0.177 \\
$\mu_{m_b}$ - & -0.446 & -0.022 & -0.015 & -0.002 & -0.021 & -0.119 & -0.015 & -0.159 \\
$\mu_{m_c}$ + & 1.011 & 0.004 & 0.076 & 0.000 & 0.040 & 0.049 & 0.000 & 0.375 \\
$\mu_{m_c}$ - & -0.253 & -0.005 & -0.022 & -0.003 & -0.010 & -0.194 & -0.010 & -0.096 \\
$\mu_{\alpha_s}$ + & 0.580 & 0.002 & 0.004 & 0.001 & 0.018 & 0.027 & 0.015 & 0.014 \\
$\mu_{\alpha_s}$ - & -0.443 & 0.000 & -0.002 & -0.001 & -0.010 & -0.033 & -0.010 & -0.018 \\
Scale Variations + & 1.325 & 0.020 & 0.078 & 0.001 & 0.061 & 0.098 & 0.021 & 0.415 \\
Scale Variations - & -0.678 & -0.022 & -0.027 & -0.004 & -0.025 & -0.230 & -0.021 & -0.186 \\
Total + & 1.353 & 0.022 & 0.078 & 0.001 & 0.062 & 0.102 & 0.021 & 0.420 \\
Total - & -0.730 & -0.024 & -0.027 & -0.004 & -0.029 & -0.231 & -0.021 & -0.197 \\
\bottomrule
\end{tabular}
}
    \label{tab:default_fit_result}
\end{table}

\subsection{Likelihood profiles and the role of the scale boundaries}
\label{sec:profiles}
\begin{figure}
    \centering
    \includegraphics[width=\linewidth]{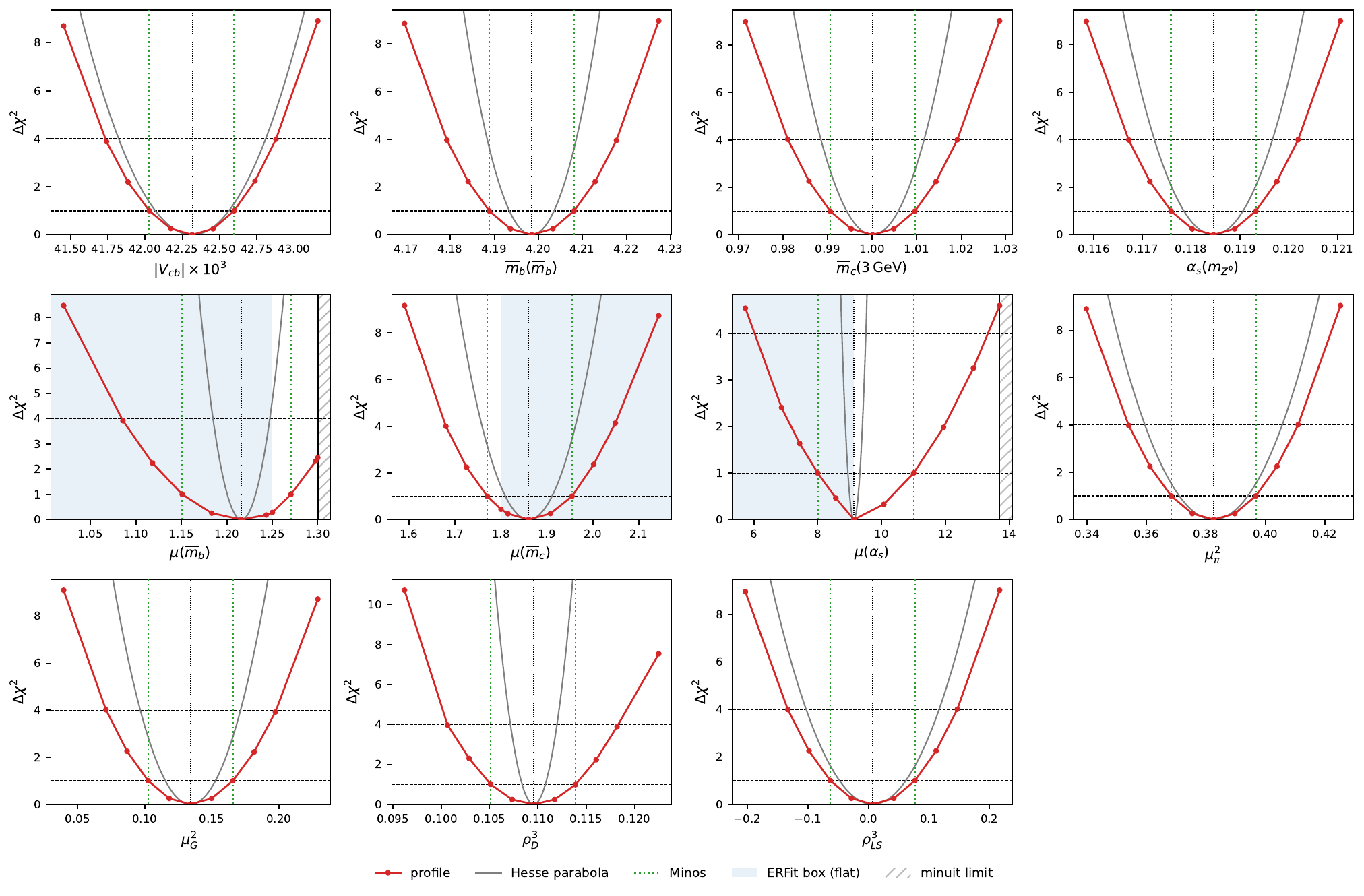}
    \caption{Profile of the $\chi^2$ in each fitted parameter of the default
    $\mathcal{O}(1/m_b^3)$ fit with free scales. At each point the parameter shown is fixed
    and all others are re-minimised. The grey curve is the parabola implied by the Hessian
    uncertainty, the dotted vertical lines mark the Minos interval, the shaded band is the
    flat region of the Rfit penalty and the hatched region lies beyond the hard limit imposed
    on the parameter. Horizontal lines indicate $\Delta\chi^2 = 1$ and $4$.}
    \label{fig:profiles_default_full}
\end{figure}

The imposed boundaries on the scale parameters can have an effect on the estimated uncertainties through the Hesse approximation if the central values are close to the boundary. To validate the fit uncertainty, we have profiled the likelihood in every fitted parameter: the parameter is fixed
on a grid around the minimum and all remaining parameters are re-minimised at each point. The $\Delta\chi^2$ along that direction is shown in  Fig.~\ref{fig:profiles_default_full}. The penalty on the scales remains
active and is indicated by the shaded
band, the region beyond the hard limit imposed on the parameter is indicated by the hatched wall.

Three features are worth noting. First, the scale $\scaleas$ is located at the edge of
the a-priori range. Second, the Hessian error on $\scaleas$ is not a confidence interval. The profile makes the
size of the discrepancy explicit: the Hessian returns an uncertainty a factor
$\numScaleasMbThreeRatio$ smaller than the interval at which the profile crosses
$\Delta\chi^2 = 1$, and the latter is strongly asymmetric. The same comparison for the physical
parameters shows disagreements between the Hessian error and the profile interval of up to
a factor $\numRhoDMbThreeRatio$ for those parameters that correlate with $\vec\mu$, notably
$\mcMS$ and $\rhoD$. With the scales fixed at
their nominal values the Hessian error, the Minos interval and the profile agree to the
percent level for every parameter. The non-Gaussianity is therefore a property of the
free-scale configuration, and specifically of the kink in the penalty at which the minimum
sits, and not a generic feature of the fit.

Third, the profiles show which constraint is active for each scale. Here the edge of the flat
Rfit region and the hard limit imposed on the parameter must be kept apart: for $\scalembkin$
they are $\numBoxMbkinHigh$ and $\numLimitMbkinHigh\,\mathrm{GeV}$, and both are reached within
the scanned range, so that the profile is visibly asymmetric between them. For $\scalemcMS$
the box starts at $\numBoxMcMSLow\,\mathrm{GeV}$ while the hard limit is at
$\numLimitMcMSLow\,\mathrm{GeV}$, and the minimum sits just inside the lower edge of the box.

These features lead to the disagreement between the estimated Hessian and Minos uncertainties: The curvature at the minimum is dominated by the boundary, and the Hessian reports an interval that is too narrow. On the flat side the $\chi^2$ rises much more slowly than a parabola through the minimum would suggest, and only the profile captures this. The grey parabolas in Fig.~\ref{fig:profiles_default_full} show the Hessian expectation directly against the actual profiles: wherever the two curves separate, the Hessian error should not be read as a confidence interval, and we therefore quote the Minos intervals throughout.

\subsection{Fit quality and leave-one-out Validation of the Experimental Measurements}
\label{sec:leave_one_out}

The p-value obtained when fitting the full set of experimental data described in Sec.~\ref{sec:inputs} is \numPvalueDefault{}, indicating that either the theoretical model is incomplete or that the uncertainties of the experimental or theoretical inputs have been underestimated. To investigate the former, we perform a leave-one-out validation, in which the moments measured by one experiment are removed one at a time. A variation in the fit quality upon the removal of a particular dataset might suggest that the corresponding measurement is in tension with the rest, while the theoretical model is otherwise capable of describing the remaining data consistently. 

The results of this validation are summarized in Tab.~\ref{tab:leave_on_out_chi2}, which clearly shows that one of the experimental datasets has a dominant impact on the overall fit quality. When excluding the Belle $q^2$ moment measurement, the theoretical model describes the experimental data well without the need for additional theoretical uncertainties, yielding a p-value of \numPvalueNoBelleqsq. A tension of the Belle $q^2$ moments with the $E_\ell$ and $M_X$ moments was previously noted in \cite{Finauri:2023kte}. Excluding both $q^2$ data sets also improves the p-value. Similarly, removing either the BaBar $E_l$ or the Belle $M_X^2$ moments improves the p-value, although less pronounced. To clarify this situation, new correlated measurements of all the moments are needed. 

The leave-one-out validation suggests that the data is internally inconsistent, but does not rule out underestimated theory uncertainties.  That is, the HQE at $\mathcal{O}(1/m_b^3)$ with the present perturbative accuracy could be inadequate to describe the experimental data. Previous analyses in Refs.~\cite{Bernlochner:2022ucr, Carvunis:2025vab,Finauri:2023kte} obtain good fit quality when accounting for an ad hoc modelling of missing higher-order HQE parameters using a correlation model. Typically, a theory covariance matrix is added which is obtained by varying $\rho_D^3$ and $\mu_G^2$ by $20$--$30\%$, which for several moments exceeds the experimental uncertainty. A good $p$-value is therefore close to guaranteed by construction.
In the following, we discuss the convergence and the impact of these missing higher-order HQE parameters in detail to obtain an uncertainty estimate. In our treatment including these effects does not improve the p-value. 
Here we do not account for such effects, instead we present an analysis at fixed order in the HQE. As such, our fit ansatz is much more rigid resulting in the poor p-value. We conclude that in a fixed-order HQE analysis with no ad hoc power-correction uncertainties, the current moment dataset cannot be described consistently. 
\begin{table}
    \centering
    \caption{Goodness of fit for the leave-one-out validation described in the text. }
    \label{tab:leave_on_out_chi2}
    \begin{tabular}{lrrrr}
\toprule
 & $\chi^2$ & d.o.f. & p-value & $\chi^2$ / d.o.f. \\
\midrule
Default & 119.5 & 67 & $<10^{-4}$ & 1.78 \\
All $q^2$ excluded & 51.8 & 40 & 0.0996 & 1.30 \\
Babar $E_l$ excluded & 77.4 & 57 & 0.0372 & 1.36 \\
Babar $M_X$ excluded & 113.0 & 59 & $<10^{-4}$ & 1.92 \\
Belle $E_l$ excluded & 109.9 & 61 & 0.0001 & 1.80 \\
Belle $M_X$ excluded & 87.8 & 60 & 0.0112 & 1.46 \\
Belle $q^2$ excluded & 62.6 & 55 & 0.2257 & 1.14 \\
Belle II $q^2$ excluded & 94.2 & 52 & 0.0003 & 1.81 \\
CDF $M_X$ excluded & 115.1 & 65 & 0.0001 & 1.77 \\
Cleo $M_X$ excluded & 113.3 & 63 & 0.0001 & 1.80 \\
Delphi $E_l$ excluded & 117.1 & 64 & $<10^{-4}$ & 1.83 \\
Delphi $M_X$ excluded & 119.5 & 65 & $<10^{-4}$ & 1.84 \\
\bottomrule
\end{tabular}

\end{table}

\subsection{Convergence and the Impact of Missing Higher Orders in the HQE}
\label{sec:convergence}

To evaluate the convergence of the Heavy Quark Expansion and ensure the reliability of the extracted $|V_{cb}|$, we employ a systematic truncation-testing strategy. We perform independent global fits at successive orders of $1/m_b$, specifically at $\mathcal{O}(1/m_b^3)$, $\mathcal{O}(1/m_b^4)$, and $\mathcal{O}(1/m_b^5)$, monitoring the stability of the central values as the series is extended. For this study, we include the HQE parameters at $\mathcal{O}(1/m_b^4)$, and $\mathcal{O}(1/m_b^5)$, called $m_i$ and $r_i$, respectively (defined in Appendix \ref{app:HQE_def}). We then use Gaussian priors $\mathcal{G}(0,\Lambda_\mathrm{QCD}^n)$ with $\Lambda_\mathrm{QCD}=0.3\,\mathrm{GeV}$ and $n=4,5$ based on the order in $1/m_b$ to set up the fit. 

\subsubsection{Extraction of higher-order coefficients}

\begin{figure}
    \centering
    \includegraphics[width=1\linewidth]{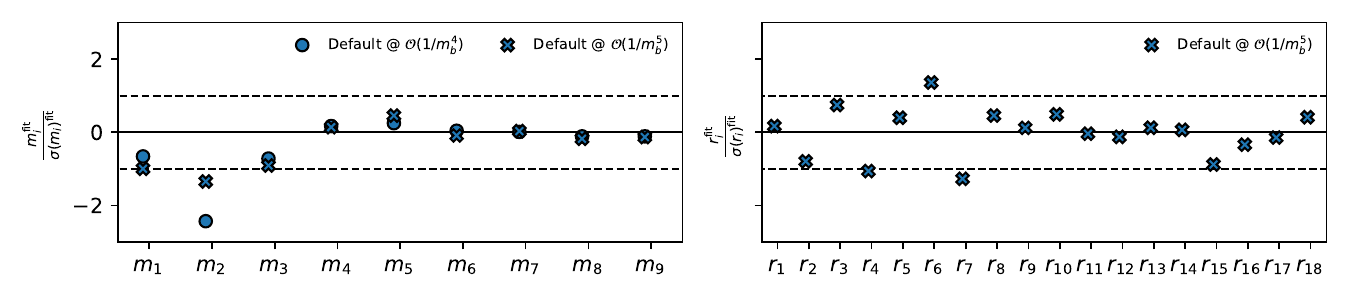}
    \caption{Significance of the higher-order HQE coefficients. The coefficients are constrained to $\Lambda_\mathrm{QCD}^n$, where $n$ denotes the order in the HQE and $\Lambda_\mathrm{QCD} = 0.3\,\mathrm{GeV}$. }
    \label{fig:pulls_on_higher_order_coeff}
\end{figure}

The pulls of the higher-order coefficients are shown in Fig.~\ref{fig:pulls_on_higher_order_coeff}, which shows no significant influence of the applied constraints on the fit results. 

The variations in the higher-order terms partially compensate for the poorer overall fit quality. Specifically, we note that if we exclude the \textit{Belle $q^2$} moments, which yields an improved fit quality, these coefficients play only a minor role as the data are already well described at $\mathcal{O}(1/m_b^3)$. This suggests that in our nominal fit, the higher-order coefficients may be used by the fit to absorb tensions in the experimental data, rather than reflecting genuine sensitivity of the full dataset to these parameters.

\begin{figure}
    \centering
    \includegraphics[width=1\linewidth]{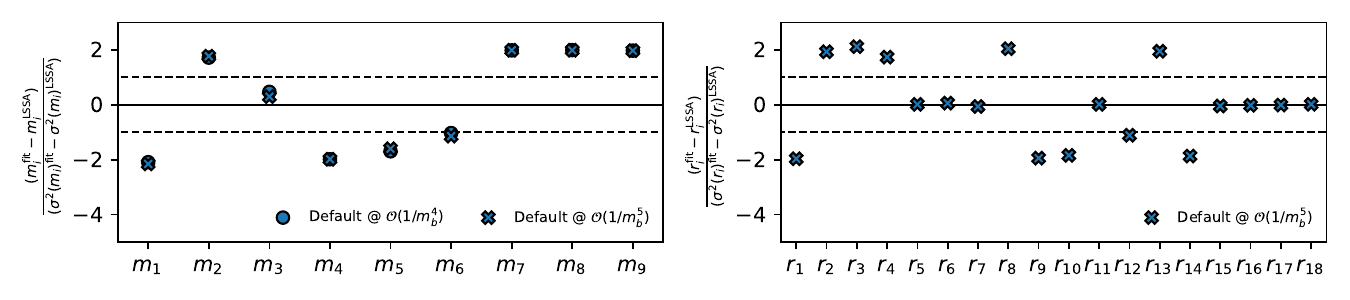}
    \caption{Comparison of the extracted higher-order coefficients with the predictions based on the LLSA as described in the text.}
    \label{fig:LLSA_higher_order_coeff}
\end{figure}

In Fig.~\ref{fig:LLSA_higher_order_coeff}, the LLSA predictions are compared with the values extracted for $m_i, r_i$ from the $1/m_b^{4,5}$ fits. These LLSA predictions are obtained following ~\cite{Heinonen:2014dxa, Mannel:2023yqf} (see Sec.~\ref{sec:semilep}) and using $M_Q=5.27926(26)$ GeV, $M_{1/2}=5.670(16)$ GeV and $M_{3/2}=5.7266(15)$ GeV \cite{ParticleDataGroup:2024cfk}. In addition, we use $\mu_\pi^2$ and $\mu_G^2$ from our fit at $\mathcal{O}(1/m_b^3)$. As discussed, the LLSA is based on the lowest lying states only, which induces an uncertainty which may be as large as $\sim50\%$ \cite{Heinonen:2014dxa}. To be conservative, we account for this large uncertainty\footnote{For coefficients that vanishing in the LLSA, we use an uncertainty of 0.05 GeV$^n$ with $n=4,5$ at $\mathcal{O}(1/m_b^{4,5})$.}.
 We observe that for most $1/m_b^4$ and $1/m_b^5$ HQE parameters there are significant differences between the LLSA predictions and values extracted from the fits. This may be attributed to the fit absorbing tensions of the experimental data not the higher-order coefficients. Moreover, higher-order HQE parameters are more likely to receive significant contributions from other states besides the lowest-lying ones in the LLSA. On the theory side, it would be interesting to expand the LLSA and include subleading contributions to better understand higher-order corrections. In addition, it would be interesting to repeat this analysis including fourth moments of the $q^2$ and lepton energy moments as those are expected to be sensitive to power corrections. The latter are currently not available.

\subsubsection{Additional uncertainty from higher-order terms}\label{sec:hounc}
To quantify the theoretical uncertainty associated with the truncation of the OPE, we adopt the following approach: while our baseline result is derived at $\mathcal{O}(1/m_b^3)$, we add an additional uncertainty obtained by incorporating the maximum difference between the $\mathcal{O}(1/m_b^3)$ and $\mathcal{O}(1/m_b^{4,5})$ fit results, e.g.
\begin{equation}
    \sigma_\mathrm{h.o.}(X) = \max_{n\in\{4,5\}}\left|X_{\mathcal{O}(1/m_b^n)} - X_{\mathcal{O}(1/m_b^3)}\right| \ .
\end{equation}
This additional error budget serves as a data-driven estimate for the impact of higher-order effects, ensuring that our final precision accounts for the potential contributions of the $1/m_b^4$ terms and beyond. 

\begin{figure}
    \centering
    \includegraphics[width=1\linewidth]{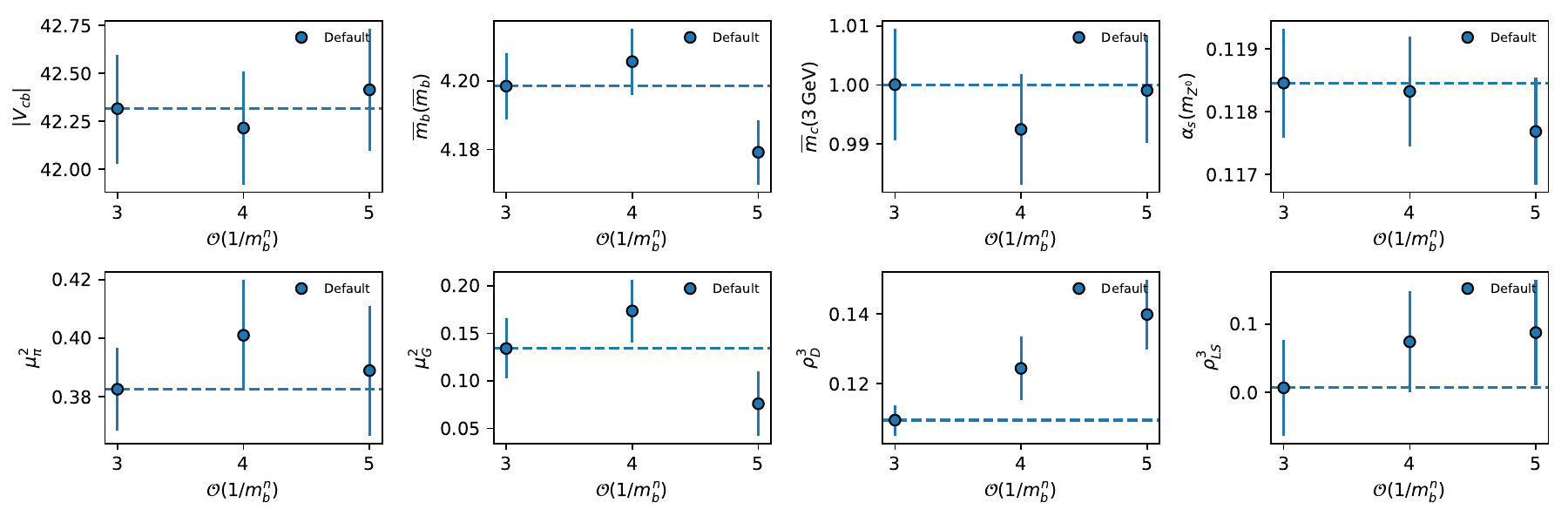}
    \caption{Extracted parameters of interest at orders $\mathcal{O}(1/m_b^n)$ with $n = \{ 3, 4, 5\}$ in the HQE. The horizontal lines indicate the central values obtained at $\mathcal{O}(1/m_b^3)$ and are shown for reference.}
    \label{fig:convergence_parameters}
\end{figure}

The parameters of interest obtained at different orders of the HQE are shown in Fig.~\ref{fig:convergence_parameters} and Tab.~\ref{tab:convergence_parameters}. We find that $|V_{cb}|$ and $\mupi$ are stable: across $n=3,4,5$ they vary by $\numRelSpreadVcb\%$ and $\numRelSpreadMupi\%$ respectively, comparable to their fit uncertainties. For $\muG$, we observe large shifts going to higher orders, up to $\numRelSpreadMuG\%$ (see also the discussion in Sec.~\ref{sec:valhqe}). The $\rhoD$ parameter increases by about $\numRelStepRhoD\%$ at each order. 

Our final results for the fitted parameters and their higher-order uncertainty are given in Tab.~\ref{tab:default_fit_result_with_ho}. The correlation matrix is given in Tab.~\ref{tab:default_correlaton_matrix}\footnote{The correlation coefficients presented here are derived from the Hesse approximation at the minimum and has to be interpreted cautiously.}.

Specifically, we obtain 
\newcommand{\VcbDefaultWithHO}{\ensuremath{|V_{cb}| = \left(\numVcb \pm \numVcbErr|_{\rm fit} \pm \numVcbSigmaHO|_{\rm h.o.}\right) \times 10^{-3} = (\numVcb \pm \numVcbTotal) \times 10^{-3}}}%
\begin{equation}\label{eq:vcbfinal}
   \VcbDefaultWithHO ,
\end{equation}
where we added the uncertainties from the fit and higher-order terms (h.o.) in quadrature. 

\newcommand{\TotalBRDefaultWithHO}{\ensuremath{\mathcal{B}(\bar B \to X_c l \bar{\nu}_l) = (\numBRCentral \pm \numBRErr|_{\rm fit} \pm \numBRSigmaHO|_{\rm h.o.})\,\%}}

For the total inclusive semileptonic branching ratio, we find 
$\mathcal{B} = \numBRMbThree\%$, $\numBRMbFour\%$ and $\numBRMbFive\%$ at
$\mathcal{O}(1/m_b^{3})$, $\mathcal{O}(1/m_b^{4})$ and $\mathcal{O}(1/m_b^{5})$.  
Our default determination then reads,
\begin{equation}\label{eq:brancfit}
   \TotalBRDefaultWithHO{} \ ,
\end{equation}
where we averaged over $B^0$ and $B^+$ and used the \texttt{kolya} default lifetimes $\tau_{B^+} = 1.638$ ps and $\tau_{B^0}=1.519$ ps to convert the semileptonic width to the branching ratio. Below, we also quote the branching ratio using the theoretical lifetime. 

\begin{table}[]
    \centering
    \caption{Extracted parameters of interest and the predicted semileptonic branching ratio $\mathcal{B}$ at orders $\mathcal{O}(1/m_b^n)$ with $n = \{ 3, 4, 5\}$ in the HQE.}
    \resizebox{\textwidth}{!}{\begin{tabular}{lrrrr}
\toprule
 & $\mathcal{O}(1/m_b^n)$ & $\mathcal{O}(1/m_b^3)$ & $\mathcal{O}(1/m_b^4)$ & $\mathcal{O}(1/m_b^5)$ \\
 & Parameter &  &  &  \\
\midrule
\multirow[t]{8}{*}{Default} & $|V_{cb}|$ & 42.316 ± 0.284 & 42.216 ± 0.296 & 42.415 ± 0.318 \\
 & $\overline{m}_{b}(\overline{m}_{b})$ & 4.199 ± 0.010 & 4.206 ± 0.010 & 4.179 ± 0.009 \\
 & $\overline{m}_{c}(3\,\mathrm{GeV})$ & 1.000 ± 0.010 & 0.992 ± 0.009 & 0.999 ± 0.009 \\
 & $\alpha_s(m_{Z^0})$ & 0.118 ± 0.001 & 0.118 ± 0.001 & 0.118 ± 0.001 \\
 & $\mu_\pi^{2}$ & 0.383 ± 0.014 & 0.401 ± 0.019 & 0.389 ± 0.022 \\
 & $\mu_G^{2}$ & 0.134 ± 0.032 & 0.173 ± 0.033 & 0.076 ± 0.034 \\
 & $\rho_D^{3}$ & 0.110 ± 0.004 & 0.124 ± 0.009 & 0.140 ± 0.010 \\
 & $\rho_{LS}^{3}$ & 0.007 ± 0.070 & 0.074 ± 0.074 & 0.088 ± 0.077 \\
\cline{1-5}
\bottomrule
\end{tabular}
}
    \label{tab:convergence_parameters}
\end{table}

\begin{table}[]
    \centering
    \caption{Determined fit parameters and uncertainties with free scale parameters and the estimated uncertainty from missing higher orders described in the text.}
    \resizebox{\textwidth}{!}{\begin{tabular}{lrrrrrrrrrrr}
\toprule
 & $|V_{cb}|$ & $\overline{m}_{b}(\overline{m}_{b})$ & $\overline{m}_{c}(3\,\mathrm{GeV})$ & $\alpha_s(m_{Z^0})$ & $\mu(\overline{m}_{b}(\overline{m}_{b}))$ & $\mu(\overline{m}_{c}(3\,\mathrm{GeV}))$ & $\mu(\alpha_s)$ & $\mu_\pi^{2}$ & $\mu_G^{2}$ & $\rho_D^{3}$ & $\rho_{LS}^{3}$ \\
\midrule
Central & 42.316 & 4.199 & 1.000 & 0.118 & 1.216 & 1.860 & 9.130 & 0.383 & 0.134 & 0.110 & 0.007 \\
Fit Uncertainty & 0.284 & 0.010 & 0.010 & 0.001 & 0.060 & 0.092 & 1.503 & 0.014 & 0.032 & 0.004 & 0.070 \\
Hesse & 0.246 & 0.005 & 0.006 & 0.001 & 0.016 & 0.051 & 0.188 & 0.012 & 0.019 & 0.001 & 0.055 \\
Minos $-$ & -0.287 & -0.010 & -0.010 & -0.001 & -0.065 & -0.090 & -1.133 & -0.014 & -0.032 & -0.004 & -0.070 \\
Minos $+$ & 0.281 & 0.010 & 0.010 & 0.001 & 0.054 & 0.094 & 1.872 & 0.014 & 0.032 & 0.004 & 0.070 \\
Higher Order Uncertainty & 0.101 & 0.019 & 0.008 & 0.001 & 0.189 & 0.142 & 0.468 & 0.018 & 0.058 & 0.030 & 0.081 \\
\bottomrule
\end{tabular}
}
    \label{tab:default_fit_result_with_ho}
\end{table}

\begin{table}[]
    \centering
    \caption{Correlation matrix for the fit uncertainty component.}
    \label{tab:default_correlaton_matrix}
    \resizebox{\textwidth}{!}{\begin{tabular}{lrrrrrrrrrrr}
\toprule
 & $|V_{cb}|$ & $\overline{m}_{b}(\overline{m}_{b})$ & $\overline{m}_{c}(3\,\mathrm{GeV})$ & $\alpha_s(m_{Z^0})$ & $\mu(\overline{m}_{b}(\overline{m}_{b}))$ & $\mu(\overline{m}_{c}(3\,\mathrm{GeV}))$ & $\mu(\alpha_s)$ & $\mu_\pi^{2}$ & $\mu_G^{2}$ & $\rho_D^{3}$ & $\rho_{LS}^{3}$ \\
\midrule
$|V_{cb}|$ & $1.000$ & $-0.332$ & $0.072$ & $-0.006$ & $-0.016$ & $0.123$ & $-0.041$ & $0.265$ & $-0.132$ & $-0.077$ & $0.165$ \\
$\overline{m}_{b}(\overline{m}_{b})$ & $-0.332$ & $1.000$ & $0.338$ & $-0.406$ & $-0.132$ & $-0.175$ & $-0.025$ & $-0.334$ & $0.317$ & $-0.320$ & $-0.174$ \\
$\overline{m}_{c}(3\,\mathrm{GeV})$ & $0.072$ & $0.338$ & $1.000$ & $-0.382$ & $-0.210$ & $0.586$ & $-0.050$ & $0.108$ & $-0.346$ & $-0.300$ & $0.305$ \\
$\alpha_s(m_{Z^0})$ & $-0.006$ & $-0.406$ & $-0.382$ & $1.000$ & $0.408$ & $-0.048$ & $0.086$ & $-0.063$ & $0.232$ & $0.567$ & $-0.251$ \\
$\mu(\overline{m}_{b}(\overline{m}_{b}))$ & $-0.016$ & $-0.132$ & $-0.210$ & $0.408$ & $1.000$ & $0.120$ & $-0.051$ & $0.015$ & $0.005$ & $-0.333$ & $-0.034$ \\
$\mu(\overline{m}_{c}(3\,\mathrm{GeV}))$ & $0.123$ & $-0.175$ & $0.586$ & $-0.048$ & $0.120$ & $1.000$ & $-0.064$ & $0.028$ & $-0.176$ & $-0.200$ & $0.261$ \\
$\mu(\alpha_s)$ & $-0.041$ & $-0.025$ & $-0.050$ & $0.086$ & $-0.051$ & $-0.064$ & $1.000$ & $-0.013$ & $0.012$ & $-0.060$ & $0.014$ \\
$\mu_\pi^{2}$ & $0.265$ & $-0.334$ & $0.108$ & $-0.063$ & $0.015$ & $0.028$ & $-0.013$ & $1.000$ & $-0.043$ & $-0.040$ & $0.661$ \\
$\mu_G^{2}$ & $-0.132$ & $0.317$ & $-0.346$ & $0.232$ & $0.005$ & $-0.176$ & $0.012$ & $-0.043$ & $1.000$ & $0.042$ & $-0.002$ \\
$\rho_D^{3}$ & $-0.077$ & $-0.320$ & $-0.300$ & $0.567$ & $-0.333$ & $-0.200$ & $-0.060$ & $-0.040$ & $0.042$ & $1.000$ & $-0.046$ \\
$\rho_{LS}^{3}$ & $0.165$ & $-0.174$ & $0.305$ & $-0.251$ & $-0.034$ & $0.261$ & $0.014$ & $0.661$ & $-0.002$ & $-0.046$ & $1.000$ \\
\bottomrule
\end{tabular}
}
\end{table}
We compare our results with the recent analysis from \cite{Carvunis:2025vab}, which uses the same data set but a different fit setup as discussed above. They find
\begin{equation}\label{eq:latres}
    |V_{cb}| = (41.64 \pm 0.47) \times 10^{-3} \ , \quad\quad \mathcal{B}(\bar B \to X_c l \bar{\nu}_l) = (10.62 \pm 0.15)\,\% \ .
\end{equation}
Note that in their analysis also $\mathcal{O}(\alpha_{\rm em})$ effects are considered, which lower the value of $|V_{cb}|$. These results are in agreement with the first determination of $|V_{cb}|$ using the full dataset of $q^2, E_\ell$ and $M_X$ moments in \cite{Finauri:2023kte}. Our $|V_{cb}|$ is higher by about $1\sigma$, with smaller uncertainty. We note that the full branching ratio quoted above cannot be compared directly with measurements as the latter always involve a kinematical cut on the lepton energy or include the $b\to u$ contribution. In Ref.~\cite{Bernlochner:2022ucr}, different branching ratio measurements were converted to the full branching ratio using theoretical inputs, which leads to a value even lower than Eq.~\eqref{eq:latres}. New branching ratio measurements, possibly with $q^2$-cuts, are therefore highly encouraged to clarify the situation.

Finally, we comment on the quoted uncertainty. In our analysis, the scale $\scaleas$ is pushed to its boundary, such that the perturbative corrections are much smaller than at conventional scales. The reduction of the uncertainty is therefore inherent to the model as the truncation of the perturbative series is not take into account, and not due to improved theoretical inputs. As such the uncertainties in our setup and previous works are different and should be averaged or compared with caution. In our fixed-scale analysis, we found that the scale variation of ${}^{+1.4}_{-0.7}$, significantly larger than in the free-scale setup. 

\subsection{Comment on the extracted HQE parameters}\label{sec:valhqe}
It is interesting to compare our final extracted HQE parameters in Table~\ref{tab:default_fit_result_with_ho} with previous analyses. At the same time, we note, as above, the difference in our setup with respect to previous analysis. It is important to take into account the specific fit setup and conditions when using extracted HQE parameters as inputs to predict for example $B\to X_u \ell \nu$ or $B_{s,d}\to X_s \ell \ell$. For lifetime predictions, our work circumvents this issue by simultaneously analysing the lifetime and the semileptonic rate, as discussed in the next section. We restrict ourselves to comparing to \cite{Carvunis:2025vab}, which uses the same data as our analysis and includes the $\alpha_s^2$ corrections to the $q^2$ moments from \cite{Fael:2024gyw}, thereby updating \cite{Finauri:2023kte} where only $\alpha_s^2\beta_0$ terms were included. The analysis in \cite{Finauri:2023kte} is the first to include also $q^2$ moments, which were previously studied separately by \cite{Bernlochner:2022ucr}. We refer to these works for a discussion on the impact of including $q^2$ moments. 

For $\mu_\pi^2$, we observe agreement with \cite{Carvunis:2025vab} around the $1\sigma$ level, where we obtain a slightly reduced uncertainty. This uncertainty is dominated by the fit uncertainty and higher-order corrections only have a limited effect.

Our value of $\muG$ is significantly lower than in previous works. We also note a large additional uncertainty from the higher-order terms. From Table~\ref{tab:convergence_parameters}, we observe a larger shift when going from $1/m_b^4$ to $1/m_b^5$. We  note that at these orders also cross-terms like $\mu_G^2 \times \rho_D^3$ contribute, introducing an additional dependence on $\mu_G^2$ when going from $1/m_b^3$ to higher powers. The chromo-magnetic operator can be related to the mass difference 
(see also \cite{Gambino:2012rd})
\begin{equation}
\frac{3}{4}\left(m_{B^*}^2- m_{B}^2\right) = C_{\rm mag}(\scaleas)\,\mu_G^2(\scaleas)+\mathcal{O}\left(\alpha_s \mu_G^2, 1/{m_b^3}\right)\ , \quad \frac{3}{4}\left(m_{B^*}^2- m_{B}^2\right) = 0.358\,\mathrm{GeV}^2 \ ,
\end{equation}
with $C_{\rm mag}=1$. Unknown higher-order terms arise as the relation is only valid in the HQET limit\footnote{This relation is often used to put a mild external constraint on $\muG$. Here we do not add such an external constraint.}. A naive estimate gives $\mathcal{O}(20\%)$ corrections to this relation. We find a value about a factor of 3 smaller, which is in tension with the leading order relation above although corrections could be large. It would be interesting to see if for a new analysis which also included correlations between the moments, this tension remains.

Regarding $\rho_{LS}^3$, it is known that the moments have a very small sensitivity to this parameter (for RPI observables its contribution vanishes). This lack of sensitivity is reflected in the uncertainty obtained from the fit. Previous fits typically prefer $\rho_{LS}^3 < 0$, see \cite{Carvunis:2025vab}. Our default fit gives a positive central value that is consistent with zero and, within its large uncertainty, with a negative value. Including $\mathcal{O}(1/m_b^{4,5})$ terms, pushes the fitted values further up. The fixed-scale fit also gives a positive value of about $2\sigma$ away from zero.

The parameter \rhoD is of key importance for lifetime and other HQE predictions due to its large prefactor. We find a small fit uncertainty; the uncertainty from higher orders is more than seven times larger. Adding the fit and higher-order components in quadrature gives
\begin{equation}
    \rhoD = \numrhoD \pm 0.030 \ .
\end{equation}
Compared to $\rhoD = 0.164 \pm 0.018$ obtained in \cite{Carvunis:2025vab}, we find agreement at the $1.5\sigma$ level.

Finally, we briefly comment on the analysis in \cite{Bernlochner:2022ucr}. In this analysis, only $q^2$ moments were considered and the RPI basis was employed. As such, caution needs to be taken when comparing the different HQE parameters as the perturbative corrections cause shifts of order $\mu^3/m_b^3 \alpha_s$ and in addition, the analysis has no sensitivity to $\mu_\pi^2$. The key point of this analysis was to extract higher-order $1/m_b^4$ HQE parameters directly from the data. In \cite{Bernlochner:2022ucr}, a significant difference between the extracted \rhoD between the fit at $\mathcal{O}(1/m_b^3)$ and $\mathcal{O}(1/m_b^4)$ was observed, where the latter had a sizeable uncertainty. The large uncertainty of the higher-order corrections found in the current analysis reflects this as well.

\subsection{Lifetime and total rate predictions}\label{sec:lifetimepred}

\newcommand{\fourerr}[8]{\ensuremath{#1^{+#2}_{-#3}{}^{+#4}_{-#5}{}^{+#6}_{-#7}\pm#8}}

In order to predict the lifetimes, 
we depart from the treatment adopted for the semileptonic observables. In the semileptonic fit, the moments constrain $\scaleas$, $\scalembkin$, and $\scalemcMS$, and the corresponding best-fit values are subsequently used to predict the total semileptonic rate. 
This is justified because the moments and the rate are derived from the same triple-differential decay distribution and therefore share a common scale dependence. 
The total $B$-meson widths, however, receive contributions from several decay channels, including not only semileptonic transitions but also non-leptonic modes such as $b\to c\bar c s$ and $b\to c\bar u d$. Specifically, 
\begin{equation}
    \Gamma_{\rm tot} = \Gamma^{\rm SL}+ \Gamma^{\rm NL} \ ,
\end{equation}
with\footnote{In the following, we neglect the suppressed $b\to u \tau \nu$ mode.}
\begin{equation}
  \Gamma^{\rm SL} = 2\,\Gamma^{\rm SL}_{b\to c\ell\nu}+ 2\,\Gamma^{\rm SL}_{b\to u\ell\nu} + \Gamma^{\rm SL}_{b\to c\tau\nu} \ , \qquad \ell = e,\mu \ .\label{eq:gammasldef}
\end{equation}
The value of the scales of the semileptonic (SL) and non-leptonic (NL) channels is not necessarily the same. 

For the semileptonic contributions, we use our default fit scenario in which the scale parameters are treated as nuisance parameters. Although one may argue that the $b\to u$ and $b\to c\tau\nu$ parts are (also) not described by the exact same triple differential rate, we expect possible effects from this difference to be small. In addition, both these channels only contribute mildly to the total semileptonic width. 

For the non-leptonic contributions, we use the values of the scale parameters and the HQE parameters at the best fit point. However, to account for the different origins of the decays, we vary the range of the scale parameters as follows: $2.28\;{\rm GeV}<\mu_{\alpha_s}<9.12$ GeV, $0.8\;{\rm GeV}<\mu_{m_b^{\rm kin}}<1.2$ GeV and $1\;{\rm GeV}<\mu_{\overline{m}_c}<3$ GeV. This way we stay agnostic to the actual scale of the non-leptonic process and scan over a common choice for the parameter space. In both cases, we follow our fit setup visualized in Fig.~\ref{fig:scalesetting}. Our approach thereby differs from other approaches, as discussed.

The lifetime predictions are currently only available up to $1/m_b^3$ corrections. At the same time, we do have extractions of $\muG, \mupi, \rhoLS$ and $\rhoD$ from our semileptonic analysis  that includes $1/m_b^4$ and $1/m_b^5$ parameters (see Table~\ref{tab:convergence_parameters}). In the following, we use these parameters to account for an additional uncertainty. Once the higher-order contributions to the non-leptonic rate are known these can be included for a more consistent treatment of the missing orders. 

For the semileptonic rates, we find
\begin{align}\label{eq:slrates}
   \Gamma(\bar{B} \to X_c \ell \bar{\nu}_\ell) &= \left(\asym{\numGammaClnu}{\numGammaClnuFitUp}{\numGammaClnuFitDown}|_{\rm fit} \pm \numGammaClnuHO|_{\rm h.o.}\right)\;\mathrm{ps}^{-1} \ , \\
   \Gamma(\bar{B} \to X_u \ell \bar{\nu}_\ell) &= \left(\asym{\numGammaUlnu}{\numGammaUlnuFitUp}{\numGammaUlnuFitDown}|_{\rm fit} \pm \numGammaUlnuHO|_{\rm h.o.}\right)\cdot 10^{-3}\;\mathrm{ps}^{-1} \ , \\ 
   \Gamma(\bar{B} \to X_c \tau \bar{\nu}_\tau) &= \left(\asym{\numGammaCtaunu}{\numGammaCtaunuFitUp}{\numGammaCtaunuFitDown}|_{\rm fit} \pm \numGammaCtaunuHO|_{\rm h.o.}\right)\;\mathrm{ps}^{-1} \ ,
\end{align}
where the first uncertainty is the one propagated from the fit covariance by the toy ensemble and the second is the higher-order uncertainty. This is obtained as described in Sec.~\ref{sec:hounc}, by evaluating the prediction with the HQE parameters determined from the $\mathcal{O}(1/m_b^n)$ and $n=4,5$ fit and taking the maximal difference with the $\mathcal{O}(1/m_b^3)$ fit. 

We note that the $b\to u$ semileptonic rate depends on the initial flavour of the $B$ meson through weak annihilation (four-quark) effects. For completeness, we quote above the result for $\bar{B}^0$. The $B^+$ rate differs by $0.14\%$, due to weak annihilation. Combining the three channels with the
multiplicities of Eq.~\eqref{eq:gammasldef} and accounting for their correlations gives
\begin{equation}\label{eq:gammasl}
    \Gamma^{\rm SL} = \asym{\numGammaSLBzero}{\numGammaSLBzeroTotalUp}{\numGammaSLBzeroTotalDown}\;\mathrm{ps}^{-1} \ .
\end{equation}

For the non-leptonic width, we find
\begin{align}\label{eq:gammanl}
    \Gamma^{\rm NL}(B^0) &= \fourerr{\numGammaNLBzero}{\numGammaNLBzeroFitUp}{\numGammaNLBzeroFitDown}{\numGammaNLBzeroScaleUp}{\numGammaNLBzeroScaleDown}{\numGammaNLBzeroBagUp}{\numGammaNLBzeroBagDown}{\numGammaNLBzeroHO} = \asym{\numGammaNLBzero}{\numGammaNLBzeroTotalUp}{\numGammaNLBzeroTotalDown}\;\mathrm{ps}^{-1} \ , \nonumber \\
    \Gamma^{\rm NL}(B^+) &= \fourerr{\numGammaNLBplus}{\numGammaNLBplusFitUp}{\numGammaNLBplusFitDown}{\numGammaNLBplusScaleUp}{\numGammaNLBplusScaleDown}{\numGammaNLBplusBagUp}{\numGammaNLBplusBagDown}{\numGammaNLBplusHO} = \asym{\numGammaNLBplus}{\numGammaNLBplusTotalUp}{\numGammaNLBplusTotalDown}\;\mathrm{ps}^{-1} \ ,
\end{align}
where we quote the uncertainties from the fit, the scale variation, the bag parameters and the missing-higher orders. The scale variation dominates and is strongly asymmetric. This asymmetric uncertainty is a consequence of our procedure, as we start from the best fit point of the semileptonic fit and then vary the range. Since $\scaleas$ comes out large and sits at
the upper edge of its range, there is little upward variation, while lowering $\scaleas$ increases $\Gamma^{\rm NL}$ substantially. A similar effect is observed for $\scalembkin$. 

Adding Eq.~\eqref{eq:gammasl} to Eq.~\eqref{eq:gammanl} gives the total widths
\begin{align}\label{eq:totalwidth}
  \Gamma(B^0) &= (\asym{\numTauInvBzero}{\numTauInvBzeroTotalUp}{\numTauInvBzeroTotalDown})\;\mathrm{ps}^{-1} \ , &
  \Gamma(B^+) &= (\asym{\numTauInvBplus}{\numTauInvBplusTotalUp}{\numTauInvBplusTotalDown})\;\mathrm{ps}^{-1} \ ,
\end{align}
the corresponding lifetimes read
\begin{align}\label{eq:lifetimes}
  \tau_{B^0} &= (\asym{\numTauBzero}{\numTauBzeroTotalUp}{\numTauBzeroTotalDown})\;\mathrm{ps} \ , &
  \tau_{B^+} &= (\asym{\numTauBplus}{\numTauBplusTotalUp}{\numTauBplusTotalDown})\;\mathrm{ps} \ .
\end{align}
Our result is compatible with the recent HQE determination of Ref.~\cite{Egner:2024lay},
\begin{align}\label{eq:egner}
    \Gamma({B^0})=(0.636^{+0.028}_{-0.037})\, \text{ps}^{-1},\qquad\Gamma({B^+})=(0.587^{+0.025}_{-0.035})\, \text{ps}^{-1} \ ,
\end{align}
which are $\numEgnerPullBzero\sigma$ and $\numEgnerPullBplus\sigma$ larger for the $B^0$ and the $B^+$ than Eq.~\eqref{eq:totalwidth}. This shift comes from the lower central value of 
$\scaleas$ used in Ref.~\cite{Egner:2024lay}, which increases the rate. In addition, the total width strongly depends on $\rhoD$, for which we find a lower value than Ref.~\cite{Egner:2024lay}, and consequently our central value of the width is lower. Finally, we briefly note that the uncertainty of \eqref{eq:egner} is symmetric, which reflects the treatment of the scale variation.

\begin{figure}
    \centering
    \includegraphics[width=0.95\linewidth]{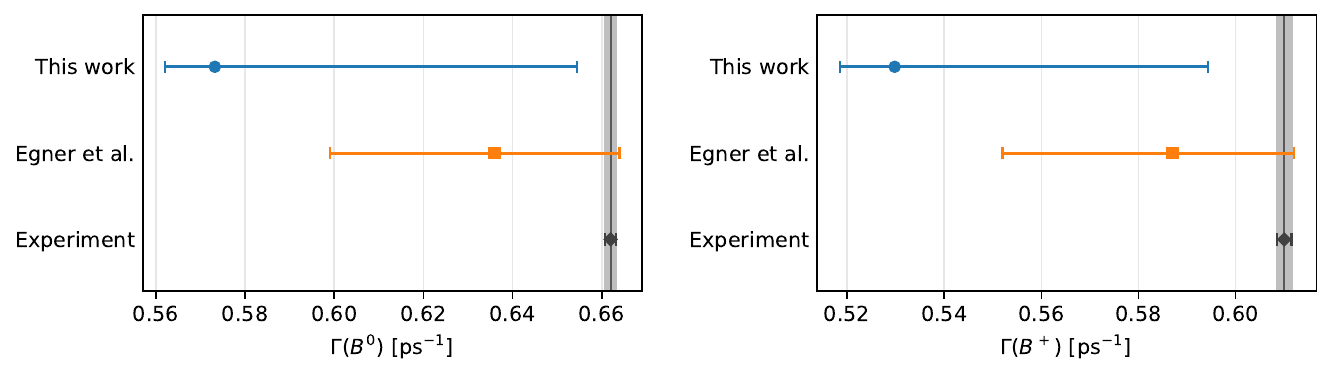}
    \caption{Total widths of the $B^0$ (left) and $B^+$ (right), showing our work~\eqref{eq:totalwidth}, the prediction of~\cite{Egner:2024lay} in
    Eq.~\eqref{eq:egner}, and the HFLAV average~\cite{HFLAV:2024ctg} given in Eq.~\eqref{eq:widthexp} as a vertical band.}
    \label{fig:width_comparison}
\end{figure}

The experimental averages for the total rates read \cite{HFLAV:2024ctg}
\begin{align}\label{eq:widthexp}
    \Gamma({B^0})|_{\rm exp}=(0.662 \pm 0.0012)\, \text{ps}^{-1},\qquad\Gamma({B^+})|_{\rm exp}=(0.610 \pm 0.0015)\, \text{ps}^{-1} \ ,
\end{align}
in agreement with the recent LHCb measurements \cite{Ozcelik:2026ichep}. Our central values are  $\numWidthDeficitBzero\%$ lower for both modes. 
Taking the upper uncertainty of
Eq.~\eqref{eq:totalwidth}, our predictions are 
$\numWidthPullBzero\sigma$ and $\numWidthPullBplus\sigma$ lower for $B^0$ and $B^+$, respectively. 
The compatibility between the predictions and the measurements stems mainly from the perturbative uncertainty of the non-leptonic width which is
large enough to cover the difference. The three determinations of the total width are compared in
Fig.~\ref{fig:width_comparison}.

For the ratio of the lifetimes, we obtain
\begin{align}\label{eq:tauratio}
    \frac{\tau_{B^+}}{\tau_{B^0}} = \fourerr{\numTauRatio}{\numTauRatioFitUp}{\numTauRatioFitDown}{\numTauRatioScaleUp}{\numTauRatioScaleDown}{\numTauRatioBagUp}{\numTauRatioBagDown}{\numTauRatioHO}
    = \asym{\numTauRatio}{\numTauRatioTotalUp}{\numTauRatioTotalDown} \ ,
\end{align}
where again we quote the fit uncertainty, the scale variation, the uncertainty from the bag parameters and the missing higher orders. In the ratio $|V_{cb}|$, the overall $m_b^5$
factor and the flavour-blind part of the width drop out, so that Eq.~\eqref{eq:tauratio} is
controlled almost entirely by the dimension-six four-quark matrix elements. The scale variation is reduced from $\numTauBzeroScaleRelDown\%$ for
$\tau_{B^0}$ to $\numTauRatioScaleRelUp\%$ in the ratio, but it does not cancel
completely, because the non-leptonic widths have a different dependence on
$\scaleas$. Comparing with Ref.~\cite{Egner:2024lay},
\begin{align}
    \frac{\tau_{B^+}}{\tau_{B^0}}=1.081^{+0.014}_{-0.016}\ , 
\end{align}
and with the experimental average
\begin{align}\label{eq:tauratioexp}
    \frac{\tau_{B^+}}{\tau_{B^0}}|_{\rm exp}=1.0832\pm 0.0034\ ,
\end{align}
all three agree well within $1\sigma$.

The agreement between the predictions and experiment suggests that the four-quark sector and the bag parameters are not the origin of differences in the individual total widths.

Finally, we give the theoretical branching ratio for $\bar{B}\to X_c \ell \bar{\nu}_\ell$ by multiplying the its semileptonic rate by the predicted lifetime, symmetrising between $B^0$ and $B^+$,
\begin{equation}\label{eq:brtheory}
       \mathcal{B}(B \to X_c \ell \bar{\nu}_\ell) = \asym{\numBRTheoryShort}{\numBRTheoryShortUp}{\numBRTheoryShortDown}\,\% \ .
\end{equation}
As both the leading term of the
lifetime and the semileptonic rate are proportional to $|V_{cb}|$, this branching ratio depends
only on $V_{ub}/V_{cb}$ and is therefore almost independent of $|V_{cb}|$. Due to the larger predicted lifetime, the central
value of the branching ratio lies above the one obtained with the measured lifetimes, Eq.~\eqref{eq:brancfit}, however, due to the large lower uncertainty both are in agreement within $1\sigma$.

\section{Conclusion}
\label{sec:conl}

We have presented a combined HQE analysis of inclusive semileptonic $\bar B \to X_c \ell \bar\nu_\ell$ moments and partial branching fractions together with the $B^+$ and $B^0$ lifetimes, using a new fit strategy. 

In our setup, the unphysical renormalisation scales $\scaleas$, $\scalembkin$ and $\scalemcMS$ are treated as fully correlated nuisance parameters and determined by the data rather than varied over a prescribed range. In addition, we do not assign an ad hoc uncertainty for missing higher-order power corrections. Instead, we perform fits at fixed order in $1/m_b$ to study the convergence of the HQE. Our nominal fit is done at order $1/m_b^3$, while we add the maximal difference with $\mathcal{O}(1/m_b^{4,5})$ as an additional uncertainty. 

Using the full set of experimental semileptonic data, we obtain
\begin{align}
    |V_{cb}| &= (\numVcb \pm \numVcbErr|_{\rm fit} \pm \numVcbSigmaHO|_{\rm h.o.})\times 10^{-3} = (\numVcb\pm\numVcbTotal)\times10^{-3} \ , \nonumber\\
    \mathcal{B}(\bar B\to X_c\ell\bar\nu_\ell) &= (\numBRCentral \pm \numBRErr|_{\rm fit} \pm \numBRSigmaHO|_{\rm h.o.})\,\% \ . 
\end{align}
The extracted HQE parameters are collected in Tab.~\ref{tab:default_fit_result_with_ho}. We note that the quoted uncertainty contains no perturbative truncation component, because the scales are fitted rather than varied this part is absorbed into the central value. This should be considered when comparing with previous analyses in Ref.~\cite{Finauri:2023kte, Carvunis:2025vab} which use the same data set.

At the same time, we stress that we obtain a poor description of the data with $p = \pValueDefault$, which suggests incompatible data or missing higher-order effects. The leave-one-out study localises much of the tension in the Belle $q^2$ moments, whose removal restores $p = \pValueNoBelleqsq$. To clarify this situation, we urge new correlated moment measurements and updated branching ratio measurements of the semileptonic $\bar B\to X_c \ell \bar{\nu}_\ell$ decay. 

We also studied the convergence of the HQE by performing fits at fixed order in $1/m_b$. We find that $|V_{cb}|$ and $\mupi$ are stable across $\mathcal{O}(1/m_b^{3,4,5})$. However, $\muG$ and $\rhoD$ show large variations at different orders, with $\rhoD$ growing monotonically by about $\numRelStepRhoD\%$ per order. These two parameters also carry the strong correlations with the fitted scales. We therefore stress the importance of using a consistent set of HQE parameters across different applications, such as calculations of semileptonic rates and moments and lifetime predictions.

In this respect, we use our setup to predict for the first time the $B^0$ and $B^+$ lifetimes directly in a simultaneous analysis. To this end, we extended the \texttt{kolya} package with the lifetime predictions of Ref.~\cite{Egner:2024lay}, as well as the $\bar B \to X_u \ell\bar\nu_\ell$ and $\bar B \to X_c\tau\bar\nu_\tau$ rates. 

For the lifetimes, we obtain 
\begin{equation}
    \tau_{B^+} = \asym{\numTauBplusShort}{\numTauBplusShortUp}{\numTauBplusShortDown}\,\mathrm{ps} \, \quad\quad \tau_{B^0} = \asym{\numTauBzeroShort}{\numTauBzeroShortUp}{\numTauBzeroShortDown}\,\mathrm{ps}
\end{equation}
 within the range of previous predictions in Ref.~\cite{Egner:2024lay}. The uncertainty is dominated by the perturbative corrections. The large asymmetry in this uncertainty is due to our setup, in which the central value is determined by the best-fit point of the semileptonic rates. Taking into account the large lower uncertainty, our predictions are below the measurements by $\numWidthPullBzero$ and $\numWidthPullBplus\sigma$. On the other hand, the lifetime ratio, Eq.~\eqref{eq:tauratio} is in good agreement with experiment. This may suggest that the remaining discrepancy between prediction and measurement lies in the  flavour-independent part of the width. 

On the theoretical side, our current setup can be improved by including the known NLO $1/m_b^2$ corrections to the two-quark operators~\cite{Mannel:2024uar,Mannel:2025fvj} and the NNLO dimension-six corrections~\cite{Moretti:2026waq} to the lifetimes, as well as the $\mathcal{O}(\alpha_{\rm em})$ corrections to the semileptonic rates \cite{Bigi:2023cbv}. 

In addition, our analysis would benefit from including several higher-order corrections that are currently unknown:
\begin{itemize}

    \item For the semileptonic moments, the complete $\alpha_s^2$ corrections to the lepton energy and hadronic mass moments are missing. 
    \item Complete $1/m_b^4$ corrections to the non-leptonic width are missing. Including these would streamline the setup for the higher-order uncertainty. 
    \item Extension of the LLSA beyond the lowest-lying states, and repetition of the higher-order extraction once the fourth lepton-energy moments become available, since these are the observables most sensitive to power corrections.
\end{itemize}

On the experimental side, new measurements of the $E_\ell$, $M_X^2$ and $q^2$ moments with the full correlation matrix among them, from a common dataset are highly anticipated. Updated inclusive branching fraction measurements, ideally also with a $q^2$ cut, are essential to truly understand the $|V_{cb}|$ discrepancy between exclusive and inclusive decays.  We look forward to measurements including experimental correlations among different moments, as this might bring some clarity over the observed tensions in the $q^2$ data.

\acknowledgments
We thank Maria Laura Piscopo for invaluable help in 
validating the lifetime implementation in \texttt{kolya}.
We thank A.\ Lenz, A.\ Rusov, T.\ Mannel, D. van Dyck, and G. Finauri for useful discussion.
The work by I.S.M.\ was supported by the Deutsche Forschungsgemeinschaft (DFG, German Research Foundation) under grant 396021762 -- TRR 257 ``Particle Physics Phenomenology after the Higgs Discovery". M.P. was supported by the German Research Foundation (DFG) Emmy-Noether Grant No. 526218088. The work of K.K.V. is supported in part
by the project Solving Beautiful Puzzles with file number VI.Vidi.223.083 of the research
programme Vidi which is financed by the Dutch Research Council (NWO). 

\clearpage 
\appendix
\section{Definition of higher order HQE elements}\label{app:HQE_def}
We employ the shorthand $\langle \bar{b}_v\,...\,b_v\rangle\equiv\langle B(v)|\bar{b}_v\,...\,b_v|B(v)\rangle$ and $g^\perp_{\mu\nu}=g_{\mu\nu}-v_\mu v_\nu$. The nine HQE elements at $1/m_b^4$ are defined as \cite{Mannel:2010wj}:
\begin{align}
    2M_B m_1&=\langle \bar{b}_v\, (iD^\rho)\,(iD^\sigma)\,(iD^\lambda)\,(iD^\delta) \,b_v\rangle\, \frac{1}{3}\left(g^\perp_{\rho\sigma}g^\perp_{\lambda\delta}+g^\perp_{\rho\lambda}g^\perp_{\sigma\delta}+g^\perp_{\rho\delta}g^\perp_{\sigma\lambda}\right)\ ,\nonumber\\
    2M_B m_2&=\langle \bar{b}_v\, \big[(iD^\rho),\, (iD^\sigma)\big]\,\big[(iD^\lambda),\, (iD^\delta)\big]\,b_v\rangle\, g^\perp_{\rho\delta}v_\sigma v_\lambda\ ,\nonumber\\
    2M_B m_3&=\langle \bar{b}_v\, \big[(iD^\rho),\, (iD^\sigma)\big]\,\big[(iD^\lambda),\, (iD^\delta)\big]\,b_v\rangle\, g^\perp_{\rho\lambda}g^\perp_{\sigma\delta}\ ,\nonumber\\
    2M_B m_4&=\langle \bar{b}_v\, \Big\{(iD^\rho),\, \Big[(iD^\sigma),\, \big[(iD^\lambda),\, (iD^\delta)\big]\Big]\Big\}\,b_v\rangle\, g^\perp_{\sigma\lambda}g^\perp_{\rho\delta}\ ,\nonumber \rule[-10pt]{0pt}{8pt}\\  
    2M_B m_5&=\langle \bar{b}_v\, \big[(iD^\rho),\, (iD^\sigma)\big]\,\big[(iD^\lambda),\, (iD^\delta)\big]\, (-i\sigma^{\alpha\beta})\,b_v\rangle\, g^\perp_{\alpha\rho}g^\perp_{\beta\delta}v_\sigma v_\lambda\ ,\nonumber\\
    2M_B m_6&=\langle \bar{b}_v\, \big[(iD^\rho),\, (iD^\sigma)\big]\,\big[(iD^\lambda),\, (iD^\delta)\big]\, (-i\sigma^{\alpha\beta})\,b_v\rangle\, g^\perp_{\alpha\sigma}g^\perp_{\beta\lambda}g^\perp_{\rho\delta}\ ,\nonumber\\
    2M_B m_7&=\langle \bar{b}_v\, \Big\{\big\{(iD^\rho),\, (iD^\sigma)\big\},\,\big[(iD^\lambda),\, (iD^\delta)\big]\Big\}\, (-i\sigma^{\alpha\beta})\,b_v\rangle\, g^\perp_{\sigma\lambda}g^\perp_{\alpha\rho}g^\perp_{\beta\delta}\ ,\nonumber\\
    2M_B m_8&=\langle \bar{b}_v\,\Big\{\big\{(iD^\rho),\, (iD^\sigma)\big\},\,\big[(iD^\lambda),\, (iD^\delta)\big]\Big\}\, (-i\sigma^{\alpha\beta}) \,b_v\rangle\, g^\perp_{\rho\sigma}g^\perp_{\alpha\lambda}g^\perp_{\beta\delta}\ ,\nonumber\\
    2M_B m_9&=\langle \bar{b}_v\, \Bigg[(iD^\rho),\, \Big[(iD^\sigma),\, \big[(iD^\lambda),\, (iD^\delta)\big]\Big]\Bigg]\, (-i\sigma^{\alpha\beta})\,b_v\rangle\, g^\perp_{\rho\beta}g^\perp_{\lambda\alpha}g^\perp_{\sigma\delta}\ .
\end{align}
The eighteen HQE parameters present at $1/m_b^5$ are defined as \cite{Mannel:2010wj}:
\begin{align}
\nonumber
2M_B r_1 &= \langle \bar{b}_v \,(i  D_\mu)\, (ivD)^3\, (i  D^\mu) \, b_v\rangle\ , \\
\nonumber 
2M_B r_2 &= \langle \bar{b}_v \,(i  D_\mu)\, (ivD)\, (i  D^\mu)\, (iD)^2 \, b_v\rangle\ , \\
\nonumber
2M_B r_3 &= \langle \bar{b}_v \,(i  D_\mu)\, (ivD)\,(i  D_\nu)\,( i
D^\mu)\,(i  D^\nu )\, b_v\rangle\ , \\  
\nonumber
2M_B r_4 &= \langle \bar{b}_v \,(i  D_\mu)\, (ivD)\, (iD)^2\,(i  D^\mu )\, b_v\rangle\ , \\  
\nonumber
2M_B r_5 &= \langle \bar{b}_v \,(iD)^2\,(ivD)\,  (iD)^2 \, b_v\rangle\ , \\  
\nonumber
2M_B r_6 &= \langle \bar{b}_v \,(i  D_\mu)\,(i  D_\nu)\, (ivD)\, (i
D^\nu)\,(i  D^\mu )\, b_v\rangle\ , \\  
\nonumber
2M_B r_7 &= \langle \bar{b}_v \,(i  D_\mu)\,(i  D_\nu)\, (ivD)\, (i
D^\mu)\,(i  D^\nu) \, b_v\rangle\ , \rule[-10pt]{0pt}{8pt}\\  
\nonumber
2M_B r_{8} &= \langle \bar{b}_v \,(i   D_\alpha) \, (ivD)^3\,( i   D_\beta)
\, (-i \sigma^{\alpha \beta })\,b_v\rangle\ , \\  
\nonumber
2M_B r_{9} &= \langle \bar{b}_v \,(i   D_\alpha) \, (ivD)\,( i   D_\beta)
\, (iD)^2 \,(-i \sigma^{\alpha \beta })\, b_v\rangle\ , \\  
\nonumber
2M_B r_{10} &= \langle \bar{b}_v \,(i   D_\mu)\, (ivD)\, (i
D^\mu)\, (i   D_\alpha) \,( i   D_\beta)  \,(-i \sigma^{\alpha \beta })\, b_v\rangle\ , \\
\nonumber
2M_B r_{11} &= \langle \bar{b}_v \,(i   D_\mu)\, (ivD)\, (i   D_\alpha)
\, (i   D^\mu)\,( i   D_\beta)  \,(-i \sigma^{\alpha \beta })\, b_v\rangle\ , \\  
\nonumber
2M_B r_{12} &= \langle \bar{b}_v \,(i   D_\alpha) \, (ivD)\, (i
D_\mu)\,( i   D_\beta) \, (i   D^\mu )\,(-i \sigma^{\alpha \beta })\, b_v\rangle\ , \\
\nonumber
2M_B r_{13} &= \langle \bar{b}_v \,(i   D_\mu)\, (ivD)\, (i   D_\alpha)
\,( i   D_\beta) \, (i   D^\mu )\,(-i \sigma^{\alpha \beta })\, b_v\rangle\ , \\  
\nonumber
2M_B r_{14} &= \langle \bar{b}_v \,(i   D_\alpha) \, (ivD)\, (iD)^2\,( i   D_\beta)  \,(-i \sigma^{\alpha \beta })\, b_v\rangle\ , \\
\nonumber
2M_B r_{15} &= \langle \bar{b}_v \,(i   D_\alpha) \,( i   D_\beta) \, (ivD)\, (iD)^2 \,(-i \sigma^{\alpha \beta })\, b_v\rangle\ , \\  
\nonumber
2M_B r_{16} &= \langle \bar{b}_v \,(i   D_\mu)\,( i   D_\alpha) \, (ivD)\,( i   D_\beta) \, (i   D^\mu )\,(-i \sigma^{\alpha \beta })\, b_v\rangle\ , \\  
\nonumber
2M_B r_{17} &= \langle \bar{b}_v \,(i   D_\alpha) \,( i   D_\mu)\, (ivD)\, (i   D^\mu)\,( i   D_\beta)  \,(-i \sigma^{\alpha \beta })\, b_v\rangle\ , \\  
2M_B r_{18} &= \langle \bar{b}_v \,(i   D_\mu)\, (i   D_\alpha) \, (ivD)\, (i   D^\mu)\,( i   D_\beta)  \,(-i \sigma^{\alpha \beta })\, b_v\rangle \ .\label{ri_definitions}
\end{align}

\section{Implementation of lifetimes in \texttt{kolya}}\label{ap:kolya}

The \verb|kolya| package has been extended to include the predicting for $B$-mesons lifetimes. In this appendix, we present the update and new functions. For an explanation of the functions in \verb|kolya| regarding the $\bar B\to X_c l\bar{\nu}_l$ observables, we refer to Ref.~\cite{Fael:2024fkt} and the code itself~\cite{fael_2024_10818195}.

\subsection{Update on the parameter classes}\label{ap:par}
The values of the physical parameters, like the quark masses, are stored in an object of \verb|parameters.physical_parameters| class
\begin{minted}{python}
>>> par = kolya.parameters.physical_parameters()
\end{minted}
This class has been extended to include the mass of the $\tau$ lepton and $W$ boson masses, the top-quark pole mass, the matching scale of the Wilson coefficients, and the scale of the bag parameters. Dimensionful quantities are given in units of GeV.  One can set the desired values for these new parameters as usual. Here, we present their names and default values:
\begin{minted}{python}
>>> par.mtau
1.777
>>> par.mW
80.3692
>>> par.mtPole
172.57
>>> par.scale_matchingEW
80.385
>>> par.scale_bag_parameters
1.5
\end{minted}
Furthermore, we introduce a class \verb|CKM_matrix| to store the values of the CKM matrix elements
\begin{minted}{python}
>>> CKM = kolya.parameters.CKM_matrix()
\end{minted}
Current default values are~\cite{UTfit:2005ras,ParticleDataGroup:2024cfk}:
\begin{minted}{python}
>>> CKM
(Vud, Vus, Vub) = (0.97435, 0.22501, 0.003732)
(Vcd, Vcs, Vcb) = (0.22487, 0.97349, 0.04183)
(Vtd, Vts, Vtb) = (0.00858, 0.04111, 0.999118)
\end{minted}
If desired, one can also use complex values for the CKM parameters. 
One can access the value of a specific element by typing for example:
\begin{minted}{python}
>>> CKM.Vcb
0.04183
\end{minted}

A new class has been added, \verb|bag_parameters|, to store the 
value of bag parameters and decay constant needed for the lifetime predictions:
\begin{minted}{python}
>>> bag = kolya.parameters.bag_parameters()
\end{minted}
In Tab.~\ref{tab:bag_param}, we present their names and default values. We evolve the Bag parameters  $\tilde{B}_i^d$ following Ref.~\cite{Kirk:2017juj} starting from the reference values at $1.5$ GeV of Ref.~\cite{Black:2024bus} to the scale $\mu_0$ which we fix at $m_b^{\rm kin}\approx 4.573$. The obtained input values are given in Tab.~\ref{tab:evolve_bag}. Correlations between the $\tilde{B}_i$ parameters are small and currently neglected. We do not evolve $\tilde{\delta}_i^{q'q}$, as they already count as corrections of $\mathcal{O}(\alpha_s)$.
\begin{table}[t!]
    \centering
    \begin{tabular}{|c|c|c|}
         \hline
         \textbf{Parameter} & \textbf{Name} & \textbf{Default value }\\
         \hline
         $\tilde{B}_1^d$  & \verb|Bd1| & 1.0026 \\
         $\tilde{B}_2^d$ & \verb|Bd2| & 0.9982 \\
         $\tilde{B}_3^d$ & \verb|Bd3| & -0.0057 \\
         $\tilde{B}_4^d$ & \verb|Bd4| & -0.0014 \\
         \hline
         $\tilde{\delta}^{q'q}_1$ & \verb|deltaqq1| & 0.0026 \\
         $\tilde{\delta}^{q'q}_2$ & \verb|deltaqq2| & -0.0018 \\
         $\tilde{\delta}^{q'q}_3$ & \verb|deltaqq3| & -0.0004 \\
         $\tilde{\delta}^{q'q}_4$ & \verb|deltaqq4| & 0.0003 \\
         \hline
         $\tilde{\delta}^{q's}_1$ & \verb|deltaqs1| & 0.0023 \\
         $\tilde{\delta}^{q's}_2$ & \verb|deltaqs2| & -0.0017 \\
         $\tilde{\delta}^{q's}_3$ & \verb|deltaqs3| & -0.0004 \\
         $\tilde{\delta}^{q's}_4$ & \verb|deltaqs4| & 0.0003 \\
         \hline
         $f_B$ & \verb|fBd| & 0.1900\\
         \hline 
    \end{tabular}
    \caption{The names and default values used in \texttt{kolya} for the bag parameters and decay constant in the \texttt{bag\_parameters} class.  
    They are taken from Refs.~\cite{Black:2024bus,Egner:2024lay,King:2021jsq} and
    defined at renormalisation scale $1.5$ GeV.}
    \label{tab:bag_param}
\end{table}

\begin{table}[t]
    \centering
    \begin{tabular}{|c|c|}
    \hline
         \textbf{Parameters} & \boldmath \textbf{Value at $\mu_0=4.573\,\text{GeV}$} \unboldmath \\
         \hline
         $\tilde{B}_1^d$ & $\phantom{-}1.0064\pm {0.0254}$ \\
         $\tilde{B}_2^d$ & $\phantom{-}1.0012\pm{0.0214}$ \\
         $\tilde{B}_3^d$ & $-0.0390\pm{0.0191}$ \\
         $\tilde{B}_4^d$ & $-0.0358\pm{0.0177}$ \\
         \hline
    \end{tabular}
    \caption{The Bag parameters values evolved to $\mu_0=4.573\, \text{GeV}$, used throughout this work. }
    \label{tab:evolve_bag}
\end{table}

\subsection{Inclusive semileptonic \boldmath $\bar B\to X_c\tau\bar{\nu}_\tau$ \unboldmath}\label{ap:btotau}
In order to implement the lifetimes of the $B$ mesons, we include the total inclusive semileptonic rate for $\bar{B}\to X_c\tau\bar{\nu}_\tau$. We employ the kinetic scheme for the bottom quark and the $\overline{\rm MS}$ scheme for the charm quark. Corrections to up to including $\mathcal{O}(1/m_b^3)$ \cite{Mannel:2017jfk,Rahimi:2022vlv} and NLO QCD corrections to the partonic contribution \cite{Jezabek:1996db,Mannel:2017jfk,Fael:2024gyw} are implemented. Both the historical and RPI basis for the HQE parameters have been implemented, and can be changed using the optional argument \verb|flag_basisPERP|. The functions\\ \verb|kolya.TotalRate_tau.TotalRate_KIN_MS| for $\Gamma( \bar{B}\to X_c\tau\bar{\nu}_\tau)$ and\newline \verb|kolya.TotalRate_tau.BranchingRatio_KIN_MS| for $\mathcal{B}( \bar{B}\to X_c\tau\bar{\nu}_\tau)$ can be used in the same way as for $\bar{B}\to X_cl\bar{\nu}_l$, see Ref.~\cite{Fael:2024fkt} for the explanation of the functions for $ \bar{B}\to X_cl\bar{\nu}_l$. 

\subsection{Inclusive semileptonic \boldmath $\bar B\to X_u l\bar{\nu}_l$ \unboldmath}\label{ap:btou}
Furthermore, we implement the total rate for the inclusive $\bar B\to X_ul\bar\nu_l$ decay, with $l=e,\mu$. This is implemented in the file \verb|TotalRate_btoulnu_SM.py| as the function \verb|X_Gamma_KIN_MS|.
The function arguments are similar to those for $\bar B\to X_cl\bar\nu_l$. 
We remind the reader of the following definition:
\begin{align}
    X(\bar B\to X_ql\bar \nu_l)\equiv\left(\frac{G_F^2|V_{qb}|^2(m_b^{\rm kin})^5A_{ew}}{192\pi^3}\right)^{-1}\Gamma(\bar B\to X_ql\bar \nu_l)\qquad q\in\{u,c\}\ ,
\end{align}
where in the decay rate reads~\cite{Fael:2019umf} 
\begin{equation}
   X(\bar B\to X_ul\bar \nu_l) =
   1 + C_F\sum_{n=1} X_n \left(\frac{\alpha_s}{\pi}\right)^n 
    - \frac{\mu_\pi^2}{2m_b^2} 
    - \frac{3\mu_G^2}{2m_b^2} 
    + \frac{3\rho_{LS}^3}{2m_b^3} 
    +  \frac{15}{2}
    \frac{\rho_D^3}{m_b^3}
    +\frac{\tau_0}{m_b^3}\, .
   \label{eq:decayrate}
\end{equation}
We note that the coefficient of $\rho_D^3$ differs from that often used in literature, namely $77/6$. This finite shift comes from the $\mathcal{O}(\alpha_s)$ mixing of the weak annihilation operators and the $\rho_D^3$ parameter as discussed in Ref.~\cite{Fael:2019umf} (see also Ref.~\cite{Gambino:2005tp}). \\

The following contributions are included for $X(\bar B\to X_ul\bar \nu_l)$:
\begin{itemize}
    \item Leading order corrections up to and including $\mathcal{O}(1/m_b^3)$, in both the historical (``perp'') and RPI basis.
    For the free quark decay, we include QCD corrections
    up to N$^3$LO~\cite{vanRitbergen:1999gs,Pak:2008cp,Fael:2023tcv,Chen:2023dsi,Chen:2026jwl}.
    Flags for turning off the NNLO and N$^3$LO corrections have been included, just as in the $\bar B\to X_cl\bar\nu_l$ case.
    \item Weak-annihilation contributions are implemented as described in Ref.~\cite{Fael:2019umf} through one additional parameter 
    \begin{align}
        \tau_0 &= 
        8 \rho_D^3 \log \left( \frac{\mu_0^2}{(m_b^{\rm kin})^2} \right)
        +16 \pi^2 f_B^2 M_B \Big(\tilde B^d_2 - \tilde B^d_1 \Big)
        \quad \text{for\ }B^+, \notag \\
        \tau_0 &= 
        8 \rho_D^3 \log \left( \frac{\mu_0^2}{(m_b^{\rm kin})^2} \right)
        +16 \pi^2 f_B^2 M_B \Big(\tilde \delta^{ud}_2 - \tilde \delta^{ud}_1 \Big)
        \quad \text{for\ }B^0,
        \label{eq:WA}
    \end{align}
    where $\mu_0$ is the renormalization scale of the Bag parameters
    in the $\Delta B = 0$ effective theory.\footnote{We assume that $\tilde B^u_i = \tilde B_i^d$.} In the current analysis, we explicitly turn off the contribution of the logarithmic term in \eqref{eq:WA} by setting \verb|flag_LogWA|=0 in \verb|Blifetimes_Total.py| since we employ $\mu_0=4.573\,\text{GeV}\approx m_b^{\rm kin}$. Although the cancellation is only exact if $\mu_0=\mbkin$, which is varied in the analysis, we checked that the effect is negligible within uncertainties.
\end{itemize}
\subsection{Non-leptonic decays and Lifetimes}
The lifetime predictions for $B_d$ and $B^+$ mesons have been implemented in the following functions:
\begin{itemize}
    \item[] \verb|Blifetimes_Total.TotalWidth_Bd_KIN_MS(par,hqe,wc,CKM,bag,Aew)|
    \item[] \verb|Blifetimes_Total.TotalWidth_Bplus_KIN_MS(par,hqe,wc,CKM,bag,Aew)|
    \item[] \verb|Blifetimes_Total.Blifetime_Bd_KIN_MS(par,hqe,wc,CKM,bag,Aew)|
    \item[] \verb|Blifetimes_Total.Blifetime_Bplus_KIN_MS(par,hqe,wc,CKM,bag,Aew)|
\end{itemize}
where the total widths are given in GeV and the inverse lifetimes in inverse picoseconds.

To evaluate the total widths we sum the contributions of $B \to X_c l \bar\nu_l$ decays
with $l=e,\mu,\tau$, the $V_{ub}$ suppressed contribution $B \to X_u l \bar\nu_l$,
with $l=e,\mu$, as well as the total width for non-leptonic decays:
\begin{itemize}
    \item The total rate of the non-leptonic decays $b\to q_1q_2q_3$ 
    is implemented in the free quark approximation. 
    We approximate the strange quark as massless, so 
    e.g.~the contributions from $b\to ccs$ and $b\to ccd$ are equal modulo the
    relative CKM factors $|V_{cs}|$ and $|V_{cd}|$.
    These contributions are universal for $B_d$ and $B^+$.
    We include corrections due to the current-current
    operators $O_{12}$ in the $\Delta B=1$ effective Hamiltonian up to NNLO~\cite{Bagan:1994zd,Bagan:1995yf,Egner:2024azu}.
    We include also the NLO effects from insertion of $O_1$ and $O_2$
    into penguin-like topologies as well as the tree-level contribution 
    from penguin operators $O_3$-$O_6$~\cite{Lenz:1997aa,Lenz:1998qp,Krinner:2013cja}.

    \item Power corrections to non-leptonic decays $b\to q_1q_2q_3$ proportional
    to two-quark operators are included at order $1/m_b^2$ and
    $1/m_b^3$.
    We use the expressions calculated in Ref.~\cite{Lenz:2020oce} (see also Refs.~\cite{Blok:1992he, Blok:1992hw,Bigi:1992ne}).
    NLO correction at order $1/m_b^2$ to two-quark operators
    have been calculated recently in Refs.~\cite{Mannel:2024uar,Mannel:2025fvj}, but are currently 
    not included.
     
    \item At order $1/m_b^3$ also four-quark operators contribute, which yield a difference between the
    $B_d$ and $B^+$ lifetimes.
    The complete expressions for the dimension-six 
    Wilson coefficients up to NLO-QCD corrections have been
    obtained in Ref.~\cite{Franco:2002fc}, 
    in the case of four-quark operators defined in HQET, 
    and in Refs.~\cite{Franco:2002fc,Beneke:2002rj} for QCD operators. Recently, the NNLO-QCD corrections for the dimension 6 operators have been calculated for the lifetime ratio $\tau(B^+)/\tau(B_d^0)$ in Ref.~\cite{Moretti:2026waq}, but currently not included.
    
    \item We do not include tiny contributions from rare decays, e.g.\ $B \to X_s \gamma$. 
\end{itemize}
We implemented a flag \verb|flag_onlyNL| (with default value 0) which can be used to only include non-leptonic contributions to the total widths or lifetimes. For an explanation of the \verb|CKM| and \verb|bag| inputs, see Appendix \ref{ap:par}. For an explanation of the \verb|par|, \verb|hqe|, \verb|wc| and \verb|Aew| we refer to Ref.~\cite{Fael:2024fkt}.

\section{Running of $\mu_G^2$}\label{ap:muGrun}
In the HQET Lagrangian the chromo-magnetic term appears in the form
\begin{equation}
    \frac{C_m(\mu)}{4m} \bar h_v G^{\mu\nu} \sigma_{\mu\nu} h_v\ .
\end{equation}
Let us consider first the running of the Wilson coefficient $C_m(\mu)$.
The solution of the RGE is formally
\begin{equation}
    C_m(\mu) = C(\mu_0) \exp 
    \left\{ 
     \int_{\alpha_s(\mu_0)}^{\alpha_s(\mu)}
     \frac{\gamma_m}{2 \beta(\alpha_s)} \, \text{d}\alpha_s
    \right\}
    =C(\mu_0) U(\mu_0,\mu)\ 
    ,
\end{equation}
where we write the Beta function and the anomalous dimension in
the following way:
\begin{align}
    \beta(\alpha_s) &= 
    - \beta_0 \left( \frac{\alpha_s}{\pi} \right)^2
     - \beta_1 \left( \frac{\alpha_s}{\pi} \right)^3 +\dots \, ,\notag \\ 
     \gamma_m &=
     \gamma^{(0)} \frac{\alpha_s}{\pi}
     +\gamma^{(1)} \left(\frac{\alpha_s}{\pi}\right)^2 + \dots \,  .
\end{align}
In QCD, we have
\begin{align}
    \beta_0 &= \frac{1}{4}\left[ \frac{11}{3} C_A - \frac{4}{3} T_F n_f \right], \notag \\
    \beta_1 &= \frac{1}{16} \left[ 
    \frac{34}{3}C_A^2 - \left( 4 C_F + \frac{20}{3} C_A \right) T_F n_f
    \right].
\end{align}
The anomalous dimension up to two loops is given by~\cite{Czarnecki:1997dz}
\begin{align}
    \gamma^{(0)} &= \frac{C_A}{2} \notag \ ,\\ 
    \gamma^{(1)} &= \frac{17}{36} C_A^2 -\frac{13}{36} C_A T_F n_f\ .  
\end{align}
The expression of the evolution operator $U(\mu_0,\mu)$ up to NLL is the following:
\begin{align}
    U(\mu_0,\mu) &=
    \left(1 + J_1 \frac{\alpha_s(\mu)}{\pi} \right)
    \left( \frac{\alpha_s(\mu_0)}{\alpha_s(\mu)}\right)^{\frac{\gamma_0}{2\beta_0}}
    \left(1 - J_1 \frac{\alpha_s(\mu_0)}{\pi} \right)\ , \notag \\
    J_1 &= \frac{\beta_1 \gamma_0}{2 \beta_0^2}-\frac{\gamma_1}{2 \beta_0} \,.
\end{align}
When taking the matrix element of the chromo-magnetic operator, 
the combination
\begin{equation}
    C_m(\mu) \mu^2_G(\mu)
\end{equation}
must be RGE invariant, so $\mu^2_G$ should run with the inverse of the evolution
operator $U(\mu_0,\mu)$:
\begin{equation}
    \mu_G^2(\mu) = \mu_G^2(\mu_0) U^{-1}(\mu_0,\mu) = \mu_G^2(\mu_0) U(\mu,\mu_0)\ ,
\end{equation}
in fact $U(\mu_0,\mu) U^{-1}(\mu_0,\mu) = U(\mu_0,\mu) U(\mu,\mu_0) = 1$.\\
The RGE evolution of $\mu_G^2$ at NLL has been implemented in the following function in \verb|kolya|:
\begin{itemize}
    \item[] \verb|HQEparametersRGE.muG2muG(muG_scale1, scale1, scale2)|
\end{itemize}
which takes as arguments the value of $\mu_G^2(\mu_1)$ at a scale $\mu_1$ (\verb|muG_scale1| and \verb|scale1|) and returns the value of $\mu_G^2(\mu_2)$ at another scale $\mu_2$ (\verb|scale2|).

\newpage

\bibliographystyle{JHEP}
\bibliography{biblio.bib}

\providecommand{\href}[2]{#2}\begingroup\raggedright\begin{thebibliography}{100}

\bibitem{Bigi:1992su}
I.I.Y.~Bigi, N.G.~Uraltsev and A.I.~Vainshtein, \emph{{Nonperturbative corrections to inclusive beauty and charm decays: QCD versus phenomenological models}}, \href{https://doi.org/10.1016/0370-2693(92)90908-M}{\emph{Phys. Lett. B} {\bfseries 293} (1992) 430} [\href{https://arxiv.org/abs/hep-ph/9207214}{{\ttfamily hep-ph/9207214}}].

\bibitem{Bigi:1993fe}
I.I.Y.~Bigi, M.A.~Shifman, N.G.~Uraltsev and A.I.~Vainshtein, \emph{{QCD predictions for lepton spectra in inclusive heavy flavor decays}}, \href{https://doi.org/10.1103/PhysRevLett.71.496}{\emph{Phys. Rev. Lett.} {\bfseries 71} (1993) 496} [\href{https://arxiv.org/abs/hep-ph/9304225}{{\ttfamily hep-ph/9304225}}].

\bibitem{Blok:1993va}
B.~Blok, L.~Koyrakh, M.A.~Shifman and A.I.~Vainshtein, \emph{{Differential distributions in semileptonic decays of the heavy flavors in QCD}}, \href{https://doi.org/10.1103/PhysRevD.50.3572}{\emph{Phys. Rev. D} {\bfseries 49} (1994) 3356} [\href{https://arxiv.org/abs/hep-ph/9307247}{{\ttfamily hep-ph/9307247}}].

\bibitem{Manohar:1993qn}
A.V.~Manohar and M.B.~Wise, \emph{{Inclusive semileptonic B and polarized Lambda(b) decays from QCD}}, \href{https://doi.org/10.1103/PhysRevD.49.1310}{\emph{Phys. Rev. D} {\bfseries 49} (1994) 1310} [\href{https://arxiv.org/abs/hep-ph/9308246}{{\ttfamily hep-ph/9308246}}].

\bibitem{Bauer:2004ve}
C.W.~Bauer, Z.~Ligeti, M.~Luke, A.V.~Manohar and M.~Trott, \emph{{Global analysis of inclusive B decays}}, \href{https://doi.org/10.1103/PhysRevD.70.094017}{\emph{Phys. Rev. D} {\bfseries 70} (2004) 094017} [\href{https://arxiv.org/abs/hep-ph/0408002}{{\ttfamily hep-ph/0408002}}].

\bibitem{Gambino:2013rza}
P.~Gambino and C.~Schwanda, \emph{{Inclusive semileptonic fits, heavy quark masses, and $V_{cb}$}}, \href{https://doi.org/10.1103/PhysRevD.89.014022}{\emph{Phys. Rev. D} {\bfseries 89} (2014) 014022} [\href{https://arxiv.org/abs/1307.4551}{{\ttfamily 1307.4551}}].

\bibitem{Alberti:2014yda}
A.~Alberti, P.~Gambino, K.J.~Healey and S.~Nandi, \emph{{Precision Determination of the Cabibbo-Kobayashi-Maskawa Element $V_{cb}$}}, \href{https://doi.org/10.1103/PhysRevLett.114.061802}{\emph{Phys. Rev. Lett.} {\bfseries 114} (2015) 061802} [\href{https://arxiv.org/abs/1411.6560}{{\ttfamily 1411.6560}}].

\bibitem{Bordone:2021oof}
M.~Bordone, B.~Capdevila and P.~Gambino, \emph{{Three loop calculations and inclusive Vcb}}, \href{https://doi.org/10.1016/j.physletb.2021.136679}{\emph{Phys. Lett. B} {\bfseries 822} (2021) 136679} [\href{https://arxiv.org/abs/2107.00604}{{\ttfamily 2107.00604}}].

\bibitem{Bernlochner:2022ucr}
F.~Bernlochner, M.~Fael, K.~Olschewsky, E.~Persson, R.~van Tonder, K.K.~Vos et~al., \emph{{First extraction of inclusive V$_{cb}$ from q$^{2}$ moments}}, \href{https://doi.org/10.1007/JHEP10(2022)068}{\emph{JHEP} {\bfseries 10} (2022) 068} [\href{https://arxiv.org/abs/2205.10274}{{\ttfamily 2205.10274}}].

\bibitem{Finauri:2023kte}
G.~Finauri and P.~Gambino, \emph{{The q$^{2}$ moments in inclusive semileptonic B decays}}, \href{https://doi.org/10.1007/JHEP02(2024)206}{\emph{JHEP} {\bfseries 02} (2024) 206} [\href{https://arxiv.org/abs/2310.20324}{{\ttfamily 2310.20324}}].

\bibitem{Carvunis:2025vab}
A.~Carvunis, G.~Finauri, P.~Gambino, M.~Jung and S.~M{\"a}chler, \emph{{New Physics in inclusive semileptonic B decays}}, \href{https://doi.org/10.1007/JHEP01(2026)037}{\emph{JHEP} {\bfseries 01} (2026) 037} [\href{https://arxiv.org/abs/2507.22123}{{\ttfamily 2507.22123}}].

\bibitem{CDF:2005xlh}
{\scshape CDF} collaboration, \emph{{Measurement of the moments of the hadronic invariant mass distribution in semileptonic $B$ decays}}, \href{https://doi.org/10.1103/PhysRevD.71.051103}{\emph{Phys. Rev. D} {\bfseries 71} (2005) 051103} [\href{https://arxiv.org/abs/hep-ex/0502003}{{\ttfamily hep-ex/0502003}}].

\bibitem{DELPHI:2005mot}
{\scshape DELPHI} collaboration, \emph{{Determination of heavy quark non-perturbative parameters from spectral moments in semileptonic B decays}}, \href{https://doi.org/10.1140/epjc/s2005-02406-7}{\emph{Eur. Phys. J. C} {\bfseries 45} (2006) 35} [\href{https://arxiv.org/abs/hep-ex/0510024}{{\ttfamily hep-ex/0510024}}].

\bibitem{CLEO:2004bqt}
{\scshape CLEO} collaboration, \emph{{Moments of the B meson inclusive semileptonic decay rate using neutrino reconstruction}}, \href{https://doi.org/10.1103/PhysRevD.70.032002}{\emph{Phys. Rev. D} {\bfseries 70} (2004) 032002} [\href{https://arxiv.org/abs/hep-ex/0403052}{{\ttfamily hep-ex/0403052}}].

\bibitem{BaBar:2009zpz}
{\scshape BaBar} collaboration, \emph{{Measurement and interpretation of moments in inclusive semileptonic decays anti-B ---{\ensuremath{>}} X(c) l- anti-nu}}, \href{https://doi.org/10.1103/PhysRevD.81.032003}{\emph{Phys. Rev. D} {\bfseries 81} (2010) 032003} [\href{https://arxiv.org/abs/0908.0415}{{\ttfamily 0908.0415}}].

\bibitem{Belle:2006jtu}
{\scshape Belle} collaboration, \emph{{Moments of the Hadronic Invariant Mass Spectrum in $B \to X_c \ell \nu$ Decays at {BELLE}}}, \href{https://doi.org/10.1103/PhysRevD.75.032005}{\emph{Phys. Rev. D} {\bfseries 75} (2007) 032005} [\href{https://arxiv.org/abs/hep-ex/0611044}{{\ttfamily hep-ex/0611044}}].

\bibitem{Belle:2006kgy}
{\scshape Belle} collaboration, \emph{{Moments of the electron energy spectrum and partial branching fraction of B ---{\ensuremath{>}} X(c) e nu decays at Belle}}, \href{https://doi.org/10.1103/PhysRevD.75.032001}{\emph{Phys. Rev. D} {\bfseries 75} (2007) 032001} [\href{https://arxiv.org/abs/hep-ex/0610012}{{\ttfamily hep-ex/0610012}}].

\bibitem{Belle:2021idw}
{\scshape Belle} collaboration, \emph{{Measurements of $q^2$ Moments of Inclusive $B \rightarrow X_c \ell^+ \nu_{\ell}$ Decays with Hadronic Tagging}}, \href{https://doi.org/10.1103/PhysRevD.104.112011}{\emph{Phys. Rev. D} {\bfseries 104} (2021) 112011} [\href{https://arxiv.org/abs/2109.01685}{{\ttfamily 2109.01685}}].

\bibitem{Belle-II:2022evt}
{\scshape Belle-II} collaboration, \emph{{Measurement of lepton mass squared moments in B{\textrightarrow}Xc{\ensuremath{\ell}}{\ensuremath{\nu}}{\textasciimacron}{\ensuremath{\ell}} decays with the Belle II experiment}}, \href{https://doi.org/10.1103/PhysRevD.107.072002}{\emph{Phys. Rev. D} {\bfseries 107} (2023) 072002} [\href{https://arxiv.org/abs/2205.06372}{{\ttfamily 2205.06372}}].

\bibitem{Czarnecki:1997sz}
A.~Czarnecki, K.~Melnikov and N.~Uraltsev, \emph{{NonAbelian dipole radiation and the heavy quark expansion}}, \href{https://doi.org/10.1103/PhysRevLett.80.3189}{\emph{Phys. Rev. Lett.} {\bfseries 80} (1998) 3189} [\href{https://arxiv.org/abs/hep-ph/9708372}{{\ttfamily hep-ph/9708372}}].

\bibitem{Aquila:2005hq}
V.~Aquila, P.~Gambino, G.~Ridolfi and N.~Uraltsev, \emph{{Perturbative corrections to semileptonic b decay distributions}}, \href{https://doi.org/10.1016/j.nuclphysb.2005.04.031}{\emph{Nucl. Phys. B} {\bfseries 719} (2005) 77} [\href{https://arxiv.org/abs/hep-ph/0503083}{{\ttfamily hep-ph/0503083}}].

\bibitem{Dassinger:2006md}
B.M.~Dassinger, T.~Mannel and S.~Turczyk, \emph{{Inclusive semi-leptonic B decays to order 1 / m(b)**4}}, \href{https://doi.org/10.1088/1126-6708/2007/03/087}{\emph{JHEP} {\bfseries 03} (2007) 087} [\href{https://arxiv.org/abs/hep-ph/0611168}{{\ttfamily hep-ph/0611168}}].

\bibitem{Pak:2008qt}
A.~Pak and A.~Czarnecki, \emph{{Mass effects in muon and semileptonic b ---{\ensuremath{>}} c decays}}, \href{https://doi.org/10.1103/PhysRevLett.100.241807}{\emph{Phys. Rev. Lett.} {\bfseries 100} (2008) 241807} [\href{https://arxiv.org/abs/0803.0960}{{\ttfamily 0803.0960}}].

\bibitem{Pak:2008cp}
A.~Pak and A.~Czarnecki, \emph{{Heavy-to-heavy quark decays at NNLO}}, \href{https://doi.org/10.1103/PhysRevD.78.114015}{\emph{Phys. Rev. D} {\bfseries 78} (2008) 114015} [\href{https://arxiv.org/abs/0808.3509}{{\ttfamily 0808.3509}}].

\bibitem{Dowling:2008mc}
M.~Dowling, J.H.~Piclum and A.~Czarnecki, \emph{{Semileptonic decays in the limit of a heavy daughter quark}}, \href{https://doi.org/10.1103/PhysRevD.78.074024}{\emph{Phys. Rev. D} {\bfseries 78} (2008) 074024} [\href{https://arxiv.org/abs/0810.0543}{{\ttfamily 0810.0543}}].

\bibitem{Biswas:2009rb}
S.~Biswas and K.~Melnikov, \emph{{Second order QCD corrections to inclusive semileptonic b ---{\ensuremath{>}} X(c) l anti-nu(l) decays with massless and massive lepton}}, \href{https://doi.org/10.1007/JHEP02(2010)089}{\emph{JHEP} {\bfseries 02} (2010) 089} [\href{https://arxiv.org/abs/0911.4142}{{\ttfamily 0911.4142}}].

\bibitem{Mannel:2010wj}
T.~Mannel, S.~Turczyk and N.~Uraltsev, \emph{{Higher Order Power Corrections in Inclusive B Decays}}, \href{https://doi.org/10.1007/JHEP11(2010)109}{\emph{JHEP} {\bfseries 11} (2010) 109} [\href{https://arxiv.org/abs/1009.4622}{{\ttfamily 1009.4622}}].

\bibitem{Alberti:2012dn}
A.~Alberti, T.~Ewerth, P.~Gambino and S.~Nandi, \emph{{Kinetic operator effects in $\bar{B}\to X_c l \nu$ at O($\alpha_s$)}}, \href{https://doi.org/10.1016/j.nuclphysb.2013.01.005}{\emph{Nucl. Phys. B} {\bfseries 870} (2013) 16} [\href{https://arxiv.org/abs/1212.5082}{{\ttfamily 1212.5082}}].

\bibitem{Alberti:2013kxa}
A.~Alberti, P.~Gambino and S.~Nandi, \emph{{Perturbative corrections to power suppressed effects in semileptonic B decays}}, \href{https://doi.org/10.1007/JHEP01(2014)147}{\emph{JHEP} {\bfseries 01} (2014) 147} [\href{https://arxiv.org/abs/1311.7381}{{\ttfamily 1311.7381}}].

\bibitem{Fael:2018vsp}
M.~Fael, T.~Mannel and K.~Keri~Vos, \emph{{$V_{cb}$ determination from inclusive $b \to c$ decays: an alternative method}}, \href{https://doi.org/10.1007/JHEP02(2019)177}{\emph{JHEP} {\bfseries 02} (2019) 177} [\href{https://arxiv.org/abs/1812.07472}{{\ttfamily 1812.07472}}].

\bibitem{Fael:2020iea}
M.~Fael, K.~Sch{\"o}nwald and M.~Steinhauser, \emph{{Kinetic Heavy Quark Mass to Three Loops}}, \href{https://doi.org/10.1103/PhysRevLett.125.052003}{\emph{Phys. Rev. Lett.} {\bfseries 125} (2020) 052003} [\href{https://arxiv.org/abs/2005.06487}{{\ttfamily 2005.06487}}].

\bibitem{Fael:2020njb}
M.~Fael, K.~Sch{\"o}nwald and M.~Steinhauser, \emph{{Relation between the $\overline{\mathrm{MS}}$ and the kinetic mass of heavy quarks}}, \href{https://doi.org/10.1103/PhysRevD.103.014005}{\emph{Phys. Rev. D} {\bfseries 103} (2021) 014005} [\href{https://arxiv.org/abs/2011.11655}{{\ttfamily 2011.11655}}].

\bibitem{Fael:2020tow}
M.~Fael, K.~Sch{\"o}nwald and M.~Steinhauser, \emph{{Third order corrections to the semileptonic b{\textrightarrow}c and the muon decays}}, \href{https://doi.org/10.1103/PhysRevD.104.016003}{\emph{Phys. Rev. D} {\bfseries 104} (2021) 016003} [\href{https://arxiv.org/abs/2011.13654}{{\ttfamily 2011.13654}}].

\bibitem{Fael:2022frj}
M.~Fael, K.~Sch{\"o}nwald and M.~Steinhauser, \emph{{A first glance to the kinematic moments of B {\textrightarrow} X$_{c}${\ensuremath{\ell}}{\ensuremath{\nu}} at third order}}, \href{https://doi.org/10.1007/JHEP08(2022)039}{\emph{JHEP} {\bfseries 08} (2022) 039} [\href{https://arxiv.org/abs/2205.03410}{{\ttfamily 2205.03410}}].

\bibitem{Mannel:2023yqf}
T.~Mannel, I.S.~Milutin and K.K.~Vos, \emph{{Inclusive semileptonic $ b\to c\ell \overline{\nu} $ decays to order $ 1/{m}_b^5 $}}, \href{https://doi.org/10.1007/JHEP02(2024)226}{\emph{JHEP} {\bfseries 02} (2024) 226} [\href{https://arxiv.org/abs/2311.12002}{{\ttfamily 2311.12002}}].

\bibitem{Egner:2023kxw}
M.~Egner, M.~Fael, K.~Sch{\"o}nwald and M.~Steinhauser, \emph{{Revisiting semileptonic B meson decays at next-to-next-to-leading order}}, \href{https://doi.org/10.1007/JHEP09(2023)112}{\emph{JHEP} {\bfseries 09} (2023) 112} [\href{https://arxiv.org/abs/2308.01346}{{\ttfamily 2308.01346}}].

\bibitem{Mannel:2021zzr}
T.~Mannel, D.~Moreno and A.A.~Pivovarov, \emph{{NLO QCD corrections to inclusive $b \rightarrow c \ell \bar{\nu}$decay spectra up to~$1/m_Q^3$}}, \href{https://doi.org/10.1103/PhysRevD.105.054033}{\emph{Phys. Rev. D} {\bfseries 105} (2022) 054033} [\href{https://arxiv.org/abs/2112.03875}{{\ttfamily 2112.03875}}].

\bibitem{Fael:2024gyw}
M.~Fael and F.~Herren, \emph{{NNLO QCD corrections to the q$^{2}$ spectrum of inclusive semileptonic B-meson decays}}, \href{https://doi.org/10.1007/JHEP05(2024)287}{\emph{JHEP} {\bfseries 05} (2024) 287} [\href{https://arxiv.org/abs/2403.03976}{{\ttfamily 2403.03976}}].

\bibitem{Finauri:2025ost}
G.~Finauri, \emph{{Kinematic moments of $ \overline{B}\to {X}_c\ell {\overline{\nu}}_{\ell } $ to order $ \mathcal{O}\left({\Lambda}_{\textrm{QCD}}^5/{m}_b^5\right) $}}, \href{https://doi.org/10.1007/JHEP04(2025)112}{\emph{JHEP} {\bfseries 04} (2025) 112} [\href{https://arxiv.org/abs/2501.09090}{{\ttfamily 2501.09090}}].

\bibitem{Tackmann:2024kci}
F.J.~Tackmann, \emph{{Beyond Scale Variations: Perturbative Theory Uncertainties from Nuisance Parameters}},  \href{https://arxiv.org/abs/2411.18606}{{\ttfamily 2411.18606}}.

\bibitem{Egner:2024lay}
M.~Egner, M.~Fael, A.~Lenz, M.L.~Piscopo, A.V.~Rusov, K.~Sch{\"o}nwald et~al., \emph{{Total decay rates of B mesons at NNLO-QCD}}, \href{https://doi.org/10.1007/JHEP04(2025)106}{\emph{JHEP} {\bfseries 04} (2025) 106} [\href{https://arxiv.org/abs/2412.14035}{{\ttfamily 2412.14035}}].

\bibitem{Fael:2024fkt}
M.~Fael, I.S.~Milutin and K.K.~Vos, \emph{{Kolya: An open-source package for inclusive semileptonic B decays}}, \href{https://doi.org/10.21468/SciPostPhysCodeb.55}{\emph{SciPost Phys. Codeb.} {\bfseries 55} (2025) 1} [\href{https://arxiv.org/abs/2409.15007}{{\ttfamily 2409.15007}}].

\bibitem{Gambino:2016jkc}
P.~Gambino, K.J.~Healey and S.~Turczyk, \emph{{Taming the higher power corrections in semileptonic B decays}}, \href{https://doi.org/10.1016/j.physletb.2016.10.023}{\emph{Phys. Lett. B} {\bfseries 763} (2016) 60} [\href{https://arxiv.org/abs/1606.06174}{{\ttfamily 1606.06174}}].

\bibitem{Mannel:2018mqv}
T.~Mannel and K.K.~Vos, \emph{{Reparametrization Invariance and Partial Re-Summations of the Heavy Quark Expansion}}, \href{https://doi.org/10.1007/JHEP06(2018)115}{\emph{JHEP} {\bfseries 06} (2018) 115} [\href{https://arxiv.org/abs/1802.09409}{{\ttfamily 1802.09409}}].

\bibitem{Gambino:2017vkx}
P.~Gambino, A.~Melis and S.~Simula, \emph{{Extraction of heavy-quark-expansion parameters from unquenched lattice data on pseudoscalar and vector heavy-light meson masses}}, \href{https://doi.org/10.1103/PhysRevD.96.014511}{\emph{Phys. Rev. D} {\bfseries 96} (2017) 014511} [\href{https://arxiv.org/abs/1704.06105}{{\ttfamily 1704.06105}}].

\bibitem{Gambino:2020crt}
P.~Gambino and S.~Hashimoto, \emph{{Inclusive Semileptonic Decays from Lattice QCD}}, \href{https://doi.org/10.1103/PhysRevLett.125.032001}{\emph{Phys. Rev. Lett.} {\bfseries 125} (2020) 032001} [\href{https://arxiv.org/abs/2005.13730}{{\ttfamily 2005.13730}}].

\bibitem{DeSantis:2025yfm}
A.~De~Santis et~al., \emph{{Inclusive Semileptonic Decays of the Ds Meson: Lattice QCD Confronts Experiments}}, \href{https://doi.org/10.1103/snc6-cpz6}{\emph{Phys. Rev. Lett.} {\bfseries 135} (2025) 121901} [\href{https://arxiv.org/abs/2504.06064}{{\ttfamily 2504.06064}}].

\bibitem{Kellermann:2025pzt}
R.~Kellermann, Z.~Hu, A.~Barone, A.~Elgaziari, S.~Hashimoto, T.~Kaneko et~al., \emph{{Inclusive semileptonic decays from lattice QCD: Analysis of systematic effects}}, \href{https://doi.org/10.1103/kltp-1p1f}{\emph{Phys. Rev. D} {\bfseries 112} (2025) 014501} [\href{https://arxiv.org/abs/2504.03358}{{\ttfamily 2504.03358}}].

\bibitem{Heinonen:2014dxa}
J.~Heinonen and T.~Mannel, \emph{{Improved Estimates for the Parameters of the Heavy Quark Expansion}}, \href{https://doi.org/10.1016/j.nuclphysb.2014.09.017}{\emph{Nucl. Phys. B} {\bfseries 889} (2014) 46} [\href{https://arxiv.org/abs/1407.4384}{{\ttfamily 1407.4384}}].

\bibitem{fael_2024_10818195}
{Fael, Matteo and Milutin, Ilija and Vos, K. Keri}, ``{Kolya 1.0}.'' \url{https://doi.org/10.5281/zenodo.10818194}, 2024.

\bibitem{Nir:1989rm}
Y.~Nir, \emph{{The Mass Ratio m(c) / m(b) in Semileptonic B Decays}}, \href{https://doi.org/10.1016/0370-2693(89)91495-0}{\emph{Phys. Lett. B} {\bfseries 221} (1989) 184}.

\bibitem{Becher:2006qw}
T.~Becher and M.~Neubert, \emph{{Toward a NNLO calculation of the anti-B ---{\ensuremath{>}} X(s) gamma decay rate with a cut on photon energy. II. Two-loop result for the jet function}}, \href{https://doi.org/10.1016/j.physletb.2006.04.046}{\emph{Phys. Lett. B} {\bfseries 637} (2006) 251} [\href{https://arxiv.org/abs/hep-ph/0603140}{{\ttfamily hep-ph/0603140}}].

\bibitem{Mannel:2015jka}
T.~Mannel, A.A.~Pivovarov and D.~Rosenthal, \emph{{Inclusive weak decays of heavy hadrons with power suppressed terms at NLO}}, \href{https://doi.org/10.1103/PhysRevD.92.054025}{\emph{Phys. Rev. D} {\bfseries 92} (2015) 054025} [\href{https://arxiv.org/abs/1506.08167}{{\ttfamily 1506.08167}}].

\bibitem{Gremm:1996df}
M.~Gremm and A.~Kapustin, \emph{{Order 1/m(b)**3 corrections to B --{\ensuremath{>}} X(c) lepton anti-neutrino decay and their implication for the measurement of Lambda-bar and lambda(1)}}, \href{https://doi.org/10.1103/PhysRevD.55.6924}{\emph{Phys. Rev. D} {\bfseries 55} (1997) 6924} [\href{https://arxiv.org/abs/hep-ph/9603448}{{\ttfamily hep-ph/9603448}}].

\bibitem{Trott:2004xc}
M.~Trott, \emph{{Improving extractions of |V(cb)| and m(b) from the hadronic invariant mass moments of semileptonic inclusive B decay}}, \href{https://doi.org/10.1103/PhysRevD.70.073003}{\emph{Phys. Rev. D} {\bfseries 70} (2004) 073003} [\href{https://arxiv.org/abs/hep-ph/0402120}{{\ttfamily hep-ph/0402120}}].

\bibitem{NNLO_El}
{Fael, Matteo and Herren, Florian and Sch\"onwald, Kay}. in preparation.

\bibitem{Becher:2007tk}
T.~Becher, H.~Boos and E.~Lunghi, \emph{{Kinetic corrections to $B \to X_{c} \ell \bar{\nu}$ at one loop}}, \href{https://doi.org/10.1088/1126-6708/2007/12/062}{\emph{JHEP} {\bfseries 12} (2007) 062} [\href{https://arxiv.org/abs/0708.0855}{{\ttfamily 0708.0855}}].

\bibitem{Bigi:1994em}
I.I.Y.~Bigi, M.A.~Shifman, N.G.~Uraltsev and A.I.~Vainshtein, \emph{{The Pole mass of the heavy quark. Perturbation theory and beyond}}, \href{https://doi.org/10.1103/PhysRevD.50.2234}{\emph{Phys. Rev. D} {\bfseries 50} (1994) 2234} [\href{https://arxiv.org/abs/hep-ph/9402360}{{\ttfamily hep-ph/9402360}}].

\bibitem{Beneke:1994sw}
M.~Beneke and V.M.~Braun, \emph{{Heavy quark effective theory beyond perturbation theory: Renormalons, the pole mass and the residual mass term}}, \href{https://doi.org/10.1016/0550-3213(94)90314-X}{\emph{Nucl. Phys. B} {\bfseries 426} (1994) 301} [\href{https://arxiv.org/abs/hep-ph/9402364}{{\ttfamily hep-ph/9402364}}].

\bibitem{Bigi:1994ga}
I.I.Y.~Bigi, M.A.~Shifman, N.G.~Uraltsev and A.I.~Vainshtein, \emph{{Sum rules for heavy flavor transitions in the SV limit}}, \href{https://doi.org/10.1103/PhysRevD.52.196}{\emph{Phys. Rev. D} {\bfseries 52} (1995) 196} [\href{https://arxiv.org/abs/hep-ph/9405410}{{\ttfamily hep-ph/9405410}}].

\bibitem{Bigi:1996si}
I.I.Y.~Bigi, M.A.~Shifman, N.~Uraltsev and A.I.~Vainshtein, \emph{{High power n of m(b) in beauty widths and n=5 ---{\ensuremath{>}} infinity limit}}, \href{https://doi.org/10.1103/PhysRevD.56.4017}{\emph{Phys. Rev. D} {\bfseries 56} (1997) 4017} [\href{https://arxiv.org/abs/hep-ph/9704245}{{\ttfamily hep-ph/9704245}}].

\bibitem{Mannel:2026vsz}
T.~Mannel, I.S.~Milutin, R.~Verkade and K.K.~Vos, \emph{{Extending the Kinetic Mass to Higher Orders in $1/m_Q$}},  \href{https://arxiv.org/abs/2602.21139}{{\ttfamily 2602.21139}}.

\bibitem{Buchalla:1995vs}
G.~Buchalla, A.J.~Buras and M.E.~Lautenbacher, \emph{{Weak Decays beyond Leading Logarithms}}, \href{https://doi.org/10.1103/RevModPhys.68.1125}{\emph{Rev. Mod. Phys.} {\bfseries 68} (1996) 1125} [\href{https://arxiv.org/abs/hep-ph/9512380}{{\ttfamily hep-ph/9512380}}].

\bibitem{King:2021jsq}
D.~King, A.~Lenz and T.~Rauh, \emph{{SU(3) breaking effects in B and D meson lifetimes}}, \href{https://doi.org/10.1007/JHEP06(2022)134}{\emph{JHEP} {\bfseries 06} (2022) 134} [\href{https://arxiv.org/abs/2112.03691}{{\ttfamily 2112.03691}}].

\bibitem{Neubert:1992fk}
M.~Neubert, \emph{{Symmetry breaking corrections to meson decay constants in the heavy quark effective theory}}, \href{https://doi.org/10.1103/PhysRevD.46.1076}{\emph{Phys. Rev. D} {\bfseries 46} (1992) 1076}.

\bibitem{Lenz:2014jha}
A.~Lenz, \emph{{Lifetimes and heavy quark expansion}}, \href{https://doi.org/10.1142/S0217751X15430058}{\emph{Int. J. Mod. Phys. A} {\bfseries 30} (2015) 1543005} [\href{https://arxiv.org/abs/1405.3601}{{\ttfamily 1405.3601}}].

\bibitem{Lenz:2022rbq}
A.~Lenz, M.L.~Piscopo and A.V.~Rusov, \emph{{Disintegration of beauty: a precision study}}, \href{https://doi.org/10.1007/JHEP01(2023)004}{\emph{JHEP} {\bfseries 01} (2023) 004} [\href{https://arxiv.org/abs/2208.02643}{{\ttfamily 2208.02643}}].

\bibitem{Albrecht:2024oyn}
J.~Albrecht, F.~Bernlochner, A.~Lenz and A.~Rusov, \emph{{Lifetimes of b-hadrons and mixing of neutral B-mesons: theoretical and experimental status}}, \href{https://doi.org/10.1140/epjs/s11734-024-01124-3}{\emph{Eur. Phys. J. ST} {\bfseries 233} (2024) 359} [\href{https://arxiv.org/abs/2402.04224}{{\ttfamily 2402.04224}}].

\bibitem{FlavourLatticeAveragingGroupFLAG:2024oxs}
{\scshape Flavour Lattice Averaging Group (FLAG)} collaboration, \emph{{FLAG Review 2024}},  \href{https://arxiv.org/abs/2411.04268}{{\ttfamily 2411.04268}}.

\bibitem{ParticleDataGroup:2024cfk}
{\scshape Particle Data Group} collaboration, \emph{{Review of particle physics}}, \href{https://doi.org/10.1103/PhysRevD.110.030001}{\emph{Phys. Rev. D} {\bfseries 110} (2024) 030001}.

\bibitem{Hocker:2001xe}
A.~Hocker, H.~Lacker, S.~Laplace and F.~Le~Diberder, \emph{{A New approach to a global fit of the CKM matrix}}, \href{https://doi.org/10.1007/s100520100729}{\emph{Eur. Phys. J. C} {\bfseries 21} (2001) 225} [\href{https://arxiv.org/abs/hep-ph/0104062}{{\ttfamily hep-ph/0104062}}].

\bibitem{Chetyrkin:2000yt}
K.G.~Chetyrkin, J.H.~Kuhn and M.~Steinhauser, \emph{{RunDec: A Mathematica package for running and decoupling of the strong coupling and quark masses}}, \href{https://doi.org/10.1016/S0010-4655(00)00155-7}{\emph{Comput. Phys. Commun.} {\bfseries 133} (2000) 43} [\href{https://arxiv.org/abs/hep-ph/0004189}{{\ttfamily hep-ph/0004189}}].

\bibitem{Schmidt:2012az}
B.~Schmidt and M.~Steinhauser, \emph{{CRunDec: a C++ package for running and decoupling of the strong coupling and quark masses}}, \href{https://doi.org/10.1016/j.cpc.2012.03.023}{\emph{Comput. Phys. Commun.} {\bfseries 183} (2012) 1845} [\href{https://arxiv.org/abs/1201.6149}{{\ttfamily 1201.6149}}].

\bibitem{Herren:2017osy}
F.~Herren and M.~Steinhauser, \emph{{Version 3 of RunDec and CRunDec}}, \href{https://doi.org/10.1016/j.cpc.2017.11.014}{\emph{Comput. Phys. Commun.} {\bfseries 224} (2018) 333} [\href{https://arxiv.org/abs/1703.03751}{{\ttfamily 1703.03751}}].

\bibitem{Gambino:2012rd}
P.~Gambino, T.~Mannel and N.~Uraltsev, \emph{{B-{\ensuremath{>}} D* Zero-Recoil Formfactor and the Heavy Quark Expansion in QCD: A Systematic Study}}, \href{https://doi.org/10.1007/JHEP10(2012)169}{\emph{JHEP} {\bfseries 10} (2012) 169} [\href{https://arxiv.org/abs/1206.2296}{{\ttfamily 1206.2296}}].

\bibitem{HFLAV:2024ctg}
{\scshape Heavy Flavor Averaging Group (HFLAV)} collaboration, \emph{{Averages of $b$-hadron, $c$-hadron, and $\tau$-lepton properties as of 2023}}, \href{https://doi.org/10.1103/x87q-tld5}{\emph{Phys. Rev.} {\bfseries D113} (2026) 012008} [\href{https://arxiv.org/abs/2411.18639}{{\ttfamily 2411.18639}}].

\bibitem{Ozcelik:2026ichep}
{\scshape LHCb} collaboration, \emph{{Beauty decays to hidden-charm final states at LHCb: First Run 3 measurements}}, {\emph{Presented at ICHEP 2026 (52nd International Conference on High Energy Physics)} (2026) }.

\bibitem{Mannel:2024uar}
T.~Mannel, D.~Moreno and A.A.~Pivovarov, \emph{{QCD corrections at subleading power for inclusive nonleptonic b{\textrightarrow}cu{\textasciimacron}d decays}}, \href{https://doi.org/10.1103/PhysRevD.110.094011}{\emph{Phys. Rev. D} {\bfseries 110} (2024) 094011} [\href{https://arxiv.org/abs/2408.06767}{{\ttfamily 2408.06767}}].

\bibitem{Mannel:2025fvj}
T.~Mannel, D.~Moreno and A.A.~Pivovarov, \emph{{QCD corrections for subleading powers in 1/mb for the nonleptonic b{\textrightarrow}cc{\textasciimacron}s transition}}, \href{https://doi.org/10.1103/PhysRevD.111.094035}{\emph{Phys. Rev. D} {\bfseries 111} (2025) 094035} [\href{https://arxiv.org/abs/2503.18775}{{\ttfamily 2503.18775}}].

\bibitem{Moretti:2026waq}
F.~Moretti, U.~Nierste, P.~Reeck and M.~Steinhauser, \emph{{Next-to-next-to-leading QCD corrections to the $\mathbf{B^+}$-$\mathbf{B_d^0}$, $\mathbf{D^+}$-$\mathbf{D^0}$, and $\mathbf{D_s^+}$-$\mathbf{D^0}$ lifetime ratios}},  \href{https://arxiv.org/abs/2604.24841}{{\ttfamily 2604.24841}}.

\bibitem{Bigi:2023cbv}
D.~Bigi, M.~Bordone, P.~Gambino, U.~Haisch and A.~Piccione, \emph{{QED effects in inclusive semi-leptonic B decays}}, \href{https://doi.org/10.1007/JHEP11(2023)163}{\emph{JHEP} {\bfseries 11} (2023) 163} [\href{https://arxiv.org/abs/2309.02849}{{\ttfamily 2309.02849}}].

\bibitem{UTfit:2005ras}
{\scshape UTfit} collaboration, \emph{{The 2004 UTfit collaboration report on the status of the unitarity triangle in the standard model}}, \href{https://doi.org/10.1088/1126-6708/2005/07/028}{\emph{JHEP} {\bfseries 07} (2005) 028} [\href{https://arxiv.org/abs/hep-ph/0501199}{{\ttfamily hep-ph/0501199}}].

\bibitem{Kirk:2017juj}
M.~Kirk, A.~Lenz and T.~Rauh, \emph{{Dimension-six matrix elements for meson mixing and lifetimes from sum rules}}, \href{https://doi.org/10.1007/JHEP12(2017)068}{\emph{JHEP} {\bfseries 12} (2017) 068} [\href{https://arxiv.org/abs/1711.02100}{{\ttfamily 1711.02100}}].

\bibitem{Black:2024bus}
M.~Black, M.~Lang, A.~Lenz and Z.~W{\"u}thrich, \emph{{HQET sum rules for matrix elements of dimension-six four-quark operators for meson lifetimes within and beyond the Standard Model}}, \href{https://doi.org/10.1007/JHEP04(2025)081}{\emph{JHEP} {\bfseries 04} (2025) 081} [\href{https://arxiv.org/abs/2412.13270}{{\ttfamily 2412.13270}}].

\bibitem{Mannel:2017jfk}
T.~Mannel, A.V.~Rusov and F.~Shahriaran, \emph{{Inclusive semitauonic $B$ decays to order ${\cal O} (\Lambda_{QCD}^3/m_b^3)$}}, \href{https://doi.org/10.1016/j.nuclphysb.2017.05.016}{\emph{Nucl. Phys. B} {\bfseries 921} (2017) 211} [\href{https://arxiv.org/abs/1702.01089}{{\ttfamily 1702.01089}}].

\bibitem{Rahimi:2022vlv}
M.~Rahimi and K.K.~Vos, \emph{{Standard Model predictions for lepton flavour universality ratios of inclusive semileptonic B decays}}, \href{https://doi.org/10.1007/JHEP11(2022)007}{\emph{JHEP} {\bfseries 11} (2022) 007} [\href{https://arxiv.org/abs/2207.03432}{{\ttfamily 2207.03432}}].

\bibitem{Jezabek:1996db}
M.~Jezabek and L.~Motyka, \emph{{Tau lepton distributions in semileptonic B decays}}, \href{https://doi.org/10.1016/S0550-3213(97)00341-6}{\emph{Nucl. Phys. B} {\bfseries 501} (1997) 207} [\href{https://arxiv.org/abs/hep-ph/9701358}{{\ttfamily hep-ph/9701358}}].

\bibitem{Fael:2019umf}
M.~Fael, T.~Mannel and K.K.~Vos, \emph{{The Heavy Quark Expansion for Inclusive Semileptonic Charm Decays Revisited}}, \href{https://doi.org/10.1007/JHEP12(2019)067}{\emph{JHEP} {\bfseries 12} (2019) 067} [\href{https://arxiv.org/abs/1910.05234}{{\ttfamily 1910.05234}}].

\bibitem{Gambino:2005tp}
P.~Gambino, G.~Ossola and N.~Uraltsev, \emph{{Hadronic mass and q**2 moments of charmless semileptonic B decay distributions}}, \href{https://doi.org/10.1088/1126-6708/2005/09/010}{\emph{JHEP} {\bfseries 09} (2005) 010} [\href{https://arxiv.org/abs/hep-ph/0505091}{{\ttfamily hep-ph/0505091}}].

\bibitem{vanRitbergen:1999gs}
T.~van Ritbergen, \emph{{The Second order QCD contribution to the semileptonic b ---{\ensuremath{>}} u decay rate}}, \href{https://doi.org/10.1016/S0370-2693(99)00407-4}{\emph{Phys. Lett. B} {\bfseries 454} (1999) 353} [\href{https://arxiv.org/abs/hep-ph/9903226}{{\ttfamily hep-ph/9903226}}].

\bibitem{Fael:2023tcv}
M.~Fael and J.~Usovitsch, \emph{{Third order correction to semileptonic $b\to u$ decay: Fermionic contributions}}, \href{https://doi.org/10.1103/PhysRevD.108.114026}{\emph{Phys. Rev. D} {\bfseries 108} (2023) 114026} [\href{https://arxiv.org/abs/2310.03685}{{\ttfamily 2310.03685}}].

\bibitem{Chen:2023dsi}
L.-B.~Chen, H.T.~Li, Z.~Li, J.~Wang, Y.~Wang and Q.-f.~Wu, \emph{{Analytic third-order QCD corrections to top-quark and semileptonic b{\textrightarrow}u decays}}, \href{https://doi.org/10.1103/PhysRevD.109.L071503}{\emph{Phys. Rev. D} {\bfseries 109} (2024) L071503} [\href{https://arxiv.org/abs/2309.00762}{{\ttfamily 2309.00762}}].

\bibitem{Chen:2026jwl}
L.~Chen, X.~Chen, X.~Guan and Y.-Q.~Ma, \emph{{Heavy-to-light Structure Functions at $\mathcal{O}(\alpha_s^3)$ in QCD}},  \href{https://arxiv.org/abs/2602.11879}{{\ttfamily 2602.11879}}.

\bibitem{Bagan:1994zd}
E.~Bagan, P.~Ball, V.M.~Braun and P.~Gosdzinsky, \emph{{Charm quark mass dependence of QCD corrections to nonleptonic inclusive B decays}}, \href{https://doi.org/10.1016/0550-3213(94)90591-6}{\emph{Nucl. Phys. B} {\bfseries 432} (1994) 3} [\href{https://arxiv.org/abs/hep-ph/9408306}{{\ttfamily hep-ph/9408306}}].

\bibitem{Bagan:1995yf}
E.~Bagan, P.~Ball, B.~Fiol and P.~Gosdzinsky, \emph{{Next-to-leading order radiative corrections to the decay $b\to c c s$}}, \href{https://doi.org/10.1016/0370-2693(95)00437-P}{\emph{Phys. Lett. B} {\bfseries 351} (1995) 546} [\href{https://arxiv.org/abs/hep-ph/9502338}{{\ttfamily hep-ph/9502338}}].

\bibitem{Egner:2024azu}
M.~Egner, M.~Fael, K.~Sch{\"o}nwald and M.~Steinhauser, \emph{{Nonleptonic B-meson decays to next-to-next-to-leading order}}, \href{https://doi.org/10.1007/JHEP10(2024)144}{\emph{JHEP} {\bfseries 10} (2024) 144} [\href{https://arxiv.org/abs/2406.19456}{{\ttfamily 2406.19456}}].

\bibitem{Lenz:1997aa}
A.~Lenz, U.~Nierste and G.~Ostermaier, \emph{{Penguin diagrams, charmless B decays and the missing charm puzzle}}, \href{https://doi.org/10.1103/PhysRevD.56.7228}{\emph{Phys. Rev. D} {\bfseries 56} (1997) 7228} [\href{https://arxiv.org/abs/hep-ph/9706501}{{\ttfamily hep-ph/9706501}}].

\bibitem{Lenz:1998qp}
A.~Lenz, U.~Nierste and G.~Ostermaier, \emph{{Determination of the CKM angle gamma and |V(ub) / V(cb)| from inclusive direct CP asymmetries and branching ratios in charmless B decays}}, \href{https://doi.org/10.1103/PhysRevD.59.034008}{\emph{Phys. Rev. D} {\bfseries 59} (1999) 034008} [\href{https://arxiv.org/abs/hep-ph/9802202}{{\ttfamily hep-ph/9802202}}].

\bibitem{Krinner:2013cja}
F.~Krinner, A.~Lenz and T.~Rauh, \emph{{The inclusive decay $b \to c\bar{c}s$ revisited}}, \href{https://doi.org/10.1016/j.nuclphysb.2013.07.028}{\emph{Nucl. Phys. B} {\bfseries 876} (2013) 31} [\href{https://arxiv.org/abs/1305.5390}{{\ttfamily 1305.5390}}].

\bibitem{Lenz:2020oce}
A.~Lenz, M.L.~Piscopo and A.V.~Rusov, \emph{{Contribution of the Darwin operator to non-leptonic decays of heavy quarks}}, \href{https://doi.org/10.1007/JHEP12(2020)199}{\emph{JHEP} {\bfseries 12} (2020) 199} [\href{https://arxiv.org/abs/2004.09527}{{\ttfamily 2004.09527}}].

\bibitem{Blok:1992he}
B.~Blok and M.A.~Shifman, \emph{{The Rule of discarding 1/N(c) in inclusive weak decays. 2.}}, \href{https://doi.org/10.1016/0550-3213(93)90505-J}{\emph{Nucl. Phys. B} {\bfseries 399} (1993) 459} [\href{https://arxiv.org/abs/hep-ph/9209289}{{\ttfamily hep-ph/9209289}}].

\bibitem{Blok:1992hw}
B.~Blok and M.A.~Shifman, \emph{{The Rule of discarding 1/N(c) in inclusive weak decays. 1.}}, \href{https://doi.org/10.1016/0550-3213(93)90504-I}{\emph{Nucl. Phys. B} {\bfseries 399} (1993) 441} [\href{https://arxiv.org/abs/hep-ph/9207236}{{\ttfamily hep-ph/9207236}}].

\bibitem{Bigi:1992ne}
I.I.Y.~Bigi, B.~Blok, M.A.~Shifman, N.G.~Uraltsev and A.I.~Vainshtein, \emph{{A QCD 'manifesto' on inclusive decays of beauty and charm}},  in \emph{{7th Meeting of the APS Division of Particles Fields}}, pp.~610--613, 11, 1992 [\href{https://arxiv.org/abs/hep-ph/9212227}{{\ttfamily hep-ph/9212227}}].

\bibitem{Franco:2002fc}
E.~Franco, V.~Lubicz, F.~Mescia and C.~Tarantino, \emph{{Lifetime ratios of beauty hadrons at the next-to-leading order in QCD}}, \href{https://doi.org/10.1016/S0550-3213(02)00262-6}{\emph{Nucl. Phys. B} {\bfseries 633} (2002) 212} [\href{https://arxiv.org/abs/hep-ph/0203089}{{\ttfamily hep-ph/0203089}}].

\bibitem{Beneke:2002rj}
M.~Beneke, G.~Buchalla, C.~Greub, A.~Lenz and U.~Nierste, \emph{{The $B^+ -B^0_d$ Lifetime Difference Beyond Leading Logarithms}}, \href{https://doi.org/10.1016/S0550-3213(02)00561-8}{\emph{Nucl. Phys. B} {\bfseries 639} (2002) 389} [\href{https://arxiv.org/abs/hep-ph/0202106}{{\ttfamily hep-ph/0202106}}].

\bibitem{Czarnecki:1997dz}
A.~Czarnecki and A.G.~Grozin, \emph{{HQET chromomagnetic interaction at two loops}}, \href{https://doi.org/10.1016/S0370-2693(97)00587-X}{\emph{Phys. Lett. B} {\bfseries 405} (1997) 142} [\href{https://arxiv.org/abs/hep-ph/9701415}{{\ttfamily hep-ph/9701415}}].

\end{thebibliography}\endgroup

\end{document}